\documentclass[12pt]{article}
\usepackage{amsmath,amssymb,graphicx,multirow,oldgerm,euscript,mathrsfs}
\def\3half{\textstyle\frac32}

\begin{document}
\begin{titlepage}
\begin{flushright}
JLAB-THY-04-42 \\
 ${\rm ECT}^{*}$-04-11\\
\end{flushright}
\renewcommand{\thefootnote}{\fnsymbol{footnote}}
\vspace{1.0em}
\begin{center}
{\bf \LARGE{Direct $\boldsymbol{CP}$ Violation, Branching Ratios and  Form Factors $\boldsymbol{B  \rightarrow \pi}$,
 $\boldsymbol{B  \rightarrow K}$  in $\boldsymbol{B}$ Decays}}
\end{center}
\vspace{1.0em}
\begin{center}
\begin{large}
O. Leitner$^{1}$\footnote{leitner@ect.it}, 
X.-H. Guo$^{2}$\footnote{xhguo@bnu.edu.cn},
A.W. Thomas$^{3}$\footnote{awthomas@jlab.org} \\
\end{large}
\vspace{1.3em}
$^1$ ${\rm{ECT}}^{*}$, Strada delle Tabarelle, 286, 
38050 Villazzano (Trento), Italy \\
\vspace{0.5em}
 $^2$  Key Laboratory of Radiation Beam Technology
and Material Modification of National Ministry of Education, and
Institute of Low Energy Nuclear Physics, Beijing Normal
University, Beijing 100875, China\\
$^3$  Jefferson Lab, 12000 Jefferson Avenue, Newport News, VA 23606, USA   
\end{center}
\vspace{3.0em}
\begin{abstract}
\vspace{1.0em}
The $B \to \pi$ and $B \to K$ transitions involved in hadronic $B$ decays are 
investigated in a phenomenological way through the framework of QCD factorization. 
By comparing our results  with experimental branching ratios from the BELLE, BABAR 
and CLEO Collaborations for all the $B$ decays including either a pion or a kaon, 
we  propose  boundaries for  the transition form factors 
$B \to \pi$ and $B \to K$ 
depending on the CKM matrix element parameters $\rho$ and $\eta$. From this analysis, 
the form factors required to reproduce  the experimental data for  branching ratios 
are  $F^{B \to \pi}= 0.31 \pm 0.12$ and  $F^{B \to K}= 0.37\pm 0.13$. 
We calculate the direct $CP$ violating asymmetry parameter, 
$a_{CP}$, for $B \to \pi^{+} \pi^{-} \pi$ and $B \to \pi^{+} \pi^{-} K$ decays, in the 
case where $\rho-\omega$ mixing effects are taken into account. Based on these results, 
we find that the direct $CP$ asymmetry for $B^{-} \to \pi^{+} \pi^{-} \pi^{-}$, 
$\bar{B}^{0} \to \pi^{+} \pi^{-} 
\pi^{0}$, $B^{-} \to \pi^{+} \pi^{-} K^{-}$, 
and $\bar{B}^{0} \to \pi^{+} \pi^{-} 
\bar{K}^{0}$, reaches its maximum when the invariant mass $\pi^{+} \pi^{-}$ is in 
the vicinity of the $\omega$ meson mass. The inclusion of  $\rho-\omega$ mixing 
provides an opportunity to erase, without ambiguity, the phase uncertainty 
mod$(\pi)$ in the determination of the CKM angles $\alpha$ in case of $b\to u$ 
and $\gamma$ in case of $b \to s$.
\end{abstract}
\vspace{7.em}
\end{titlepage}
\newpage

%
%
%
%
%
%
%
%
%
%
%
\section{Introduction}

The violation of $CP$\footnote{$C$ and $P$ denote respectively 
the charge conjugation
and parity transformation.} symmetry in the framework of the 
Standard Model is supposed  to arise  {}from the well-known 
Cabibbo-Kobayashi-Maskawa (CKM)~\cite{Cabibbo:1963yz,Kobayashi:1973fv} 
matrix through a  ``weak'' phase, $\phi$. This  matrix is based on 
the quark flavors, charged currents and  weak 
interaction properties {\it within} the Standard 
Model. In order to make that picture as accurate as 
possible and to look for 
complementary  insight from New Physics (NP), $B$ 
meson decays~\cite{Carter:1980tk,Bigi:1981qs,Guo:2000mz} 
have been investigated extensively, for many years, in  
both  experimental and theoretical ways.

On the experimental side, branching ratios (of order  $O(10^{-6})$)
for  $B$ charmless or charm (exclusive non leptonic) decays have been measured more and more 
accurately thanks to three main $e^+ e^-$ colliders (list non exhaustive) operating at 
the $\Upsilon (4S)$ resonance: CLEO (symmetric-energy $B$-factory), BELLE and BABAR 
(asymmetric-energy $B$-factories). Despite a small number of 
discrepancies in some of the measurements
of branching ratios, these $B$-facilities provide direct 
and/or indirect constraints on the 
unitarity triangle (UT) parameters $\rho$ and $\eta$. The latter parameters (known at the level
of accuracy of $10\%$) are the corner stone of the whole $CP$ violation theory. They are subject
to tremendous efforts into their experimental determination since all of the $B$-branching ratios
($CP$ violating asymmetries as well) are directly related to them within the theoretical formalism.
The experimental determination of $CP$ violation (direct or indirect) is not as easy
as for  branching ratios because of the competition (sum or difference) between different
topologies suffering from large hadronic uncertainties. However, since its discovery in 
neutral kaon decay in 1964~\cite{Christenson:1964fg} and in neutral $B$ decay in 
2001~\cite{Aubert:2001nu,Abe:2001xe}, its measurement has been regularly
improved. The main quantities which are directly measured are $\epsilon, \epsilon^{\prime}$ for 
kaon decays and $\sin 2 \alpha, \sin 2 \beta$ for $B$ decays. Altogether they mainly restrict the
allowed range of values~\cite{Charles:2004jd} for the parameters $\rho$ and $\eta$ [$0.077<\rho<0.266, 
\;\;  0.298<\eta<0.406$ at 95\%  confidence level and $0.130<\rho<0.226,\;\;  0.324<\eta<0.377$ 
at 68\%  confidence level].

On the theoretical side, direct $CP$ violating asymmetries in $B$ decays occur through the 
interference of, at least, two amplitudes with different weak phase, $\phi$, and strong 
phase, $\delta$. The extraction of the weak phase $\phi$ (which is determined by a combination
of CKM matrix elements) is made through the measurement of a $CP$ violating asymmetry. However,
one must know the strong phase $\delta$ which is not still well determined in any theoretical
framework. $B$ non-leptonic decay amplitudes involve hadronic matrix elements,
$\langle M_1 M_2 | O_i| B \rangle$, and  their computation is not trivial since it requires that one 
factorizes hadronic matrix elements of four quark operators. 
Assuming that they are saturated by a vacuum
intermediate state, it reduces to the product of two matrix
elements of two quark currents, corresponding to the weak
transition form factor for $B \rightarrow M_1$ ($M_2$) and the
decay constant of $M_2$ ($M_1$). In addition to this ``naive'' image 
(naive factorization)~\cite{Fakirov:1978ta,Cabibbo:1978zv,Dugan:1991de} justified in 
the large $N_c$-limit, radiative non-factorizable corrections (at  order  $O(\alpha_s)$) 
coming from the light quark spectator of the $B$ meson are included in the QCD factorization~\cite{
Beneke:2001ev,Sachrajda:2001uv,Beneke:1999br,Neubert:2001cq,Neubert:2000kk,Beneke:2000pw,Beneke:2002nj,
Beneke:1999gt,Beneke:2000fw} framework\footnote{For completeness,
let us mention the perturbative hard scattering approach (PQCD)~\cite{Li:1996iu,Keum:2000ph,Keum:2000ms,Keum:2003qi} 
and soft collinear effective theory (SCET)~\cite{Bauer:2002sh,Bauer:2001cu}.}, where the main uncertainty 
comes from the $O(\Lambda_{QCD}/m_B)$ terms.

It is known that in order to have a large signal for $CP$   violation, we have
to appeal to some phenomenological mechanism to obtain a large strong phase $\delta$. 
In this regard, the isospin
symmetry violating mixing between $\rho^{0}$ and $\omega$ can be extremely important,
since it can lead to a large $CP$ violation in $B$ decays such as 
$B \to \rho^0(\omega) Y \to \pi^+ \pi^- Y$  ($Y$ defines a meson) because the strong 
phase passes through $90^{o}$ at the $\omega$ resonance~\cite{Enomoto:1996cv,Gardner:1998yx,
Guo:1998eg,Guo:1999ip}. In addition, as the $\omega$ is very narrow, one is essentially
guaranteed that the strong phase will pass through $\pi/2$ very close to the $\omega$-mass regardless
of other background sources of strong phase.

In this paper, we first investigate  in a phenomenological way, the dependence on the form
factors $F^{B \to \pi}$ and $F^{B \to K}$ of all the branching ratios for $B$ decaying
into $\pi X$ or $K X$, where $X$ is either a pseudo-scalar ($\pi, K, \eta^{(\prime)}$), 
or a vector ($\rho, \omega, K^{*}, \phi$) meson. At the same time, we determine, by making 
comparison between experimental and theoretical results,  a range  of values for the 
hard-scattering (annihilation) phases, $\varphi_{H} (\varphi_{A})$ and parameters $\varrho_{H} 
(\varrho_{A})$, respectively. These phases, $\varphi_{H,A}$, and  parameters, $\varrho_{H,A}$, arise
in QCD factorization because of divergent endpoint integrals when the hard scattering
and annihilation contributions are calculated. The analysis is performed  
using all the latest data for $b \to s$ and $b \to u$ 
transitions, concentrating on the CLEO, BABAR and BELLE branching ratio results. 
Based on this investigation, we study the sign of $\sin \delta$, the ratio between tree and 
penguin contributions and finally the direct  $CP$ violating asymmetry in $B$ decays 
such as $B \to \pi^+ \pi^- Y$, where $Y=\{\pi^{0,\pm},K^{0,\pm}\}$.

The reminder of this paper is organized as it follows. 
In section 2, we present the form
of the low energy effective Hamiltonian based on the 
operator product expansion. In section 3
we recall the naive factorization procedure and  
give the QCD factorization formalism based
on an expansion of naive factorization in 
term of $\Lambda_{QCD}/m_B$. Section 4 focuses on 
the $\rho^{0} - \omega$, $\eta - \eta^{\prime}$ 
mixing schemes as well as on the $CP$ violation
formalism. In sections 5 and 6, the CKM matrix 
and all the necessary input parameters (form 
factor, decay constant, masses) are discussed in detail, 
respectively. The following section 
is devoted to results and discussions for $B$ branching 
ratios and direct $CP$ violating 
asymmetries. Finally, in the last section, we  draw some conclusions.

%
\section{Effective  Hamiltonian}\label{sec10.2}
%
%

In any phenomenological treatment of the weak decays of hadrons, the starting point is the 
weak effective Hamiltonian at low energy~\cite{Buras:1999rb, Buras:1998ra, Buchalla:1996vs,
Stech:1997ij, Buras:1995iy}. It is obtained by integrating out the heavy fields (e.g. the top
quark, $W^{\pm}$ and $Z$ bosons) from the Standard Model Lagrangian. It can be written as,
%
\begin{equation}\label{eq1}
{\cal H}_{eff}^{\bigtriangleup B=1}=\frac {G_{F}}{\sqrt 2}
\sum_{i} V_{CKM} C_{i} (\mu)  O_i (\mu) \ ,
\end{equation}
where $G_{F}$ is the Fermi constant, $V_{CKM}$ is the CKM matrix element, 
$C_{i}(\mu)$ are the Wilson coefficients, $O_i(\mu)$ are the operators entering 
the Operator Product Expansion (OPE) and $\mu$ represents the renormalization scale.
 In the present case, since we take into account both tree and penguin operators,  
the matrix elements of the effective weak Hamiltonian reads as,
%
\begin{multline}\label{eq2}
\langle M_1 M_2|{\cal H}_{eff}^{\bigtriangleup B=1}| B \rangle= \\
\frac {G_{F}}{\sqrt 2} \Biggl[  V_{ub}V_{uq}^{*} 
\sum_{i=1}^{2} C_{i}(\mu)\langle M_1 M_2 |O_{i}^{q}|B
 \rangle(\mu) 
- V_{tb}V_{tq}^{*} \sum_{i=3}^{10} C_{i}(\mu)\langle M_1 M_2 |O_{i}|B 
\rangle(\mu) \Biggr] + h.c.\ ,
\end{multline}
where $q=d \; {\rm or} \; s$ according to the transition $b \to u$ or $b \to s$. 
 $\langle M_1 M_2 |O_{i}| B \rangle$ are the hadronic matrix elements, and   $M_i M_j$  
indicates  either  a pseudo-scalar and a vector  in the final state, or  two pseudo-scalar 
mesons in the final state.  The matrix elements  describe the transition between initial  
and final states at scales lower than $\mu$ and include, up to now, the main  uncertainties 
in the calculation because  they involve   non-perturbative physics.
The Operator Product Expansion   is 
used  to separate the calculation of the amplitude, $A(M \rightarrow F)\propto C_{i}(\mu) \langle F 
| O_{i} | M \rangle (\mu)$,  into two distinct physical regimes. One is called {\it hard} or
short-distance physics, represented by $C_{i}(\mu)$ and calculated  by a perturbative approach. 
The other  is called {\it soft} or long-distance physics. This part is described by 
$O_{i}(\mu)$, and is derived by using a non-perturbative approach such as the $1/N_c$ 
expansion~\cite{Novikov:1985rf}, QCD sum rules~\cite{Shifman:1979bx, Shifman:1979by} or 
hadronic sum rules.

\noindent The operators, $O_{i}(\mu)$, can be understood as local operators which govern 
effectively a given decay, reproducing the weak interaction of quarks in a point-like
 approximation. The definitions 
of the operators $O_{i}$~\cite{Buras:1999rb,Buras:1998ra,Buchalla:1996vs} are recalled for completeness,
\newline
\noindent - Current-current operators:
\begin{eqnarray}\label{eq3}
O_1^p =& (\bar p_{\alpha} b_{\alpha})_{V-A} (\bar q_{\beta} p_{\beta})_{V-A}\ ,
\hspace{1.65cm}& \hspace{-0.15cm}
O^p_2 = (\bar p_{\alpha}  b_{\beta})_{V-A} (\bar q_{\beta} p_{\alpha})_{V-A}\ , 
\end{eqnarray}
\noindent - QCD-penguin operators:
\begin{eqnarray}\label{eq4}
O_3  =& (\bar q_{\alpha} b_{\alpha})_{V-A} \sum{}_{\!q^{\prime}}\,(\bar q^{\prime}_{\beta} q_{\beta})_{V-A}\ ,
\hspace{0.7cm}& \hspace{-0.15cm}
O_4 = (\bar q_{\alpha} b_{\beta})_{V-A} \sum{}_{\!q^{\prime}}\,(\bar q^{\prime}_{\beta} q_{\alpha})_{V-A}\ ,
\nonumber \\
O_5  =& (\bar q_{\alpha} b_{\alpha})_{V-A} \sum{}_{\!q^{\prime}}\,(\bar q^{\prime}_{\beta} q_{\beta})_{V+A}\ , 
\hspace{0.7cm}&\hspace{-0.15cm}
O_6 = (\bar q_{\alpha} b_{\beta})_{V-A} \sum{}_{\!q^{\prime}}\,(\bar q^{\prime}_{\beta} q_{\alpha})_{V+A}\ ,
\end{eqnarray}
\noindent - Electroweak penguin operators:
\begin{eqnarray}\label{eq5}
O_7  =& (\bar q_{\alpha} b_{\alpha})_{V-A} \sum{}_{\!q^{\prime}}\,{\textstyle\frac32} e_q^{\prime} 
(\bar q^{\prime}_{\beta} q_{\beta})_{V+A}\ , 
\hspace{0.09cm}&  \hspace{-0.15cm}
O_8 = (\bar q_{\alpha} b_{\beta})_{V-A} \sum{}_{\!q^{\prime}}\,{\textstyle\frac32} e_q^{\prime}
(\bar q^{\prime}_{\beta} q_{\alpha})_{V+A}\ , \nonumber \\
O_9 =& (\bar q_{\alpha} b_{\alpha})_{V-A} \sum{}_{\!q^{\prime}}\,{\textstyle\frac32} e_q^{\prime}
(\bar q^{\prime}_{\beta} q_{\beta})_{V-A}\ , 
\hspace{0.1cm}& \hspace{-0.3cm}
O_{10} = (\bar q_{\alpha} b_{\beta})_{V-A} \sum{}_{\!q^{\prime}}\,{\textstyle\frac32} e_q^{\prime}
(\bar q^{\prime}_{\beta} q_{\alpha})_{V-A}\ , 
\end{eqnarray}
where $(\bar q_1 q_2)_{V\pm A}=\bar q_1\gamma_\mu(1\pm\gamma_5)q_2$, 
$\alpha,\beta$ are colour indices, $e_q^{\prime}$ are the electric charges of the quarks 
in units of $|e|$, and a summation over all the active quarks, 
$q^{\prime}=u,d,s,c$, is implied. In Eq.~(\ref{eq3}) $p$ denotes the quark $u$ or $c$ 
and $q$ denotes the quark $u$ or $s$, according to the given transition $b \to u$ or $b \to s$. 
Finally, expressions for the operators $O_{7\gamma}$ and $O_{8g}$  are,
\begin{eqnarray}
O_{7\gamma} &=& \frac{-e}{8\pi^2}\,m_b\, 
\bar s\sigma_{\mu\nu}(1+\gamma_5) F^{\mu\nu} b\ , \nonumber \\
O_{8g} &=&   \frac{-g_s}{8\pi^2}\,m_b\, 
\bar s\sigma_{\mu\nu}(1+\gamma_5) G^{\mu\nu} b\ . 
\label{eq6}
\end{eqnarray}
In Eq.~(\ref{eq6}) the definition of the dipole operators $O_{7\gamma}$ and $O_{8g}$
corresponds to the sign convention applied for the gauge-covariant derivative, $i D^{\mu} 
= i \delta^{\mu} + g_{s}  A^{\mu}_{a} t_{a}$.

The Wilson coefficients~\cite{Buras:1998ra}, $C_{i}(\mu)$,  represent the physical contributions
from scales higher than $\mu$ (through OPE one can separate short-distance and long-distance
contributions). Since  QCD has the property of  asymptotic freedom, they can be calculated in perturbation
theory. The Wilson coefficients include the contributions of all heavy particles, such as 
the top quark, the $W^{\pm}$ bosons, and the charged Higgs boson. Usually, the scale $\mu$ is 
chosen to be of order   ${\cal O}(m_{b})$ for $B$ decays. Their calculation will be discussed in
the following section.
%
%
\section{Factorization}
%
%
\subsection{Naive factorization}
%
%
The computation of  the hadronic matrix elements, $\langle M_1 M_2 |O_{i}| B \rangle$,
is not  trivial and requires some assumptions. The first general method is the so-called 
``factorization''\footnote{The $SU(N_{c})$ colour-algebra relation 
\begin{equation}\label{eq7}
T_{\alpha \beta}^{a} T_{\gamma \delta}^{a}= \frac12 \left( \delta_{\alpha\delta} \delta_{\beta \gamma}
-\frac{1}{N_{c}^{eff}} \delta_{\alpha\delta} \delta_{\beta \gamma}\right)\ ,\nonumber
\end{equation}
is required for the factorization procedure. By reordering the colour indices and 
including the  octet-octet current operators (yielding from Fierz transformation)
through   the variable $\xi_{i}$ (since they are non-factorizable), the result  takes 
into account the colour-allowed and colour-suppressed contributions which can 
occur in the decay at  the tree level. $N_{c}^{eff}$ ($N_{c}^{eff}$ is the effective number of  colours)
 is defined as a parameter which, by assumption, 
includes all hadronization effects which  cannot be  factorized completely and is written   as,
\begin{equation}\label{eq8}
\frac{1}{(N_{c}^{eff})_{i}} = \frac{1}{3} + \xi_{i}\ , {\rm with} \; i=1, \cdots ,10 \ ,\nonumber
\end{equation}
where  it is assumed that $N_{c}^{eff}$ is the same for all operators $O_{i}$.}
procedure~\cite{Fakirov:1978ta, Cabibbo:1978zv, Dugan:1991de}, 
in which one  approximates  the matrix element as a product of a transition matrix element
between a $B$ meson and one final state meson and  a matrix element which describes the 
creation of the second meson from the vacuum. In other words, the hadronic matrix elements
are expressed in terms of  form factor, $F^{B \to M_{i}}$, times  decay constant, $f_{M_{j}}$.  
This can be formulated as,
\begin{align}\label{eq9}
\langle M_1 M_2 |O_{i}| B \rangle 
=& \langle M_1  | J_{2i}^{\mu}| 0 \rangle\ \langle M_2  | J_{\mu,1i}| B \rangle   
\propto f_{M_{1}} F^{B \to M_{2}}\ , \nonumber \\
{\rm or}\;\; \langle M_1 M_2 |O_{i}| B \rangle 
=& \langle M_2  | J_{4i}^{\mu}| 0 \rangle\ \langle M_1  | J_{\mu,3i}| B \rangle
\propto f_{M_{2}} F^{B \to M_{1}}\ ,
\end{align}
where the $J_{\mu,ji}$ are the transition currents such as $\bar{q}_j \gamma_\mu (1-\gamma_5)q_i$. 
This approach is known as {\it naive} factorization since it factorizes  
$\langle M_1 M_2 |O_{i}| B \rangle$ into a simple product of two quark matrix elements.
Therefore, the amplitude for a given decay can be finally written  as,
\begin{multline}\label{eq10}
A(B \rightarrow M_1 M_2)  \propto  \frac{G_F}{\sqrt{2}}\Biggl[ \sum_{i=1}^{10}  
V_{CKM} C_{i}(\mu) \langle M_{1}M_{2}  | O_{i}| B \rangle(\mu) \Biggr] \\
 \propto  \frac{G_F}{\sqrt{2}} \Biggl[ \sum_{i=1}^{10}  V_{CKM}  C_{i}(\mu) 
\langle M_{1}  | J_{2i}^{\mu}| 0 \rangle \langle M_{2}  | J_{\mu,1i}| B \rangle  \Biggr]\ .
\end{multline}
A possible  justification for this approximation has been given by Bjorken~\cite{Bjorken:1989kk}:
the  heavy quark decays are very energetic, so the quark-antiquark pair in a meson in the  final 
state moves very fast away from the localized weak interaction. The hadronization of the 
quark-antiquark pair occurs far away from the remaining  quarks. Then, the meson can be factorized
out and the interaction between the quark pair in the meson and the remaining quark is  tiny.

The main uncertainty in this approach is that  the  final state interactions (FSI)  are  
neglected. Corrections associated with the factorization hypothesis are parameterized
 and hence there may be large uncertainties~\cite{Quinn:1999yq}. 
Moreover, the $\mu$-scale dependence of the amplitude does not match on the two sides of 
Eq.~(\ref{eq10}). Finally, the lack of a dynamical mechanism for creating a strong phase
may lead to  discrepancies in comparison with experimental data. 
In spite of this, there are indications that the approach  should give, at least,  a good estimate of 
the magnitude of the $B$ decay amplitude in many cases~\cite{Cheng:1994zx, Cheng:1997xy}.

Working  in the Quark Model scheme, the matrix elements for $B \to M_{i}$ and $B \to M_{i}^{*}$ 
(where $M_{i}$ and $M_{i}^{*}$ denote pseudo-scalar and vector mesons, respectively) can be 
decomposed as  follows for the pseudo-scalar-pseudo-scalar transition~\cite{Bauer:1987bm,Wirbel:1985ji}:
\begin{equation}\label{eq11}
{ \langle M_{i}|J_{\mu}|B \rangle} =\left( p_{B} + p_{M_{i}}- \frac{m_{B}^{2}-m_{M_{i}}^{2}}{k^{2}}
k \right)_{\mu} { F_{1}(k^{2})}+\frac{m_{B}^{2}-m_{M_{i}}^{2}}{k^{2}}k_{\mu}{ F_{0}(k^{2})}\ , 
\end{equation}
where $J_{\mu}$ is the weak current defined as $J_{\mu}=\bar{q} \gamma_{\mu} (1-\gamma_{5})b$ with
 $q={u,d,s}$ and $k=P_{B}-P_{M_{i}(M_{i}^{*})}$. $F_{0}, F_{1}$ are the form factors related to the 
transition $0^{-} \to 0^{-}$. In the helicity basis, they represent, respectively,  the transition amplitudes 
corresponding to the exchange of a vector and a scalar boson in the t-channel. For the vector 
transition, one has:
\begin{multline}\label{eq12}
{ \langle M^{\star}_{i}|J_{\mu}|B \rangle}=\frac{2}{m_{B}+m_{M_{i}^{\star}}} \epsilon_{\mu \nu \rho \sigma}
\epsilon^{\star \nu} p_{B}^{\rho} p_{M_{i}^{\star}}^{\sigma}  { V(k^{2})} +i \Biggl\{ \epsilon_{\mu}^{\star}(m_{B}
+m_{M_{i}^{\star}}){ A_{1}(k^{2})}  \\
- \frac{\epsilon^{\star}  \cdot k}{m_{B}+m_{M_{i}^{\star}}} (P_{B}+P_{M_{i}^{\star}})_{\mu}{ A_{2}(k^{2})}- \frac
{\epsilon^{\star}  \cdot k}{k^{2}}2m_{M_{i}^{\star}} \cdot k_{\mu}{  A_{3}(k^{2})}
\Biggr\}  \\
+i \frac{\epsilon^{\star} \cdot k}{ k^{2}}2m_{M_{i}^{\star}} \cdot k_{\mu}{  A_{0}(k^{2})}\ , 
\end{multline}
where $\epsilon_{\mu}$ is the polarization of $M_{i}^{*}$ with the condition $\epsilon_{\mu} 
\cdot P_{M_{i}^{*}}^{\mu}=0$. $A_{0}, A_{1}, A_{2}, A_{3}$ and $V$ are the form factors which 
describe the axial and vector transitions $0^{-} \to 1^{-}$, respectively. $A_{1}(k^{2})$ and 
$A_{2}(k^{2})$ are related to the $1^{+}$ intermediate states whereas $A_{0}(k^{2})$ refers to
the $0^{+}$ state. As regards $V(k^{2})$, it can be understood as the $1^{-}$ intermediate state
in the transition $0^{-} \to 1^{-}$. Finally, in order to cancel the poles at $k^{2}=0$, the 
form factors respect the conditions:
\begin{equation}\label{eq13}
F_{1}(0) = F_{0}(0), \;\;  A_{3}(0) = A_{0}(0)\ ,
\end{equation}
as well as  the following relation:
\begin{gather}\label{eq14}
A_{3}(k^{2})  = \frac {m_{B}+m_{M_{i}^{\star}}}{2m_{M_{i}^{\star}}}A_{1}(k^{2})
- \frac {m_{B}-m_{M_{i}^{\star}}} {2m_{M_{i}^{\star}}}A_{2}(k^{2})\ .
\end{gather}
Regarding the matrix element $\langle M_{i}(q) | \bar{q}_{1} \gamma_{\mu}(1 - 
\gamma_{5}) q_{2}| 0 \rangle$, it simplifies as follows:
\begin{align}
{ \langle M_{i}(q) | \bar{q}_{1} \gamma_{\mu} \gamma_{5} q_{2}| 0 \rangle} & = - i f_{P} q_{\mu}\ , \;\; 
{\rm for \;\; a \; pseudo\!\!-\!\!scalar\;\; meson}\ , \nonumber\\
{ \langle M_{i}(q) | \bar{q}_{1} \gamma_{\mu} q_{2} | 0 \rangle}  & = f_{V} m_{V} \epsilon_{\mu}\ , \;\;
 {\rm for \;\; a \;  vector \;\; meson}\ ,
\end{align}
where $f_{p}$ and $f_{v}$ define the decay constant for pseudo-scalar and vector mesons, respectively.

\noindent Next, the Wilson coefficients, $C_{i}(\mu)$,~\cite{Buras:1999rb,Buras:1998ra,Buchalla:1996vs} 
are calculated to the next to leading order 
(NLO) and $C_{i}(\mu)$ is given by,
\begin{equation}\label{eq15}
C_{i}(\mu)= U(\mu,M_{W}) C_{i}(M_{W})\ ,
\end{equation}
where $U(\mu,M_{W})$ describes the QCD evolution. The strong interaction being independent of 
quark flavour, the $C_{i}(\mu)$ are the same for all $B$ decays. The matrix elements of operators
$O_{i}(\mu)$ are renormalized to the one loop order and calculated at the scale $\mu=m_{b}$. Then,
by making a matching between the full and  effective theories,
\begin{equation}\label{eq16}
C_{i}(m_{b}) O_{i}(m_{b})= C_{i}^{\prime} \langle O_{i} \rangle^{tree}\ ,
\end{equation}
where $\langle O_{i} \rangle^{tree}$ are the matrix elements at the tree level, the 
``effective'' Wilson coefficients  are obtained.
It is  more convenient to redefine them (the values of $C_i$ are given in Table 1) such as:
\begin{eqnarray}\label{eq17}
   a_i  &= & \left( C_i + \frac{C_{i\pm 1}}{N_c^{eff}} \right) N_i  + P_i \ ,
\end{eqnarray}
where the upper (lower) signs apply when $i$ is odd (even) and
$N_i$ is equal to one for all the cases. $P_i$ involves the QCD
penguin and electroweak penguin corrections for $i=3-6$ and
$i=7-10$, respectively.  $P_i$ takes the following
form~\cite{Stech:1997ij,Deshpande:1995pw,Fleischer:1994gr,Fleischer:1993gp},
\begin{equation}\label{eq18}
  P_i   = \left\{\,\,
   \begin{array}{ll}
    {\displaystyle 0} \,; & \qquad
     i=1,2, \\[0.4cm] 
    {\displaystyle P_{s} \left( \frac{1}{N_{c}^{eff}} - \frac13 \right)} \,; & \qquad
     i=3,5, \\[0.4cm] 
    {\displaystyle P_{s} \left( 1- \frac{1}{3 N_{c}^{eff}} \right)} \,; & \qquad
     i=4,6, \\[0.4cm] 
     {\displaystyle P_{e}} \,; & \qquad
     i=7,9, \\[0.4cm] 
     {\displaystyle \frac{P_{e}}{N_{c}^{eff}}} \,; & \qquad
     i=8,10, 
   \end{array}\right.
\end{equation}
with the following explicit expressions for $P_{s}$ and $P_{e}$:
\begin{align*}
P_{s} & =\frac{\alpha_{s}}{8\pi} C_{2}\biggl(\frac{10}{9}+G(m_{c},\mu,q^{2})\biggr)\ ,  \\
P_{e} & =\frac{\alpha_{em}}{9\pi}(3C_{1}+C_{2})\biggl(\frac{10}{9}+G(m_{c},\mu,q^{2})\biggr)\ .
\end{align*}
The function, $G(m_{c},\mu,q^{2})$, models the one-gluon(photon) exchange 
and takes the form,
\begin{eqnarray*}
 G(m_{c},\mu,q^{2})=4\int_{0}^{1}dx \; x(x-1){\rm ln} \frac{m_{c}^{2}-x(1-x)q^{2}}{\mu^{2}}\ .
\end{eqnarray*}
Here $q^{2}$ is   the typical  momentum transfer of the gluon or photon in the 
penguin diagrams.  The  explicit expression for $G(m_{c},\mu,q^{2})$ is given in 
Ref.~\cite{Kramer:1994yu}. Based on simple arguments  at the quark level, the value 
of $q^{2}$ is usually chosen in the range $0.3 < q^{2}/m_{b}^{2} < 0.5$~\cite{Enomoto:1996cv, 
Gardner:1998yx}.  We underline that the naive factorization approach does not 
take into account the final state flavour in the calculation of  QCD or 
electroweak penguin corrections. Note as well that if $N_{c}^{eff}=N_{c}=3$,  
QCD penguin corrections go to zero for $i=3,5$.
%
\subsection{QCD factorization}
%
{}Factorization in charmless $B$ decays involves 
three fundamental scales: the weak 
interaction scale, $M_{W}$,  the $b$ quark mass scale, $m_{b}$, and the strong 
interaction scale, $\Lambda_{QCD}$. The matrix 
elements $\langle M_1 M_2 | O_{i} | 
B \rangle(\mu)$ that  depend on both $m_{b}$ 
and $\Lambda_{QCD}$,  contain perturbative 
and non-perturbative effects which are not 
accurately estimated in  naive factorization. 
The aim is therefore  to obtain a good  estimate of the matrix elements
without using the latter formalism, where the matrix element of a four fermion
operator is directly replaced by the product of 
the matrix elements of two currents, 
one semi-leptonic and the other purely leptonic. 
The QCD factorization (QCDF)\footnote{For a complete
and detailed presentation of QCDF, we mainly refer 
to the papers of M. Beneke, M. Neubert, G. Buchalla and 
C.T. Sachrajda ~\cite{Beneke:1999br, Neubert:2001cq, 
Neubert:2000kk, Beneke:2000pw, Beneke:2002nj, Beneke:1999gt}.} approach, based
on the concept of color transparency  as well as on a soft collinear  factorization where the 
particle energies are bigger than the scale $\Lambda_{QCD}$,  allows us to write down the 
matrix elements  $\langle M_1 M_2 | O_{i} | B \rangle(\mu)$ at the  leading order in 
$\Lambda_{QCD}/m_{b}$.  When a heavy quark expansion such as $m_{b} \gg \Lambda_{QCD}$
 is assumed, it follows that:
\begin{eqnarray}\label{eq19}
\langle  M_{1} M_{2} | O_{i} | B \rangle (\mu) = { \langle  M_{1} | j_{1} | B \rangle}
 {  \langle  M_{2} | j_{2} | 0 \rangle}
\biggl[ 1 + \sum_{n} r_{n} \alpha_{s}^{n} + { \mathcal{O}(\Lambda_{QCD}/m_{b})}\biggr]\ ,
\end{eqnarray}
where $r_{n}$ refers to the radiative corrections in $\alpha_{s}$ and $j_{i}$ are the 
quark currents defined as usual. It is straightforward to see that  if we neglect the 
corrections at  order  $\alpha_{s}$ and work to the leading power in $\Lambda_{QCD}/m_{b}$,  
we recover the conventional naive factorization in the heavy quark limit. Despite the fact 
that most of the non-factorizable power-suppressed corrections are neglected, the hard-scattering
spectator interactions and annihilation contributions cannot be ignored since they are 
chirally-enhanced.

\begin{figure}[hpbt]
\includegraphics*[width=1.000\columnwidth]{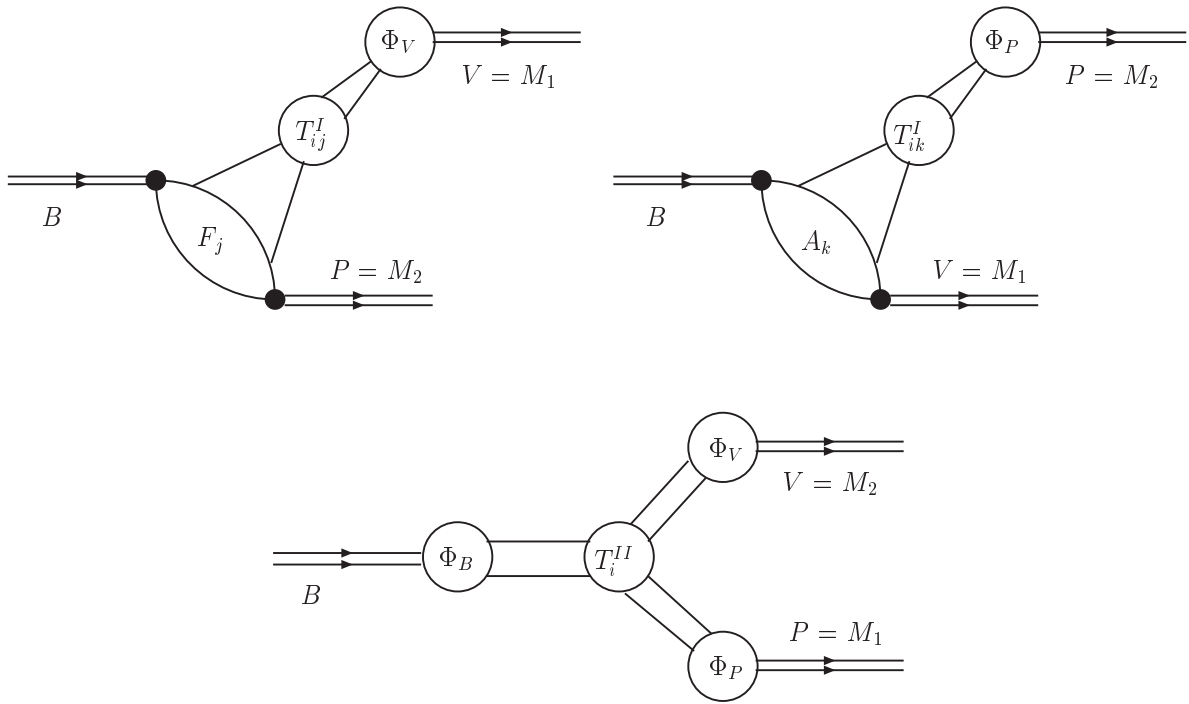}
\caption{Graphical representation of the QCD factorization formula.}
\label{fig2}
\end{figure}

We can write the matrix elements  $\langle M_1 M_2 | O_{i} | B \rangle(\mu)$, 
at the  leading order in $\Lambda_{QCD}/m_{b}$, in the QCDF approach by using a partonic 
language which gives~\cite{Beneke:1999br, Neubert:2001cq, Neubert:2000kk, Beneke:2000pw, 
Beneke:2002nj, Beneke:1999gt}:
\begin{multline}\label{eq20}
\langle  M_1 M_2 | O_{i} | B \rangle(\mu) = F_{j}^{B \rightarrow {M}_{1}}(0) \int_{0}^{1} 
d x T_{ij}^{I}(x) \phi_{{M}_{2}}(x)
+ A_{k}^{B \rightarrow {M}_{2}}(0) \int_{0}^{1} d y  T_{ik}^{I}(y) \phi_{{M}_{1}}(y) \\
+ \int_{0}^{1} d \xi  \int_{0}^{1} d x  \int_{0}^{1} d y T_{i}^{II}(\xi,x,y) \phi_{B}(\xi) 
\phi_{{M}_{2}}(x) \phi_{{M}_{1}}(y)\ ,
\end{multline}
where $\phi_{M_{i}}$ with $M_{i}=V,P,B$ are the leading twist light cone distribution 
amplitudes (LCDA) of the valence quark Fock states. The light cone  momentum fractions of 
the constituent quarks of the vector, pseudo-scalar and $B$ mesons are given respectively 
by $x,y,$ and $\xi$. The form factors for $B \to P$ and $B \to V$ semi-leptonic decays, 
evaluated at $k^{2}=0$, are denoted by $F_{j}^{B \rightarrow P}(0)$ and $A_{k}^{B \rightarrow V}(0)$.  
The hadronic decay amplitude involves both soft and hard contributions. At leading order, all 
the non-perturbative effects are assumed to be contained in the semi-leptonic form factors 
and the light cone distribution amplitudes. Then, non-factorizable interactions are dominated 
by hard gluon exchanges (in the case where the  $O(\Lambda_{QCD}/m_{b}$) terms are neglected) 
and can be  calculated  perturbatively, in order to correct the naive factorization approximation. 
 These hard scattering kernels,   $T_{ik}^{I}$ and $T_{i}^{II}$, are calculable order
by order in perturbation theory. The naive factorization terms are recovered by the leading 
terms of $T_{ik}^{I}$ and $T_{ij}^{I}$ coming from the tree level, whereas  vertex  and penguin 
corrections (Fig.~\ref{fig2})  are included at the order of $\alpha_{s}$ in $T_{ik}^{I}$ and $T_{ij}^{I}$. 
The  hard interactions (at order  $O(\alpha_{s}$)) between the spectator quark and the emitted meson 
(Fig.~\ref{fig2}), at large  gluon momentum,  are taken into account  by $T_{i}^{II}$.

The coefficients, $a_{i}^{p}(M_1 M_2)$, written as a linear combination of Wilson coefficients,
 $C_{i}(\mu)$, (see Table 1) are calculated at next-to-leading order in $\alpha_s$ and contain all the 
non-factorizable effects.  Their  general form is~\cite{Beneke:2003zv},
\begin{multline}\label{eq21}
   a_i^p(M_1 M_2) = \left( C_i + \frac{C_{i\pm 1}}{N_c} \right)
   N_i(M_2) \\
   + \,\frac{C_{i\pm 1}}{N_c}\,\frac{C_F\alpha_s}{4\pi}
   \left[ V_i(M_2) + \frac{4\pi^2}{N_c}\,H_i(M_1 M_2) \right] 
   + P_i^p(M_2) \,,
\end{multline}
where $p$ denotes the $u$ or $c$ quark. $N_i(M_2)=0$ if $i=6,8 $ with  $M_2=V$, and $N_i(M_2)= 1$ 
for all other cases. The first terms in Eq.~(\ref{eq21}) include  naive factorization followed by  
the vertex and the hard spectator interactions,  whereas the last term describes the  
penguin corrections. Contrary to naive factorization, the QCDF approach takes into 
account the flavour of  the final state through the coefficients, $a_{i}^{p}(M_1 M_2)$.

\begin{figure}[hpbt]
\begin{center}
\includegraphics*[width=0.85\columnwidth]{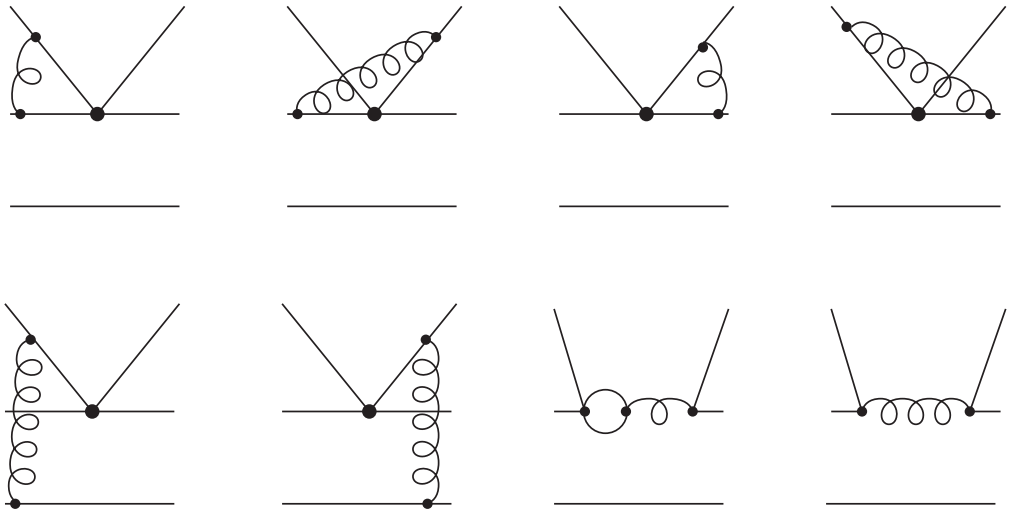}
\end{center}
\caption{Vertex corrections, (first row) and 
hard spectator scattering and penguin corrections (second row) at the order $\alpha_s$.}
\label{fig3}
\end{figure}

The vertex corrections, $V_i(M_2)$, (see Fig.~\ref{fig3}) are given by~\cite{Beneke:2003zv}, 
\begin{equation}\label{eq22}
   V_i(M_2) = \left\{\,\,
   \begin{array}{ll}
    {\displaystyle \int_0^1\!dx\,\Phi_{M_2}(x)\,
     \Big[ 12\ln\frac{m_b}{\mu} - 18 + g(x) \Big]} \,; & \qquad
     i=\mbox{1--4},9,10, \\[0.4cm] 
   {\displaystyle \int_0^1\!dx\,\Phi_{M_2}(x)\,
     \Big[ - 12\ln\frac{m_b}{\mu} + 6 - g(1-x) \Big]} \,; & \qquad
     i=5,7, \\[0.4cm]
   {\displaystyle \int_0^1\!dx\,\Phi_{m_2}(x)\,\Big[ -6 + h(x) \Big]}
    \,; & \qquad i=6,8,
   \end{array}\right.
\end{equation}
where, $\Phi_{M_2}(x)$, is the light-cone amplitude for a pseudo-scalar or vector meson
and, $\Phi_{m_2}(x)$, is the twist-3 amplitude\footnote{The expression for  $\Phi_{M_2}(x)$ and $\Phi_{m_2}(x)$
 are given in Section 6.} for the same mesons. The kernels $g(x)$ 
and $h(x)$ have the following form~\cite{Beneke:2003zv},
\begin{equation}\label{eq23}
\begin{aligned}
   g(x) &= 3\left( \frac{1-2x}{1-x}\ln x-i\pi \right) \\
   &\quad\mbox{}+ \left[ 2 \,\mbox{Li}_2(x) - \ln^2\!x
    + \frac{2\ln x}{1-x} - (3+2i\pi)\ln x - (x\leftrightarrow 1-x)
    \right] , \\ 
   h(x) &= 2 \,\mbox{Li}_2(x) - \ln^2\!x - (1+2\pi i)\,\ln x  
    - (x\leftrightarrow 1-x) \,. 
\end{aligned}
\end{equation}
Next, $H_{i}(M_1 M_2)$  (see Fig.~\ref{fig3}) describes the hard gluon exchanges between the spectator quark in 
the $B$ meson and the emitted meson, $M_{i}$ (pseudo-scalar or vector). These hard scattering 
corrections  include  chirally-enhanced contributions and take the form~\cite{Beneke:2003zv},
\begin{equation*}
    H_i(M_1M_2) =\left\{\,\,
   \begin{array}{ll}
    {\displaystyle  \frac{m_B}{\lambda_B} \int_0^1\!dx \int_0^1\!dy \left[
   \frac{\Phi_{M_2}(x)\Phi_{M_1}(y)}{\bar x\bar y}
   + r_\chi^{M_1}\,\frac{\Phi_{M_2}(x)\Phi_{m_1}(y)}{x\bar y} \right]} \,; & \qquad
     \\ 
   {\displaystyle -  \frac{m_B}{\lambda_B} \int_0^1\!dx \int_0^1\!dy \left[
   \frac{\Phi_{M_2}(x)\Phi_{M_1}(y)}{x\bar y}
   + r_\chi^{M_1}\,\frac{\Phi_{M_2}(x)\Phi_{m_1}(y)}{\bar x\bar y} 
   \right]} \,; & \qquad
      \\
   {\displaystyle \!\! 0}\,; & \qquad
   \end{array}\right.
\end{equation*}
with $\frac{m_B}{\lambda_B}= \int_{0}^{1} d \xi 
     \frac{\phi_{B}(\xi)}{\xi}$ and   where  the subscript $i$ goes as in Eq.~(\ref{eq22}).
 $H_{i}(M_1 M_2)$ is multiplied by  $B_{M_1 M_2}/A_{M_1 M_2}$ given by:
\begin{equation}\label{eq25}
   \frac{B_{M_1 M_2}}{A_{M_1 M_2}} = \left\{\,\,
   \begin{array}{ll}
    {\displaystyle \frac{f_{B} f_{M_1}}{m_B^2 F_1^{B\to M_1}(0)\,}} \,; & \qquad
    \mbox{if~} M_1=M_2=P, \\
   {\displaystyle \frac{f_{B} f_{M_1}}{m_B^2 A_0^{B\to M_1}(0)\,}} \,; & \qquad
     \mbox{if~} M_1=V,\, M_2=P,\\
   {\displaystyle \frac{f_{B} f_{M_1}}{m_B^2 F_1^{B\to M_1}(0)\,}}
    \,; & \qquad 
      \mbox{if~} M_1=P,\, M_2=V,
   \end{array}\right.
\end{equation}
with the usual definitions for $f_{B}, f_{M_{i}}, A_{0}^{B \rightarrow V}(0),
F_1^{B\to P}(0)$  and $m_{B}$. The chiral  enhancement factor is parameterized 
by the term  $r_\chi^{M_1}= 2 m_{M_1}^{2}/m_{b}(\mu)(m_{1}+m_{2})(\mu)$,
with $m_{1}$ and $m_{2}$ being the current quark masses of the valence quarks in the meson.

Finally, $P_{i}^{p}(M_{2})$  are the QCD penguin ($i=4,6$) and electroweak 
penguin ($i=8,10$) contributions. These quantities contain all of the 
non-perturbative dynamics and are the  result of the convolution of hard 
scattering kernels, $G_{M_2}(s,x)$, with meson distribution amplitudes, $\Phi_{M_i}(x)$, 
such as,
\begin{align}\label{eq26}
   G_{M_2}(s) &= \int_0^1\!dx\,G(s-i\epsilon,1-x)\,\Phi_{M_2}(x)\ , \nonumber \\
   \hat G_{M_2}(s) &= \int_0^1\!dx\,G(s-i\epsilon,1-x)\,\Phi_{m_2}(x) \,.
\end{align}
The function $G(s,x)$ is  given in Ref.~\cite{Beneke:2003zv}. The imaginary parts arising  in 
the previous penguin functions, $g(x)$, $h(x)$, and $G(s,x)$, give us three sources of strong 
re-scattering phases. If $M_{2}$ is a pseudo-scalar, the penguin contributions, 
$P_{i}^{p}(M_{2})$, for $i=4,6,8,10$, are respectively~\cite{Beneke:2003zv}:
\begin{equation*}
   P_i^{p}(M_2) = \left\{\,\,
   \begin{array}{l}
     \frac{C_F\alpha_s}{4\pi N_c}\,\Bigg\{
    C_1 \!\left[ \frac43\ln\frac{m_b}{\mu}
    + \frac23 - G_{M_2}(s_p) \right]\! 
    + C_3 \!\left[ \frac83\ln\frac{m_b}{\mu} + \frac43
    - G_{M_2}(0) - G_{M_2}(1) \right] \\
   + (C_4+C_6)\!
    \left[ \frac{4n_f}{3}\ln\frac{m_b}{\mu}
    - (n_f-2)\,G_{M_2}(0) - G_{M_2}(s_c) - G_{M_2}(1) \right] \\
   - 2 C_{8g}^{\rm eff} \int_0^1 \frac{dx}{1-x}\,
    \Phi_{M_2}(x) \Bigg\} \,;  
       \\[0.4cm] 
    \frac{C_F\alpha_s}{4\pi N_c}\,\Bigg\{
    C_1 \!\left[ \frac43\ln\frac{m_b}{\mu}
    + \frac23 - \hat G_{M_2}(s_p) \right]\!
    + C_3 \!\left[ \frac83\ln\frac{m_b}{\mu} + \frac43
    - \hat G_{M_2}(0) - \hat G_{M_2}(1) \right] \\
    + (C_4+C_6)\!
    \left[ \frac{4n_f}{3}\ln\frac{m_b}{\mu}
    - (n_f-2)\,\hat G_{M_2}(0) - \hat G_{M_2}(s_c) - \hat G_{M_2}(1)
    \right] \\ - 2 C_{8g}^{\rm eff} \Bigg\}\,;  
       \\[0.4cm]
     \frac{\alpha}{9\pi N_c}\,\Bigg\{
   (C_1+N_c C_2) \left[ \frac{4}{3}\ln\frac{m_b}{\mu} + \frac23  
   - \hat G_{M_2}(s_p) \right] - 3 C_{7\gamma}^{\rm eff} \Bigg\}\,;  
       \\[0.4cm]
   \frac{\alpha}{9\pi N_c}\,\Bigg\{
   (C_1+N_c C_2) \left[ \frac{4}{3}\ln\frac{m_b}{\mu} + \frac23  
   - G_{M_2}(s_p) \right]
   - 3 C_{7\gamma}^{\rm eff} \int_0^1 \frac{dx}{1-x}\,\Phi_{M_2}(x)
   \Bigg\}\,;  
   \end{array}\right.
\end{equation*}
$s_{q}= m_{q}^{2}/m_{b}^{2}$ is the mass ratio and can be equal 
to $s_{u}=s_{d}=0, s_{c}= m_{c}^{2}/m_{b}^{2}$ or $s_{b}=1$. All active quarks at 
the scale $\mu= O(m_{b})$ are represented by $q^{\prime}=u,d,s,c,b$. 
The other parameters are $C_{i}\equiv C_{i}(\mu)$ (in Naive Dimensional Regularization), $\alpha_s 
\equiv \alpha_{s}(\mu)$ (next to leading order), $C_{F}= (N_{c}^{2}-1)/2 N_{c}$ 
with $N_{c}=3$. If $M_{2}$ is a vector meson, $P_{i}^{p}(M_{2})$ for $i=4,6,8,10$, 
respectively, take the form~\cite{Beneke:2003zv}:
\begin{equation}\label{eq27}
   P_i^{p}(M_2) = \left\{\,\,
   \begin{array}{l}
     \frac{C_F\alpha_s}{4\pi N_c}\,\Bigg\{
    C_1 \!\left[ \frac43\ln\frac{m_b}{\mu}
    + \frac23 - G_{M_2}(s_p) \right]\! 
    + C_3 \!\left[ \frac83\ln\frac{m_b}{\mu} + \frac43
    - G_{M_2}(0) - G_{M_2}(1) \right] \\
   + (C_4+C_6)\!
    \left[ \frac{4n_f}{3}\ln\frac{m_b}{\mu}
    - (n_f-2)\,G_{M_2}(0) - G_{M_2}(s_c) - G_{M_2}(1) \right] \\
   - 2 C_{8g}^{\rm eff} \int_0^1 \frac{dx}{1-x}\,
    \Phi_{M_2}(x) \Bigg\} \,; \\[0.4cm] 
      - \frac{C_F\alpha_s}{4\pi N_c}\,\Bigg\{
    C_1\,\hat G_{M_2}(s_p)
    + C_3\,\Big[ \hat G_{M_2}(0) + \hat G_{M_2}(1) \Big] \nonumber\\
      (C_4+C_6) \left[ (n_f-2)\,\hat G_{M_2}(0)
    + \hat G_{M_2}(s_c) + \hat G_{M_2}(1) \right] \Bigg\} \,; \\[0.4cm]
     - \frac{\alpha}{9\pi N_c}\, \Bigg\{ (C_1+N_c C_2)\,
   \hat G_{M_2}(s_p)\Bigg\} \,; \\[0.4cm]
    \frac{\alpha}{9\pi N_c}\,\Bigg\{
   (C_1+N_c C_2) \left[ \frac{4}{3}\ln\frac{m_b}{\mu} + \frac23  
   - G_{M_2}(s_p) \right]
   - 3 C_{7\gamma}^{\rm eff} \int_0^1 \frac{dx}{1-x}\,\Phi_{M_2}(x)
   \Bigg\} \,.
   \end{array}\right.
\end{equation}
The vertex and penguin corrections to the hard-scattering kernels are evaluated 
at the scale $\mu \sim m_{b}$. Because the gluon is  off-shell, the strong coupling 
constant, $\alpha_{s}(\mu)$, the Wilson coefficients, $C_{i}(\mu)$, and then the 
hard-scattering contributions, $P_i^{p}(M_2)$, are evaluated at the scale $\mu_{h}
= \sqrt{\Lambda_{h} \mu}$, with $\Lambda_{h}=0.5$ GeV, rather than the scale $\mu = m_{b}$.

\begin{figure}[hpbt]
\begin{center}
\includegraphics*[width=0.85\columnwidth]{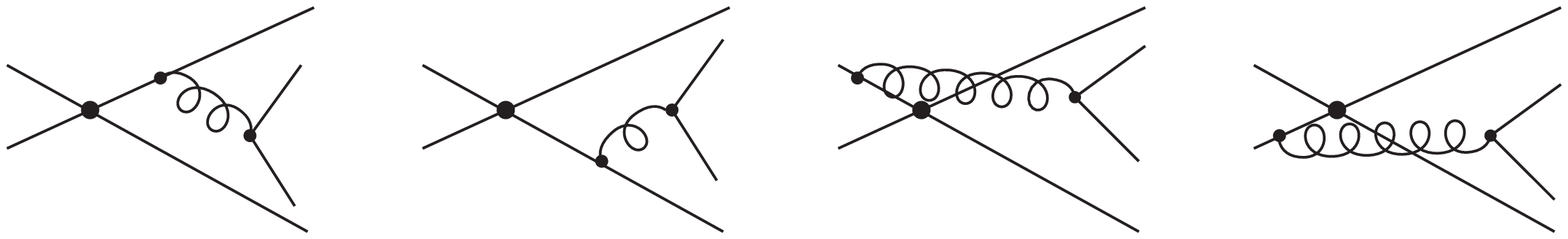}
\end{center}
\caption{Order $\alpha_s$ corrections to the weak annihilation.}
\label{fig4}
\end{figure}

It has been shown in Ref.~\cite{Keum:2000ms,Keum:2001wi} that the weak annihilation contributions 
cannot  be neglected in $B$ meson decays even though they are power suppressed in heavy-quark 
limit ($\Lambda_{QCD}/m_{b}$). Moreover, their contributions could carry large strong 
phases with QCD corrections and hence, large $CP$ violation might be obtained in $B$ 
meson decays. The annihilation contributions, at leading order in $\alpha_s$, are given by 
the diagrams  drawn in Fig.~\ref{fig4}. They do not arise in the QCD factorization 
formulation  because of their endpoint singularities. Nevertheless, their contributions 
denoted by $A_{k}^{i}(M_{1} M_{2})$ ($\equiv A_{k}^{i}$ for simplicity), are 
approximated in terms of convolutions of hard scattering kernels with light cone 
expansions for the final state mesons.  If we define $x$ as the longitudinal momentum 
fraction of the quark contained in $M_{2}$ and ${\bar y}=1-y$, the momentum fraction 
of the antiquark contained in $M_{1}$, then the diagrams  related to the annihilation 
contributions  can be expressed   as~\cite{Beneke:2003zv}:
\begin{eqnarray}\label{eq28}
   A_1^i \!\! &=& \!\! \pi\alpha_s \int_0^1\! dx dy\, 
    \left\{ \Phi_{M_2}(x)\,\Phi_{M_1}(y)
    \left[ \frac{1}{y(1-x\bar y)} + \frac{1}{\bar x^2 y} \right]
    + r_\chi^{M_1} r_\chi^{M_2}\,\Phi_{m_2}(x)\,\Phi_{m_1}(y)\,
     \frac{2}{\bar x y} \right\} ,
    \nonumber\\  
   A_2^i\!\!  &=& \!\! \pi\alpha_s \int_0^1\! dx dy\, 
    \left\{ \Phi_{M_2}(x)\,\Phi_{M_1}(y)
    \left[ \frac{1}{\bar x(1-x\bar y)} + \frac{1}{\bar x y^2} \right]
    + r_\chi^{M_1} r_\chi^{M_2}\,\Phi_{m_2}(x)\,\Phi_{m_1}(y)\,
     \frac{2}{\bar x y} \right\} ,
    \nonumber\\  
   A_3^i \!\! &=& \!\! \pi\alpha_s \int_0^1\! dx dy\,
    \left\{r_\chi^{M_1}\,\Phi_{M_2}(x)\,\Phi_{m_1}(y)\,
    \frac{2\bar y}{\bar x y(1-x\bar y)}
    - r_\chi^{M_2}\,\Phi_{M_1}(y)\,\Phi_{m_2}(x)\,
    \frac{2x}{\bar x y(1-x\bar y)} \right\} , \nonumber
\end{eqnarray}
and,
\begin{eqnarray}\label{eq29}
   A_1^f \!\! &=& \!\! A_2^f =0 \,, \nonumber\\   
   A_3^f \!\! &=& \!\! \pi\alpha_s \int_0^1\! dx dy\,
    \left\{r_\chi^{M_1}\,\Phi_{M_2}(x)\,\Phi_{m_1}(y)\,
    \frac{2(1+\bar x)}{\bar x^2 y}
    +  r_\chi^{M_2}\,\Phi_{M_1}(y)\,\Phi_{m_2}(x)\,
    \frac{2(1+y)}{\bar x y^2} \right\} , \nonumber
\end{eqnarray}
where the superscripts `$i$' and `$f$' refer to gluon emission from the initial
and final-state quarks, respectively. When $M_1$ is a vector meson and $M_2$ a pseudo-scalar, (i.e. in the case
of decays such as $B \to M_1^{V} M_2^{PS}$)
one has to change the sign of the second (twist-4) term in $A_1^i$, the first (twist-2) 
term in $A_2^i$, and the second term in $A_3^i$ and $A_3^f$. When $M_2$ is a vector meson 
and $M_1$ a pseudo-scalar (i.e. in the case
of decays such as $B \to M_1^{PS} M_2^{V}$), one only has to change the overall sign of $A_2^i$.

Taking into account the flavour structure of the various operators involved  in 
the weak annihilation topologies, the annihilation  amplitude can be written as,
\begin{eqnarray}\label{eq30}
A^{a}(B \to M_1 M_2) \propto f_{B} f_{M_1} f_{M_2} \sum_{p=u,c} \sum_{i=1,4} V_{pb}V_{ps}^{*} b_{i}\ .
\end{eqnarray}
The coefficients $b_{i}$ in Eq.~(\ref{eq30}) are  expressed in terms of linear combinations of
$A_{i}(M_{1} M_{2})$ and they take  the following form:
\begin{align}
\label{eq31}
  b_{1} & =   \frac{C_{F}}{N_{c}^{2}} C_{1} A_{1}^{i}\ , \nonumber \\
  b_{2} & =  \frac{C_{F}}{N_{c}^{2}} C_{2} A_{1}^{i}\ , \nonumber \\
  b_{3} & = \frac{C_{F}}{N_{c}^{2}} \Biggl\{ C_{3} A_{1}^{i}
 + C_{5} A_{3}^{i}  + \Big[ C_{5} + N_{c} C_{6} \Big] A_{3}^{f} \Biggr\}\ , \nonumber \\
  b_{4} & =  \frac{C_{F}}{N_{c}^{2}} \Big\{ C_{4} A_{1}^{i}
+ C_{6} A_{2}^{i} \Big\}\ , \nonumber \\
  b_{3}^{ew} & = \frac{C_{F}}{N_{c}^{2}} \Biggl\{ C_{9}A_{1}^{i}
  + C_{7} A_{3}^{i}  + \Big[ C_{7} + N_{c} C_{8} \Big] A_{3}^{f} \Biggr\}\ , \nonumber \\
  b_{4}^{ew} & =  \frac{C_{F}}{N_{c}^{2}} \Big\{ C_{10} A_{1}^{i}
 + C_{8} A_{2}^{i} \Big\}\ ,
\end{align}
where $b_{i} (\equiv b_{i}(M_{1},M_{2}))$  are respectively the current-current 
annihilation parameters arising from the hadronic matrix elements of the effective 
operators for $i=1,2$, the QCD penguin annihilation parameters for  $i=3,4$ and the 
electroweak penguin annihilation parameters for $i=3,4$ with the subscript $ew$ attached 
to  $b_{i}$. The quantities $b_{i}$ depend on the final state mesons through the terms  
$A_k^i$  defined previously.

The calculation of the hard spectator as well as the annihilation contributions
involves the twist-3 distribution amplitude, $\phi_{m_{i}}$. It happens that these
power-suppressed contributions involve divergences because of the non-vanishing
endpoint behaviour of $\phi_{m_{i}}$. This divergence, $X^{M_{i}}$, analyzed as 
(for example),
\begin{eqnarray}\label{eq32}
   \int_0^1\frac{d y}{\bar y}\,\Phi_{m_i}(y) 
   &=& \Phi_{m_i}(1)\,\int_0^1\frac{d y}{\bar y}
    + \int_0^1\frac{d y}{\bar y}\,\Big[ \Phi_{m_i}(y)-\Phi_{m_i}(1)
    \Big]\ , \nonumber\\
   &\equiv& \Phi_{m_i}(1)\,X^{M_i}
    + \int_0^1\frac{d y}{[\bar y]_+}\,\Phi_{m_i}(y) \, ,
\end{eqnarray}
is applied  in the following cases as well:
\begin{equation}\label{eq33}
   \int_0^1 \frac{dy}{y}\to X^{M_i} \,, \qquad {\rm and} \;\;
   \int_0^1\!dy\,\frac{\ln y}{y}\to -\frac{1}{2}\,(X^{M_i})^2 \,.
\end{equation}
The perturbative calculation of the hard scattering spectator and annihilation
contributions is regulated by a physical scale of order $\Lambda_{QCD}$. Therefore, treating 
the divergent endpoint parameterized by $X^{M_i}$  in a phenomenological
way, one  may take the following ansatz:
\begin{equation}\label{eq34}
   X^{M_i} = \left( 1 + \varrho^{M_{i}}\,e^{i\varphi^{M_{i}}} \right)
   \ln\frac{m_B}{\Lambda_h} \,, \qquad {\rm with} \;\;
    \qquad \Lambda_h=0.5\,\mbox{GeV} \,,
\end{equation}
where the phase $\varphi^{M_{i}}$ and the coefficient $\varrho^{M_{i}}$ give rise to  
a  dynamically generated strong interaction phase. This divergence, $X^{M_{i}}$, coming 
from the twist-3 contribution holds for both  hard spectator scattering, 
$X_{H}^{M_{i}}$, and weak annihilation,  $X_{A}^{M_{i}}$. It is expected to take the form
given in Eq.~(\ref{eq34}) since the soft interaction is regulated by a physical scale
$\Lambda_{QCD}/m_b$. Moreover a strong phase (complex part of Eq.~(\ref{eq34})) can arise
because of multiple soft scattering. In this model dependent
way of dealing with the two latter corrections, we assume that $X_{H}^{M_{i}}$
and $X_{A}^{M_{i}}$ are not universal. In other words, they depend on the 
flavour of the meson, $M_{i}$, but they do not depend on the weak vertex. Moreover, 
in order to make this dependence more efficient, in all the calculations  we use  
the full distribution amplitudes for $\phi_{M_{i}}$ and $\phi_{m_{i}}$
by taking into account the Gegenbauer expansion of the asymptotic distribution amplitude.

%
\section{Mixing scheme}
%
%
\subsection{$\boldsymbol{\rho^{0}-\omega}$ mixing scheme}
%
The direct  $CP$  violating asymmetry parameter, $a_{CP}$, is found to be small for most of 
the non-leptonic exclusive $B$ decays when either the naive or QCD factorization framework
is applied. However,  in the case of $B$ decay channels involving the $\rho^0$ meson, it appears
that  the asymmetry may be large in the vicinity of $\omega$ meson mass. To obtain a  large signal
for direct  $CP$  violation requires some mechanism to make both $\sin\delta$  and  $r$ large  (see Eq.~(\ref{eq52}) below). 
We stress that   $\rho^0-\omega$ mixing has the dual advantages that the strong phase difference
is large (passing rapidly through $90^{o}$ at the $\omega$ resonance) and well known~\cite{Enomoto:1996cv,Gardner:1998yx,
Guo:1998eg,Guo:1999ip,Gardner:1998za,Gardner:1997qk}. 
In the vector meson dominance model~\cite{Sakurai:1969ju}, the photon propagator is dressed by coupling
to  the vector mesons $\rho^0$ and $\omega$. In this regard, the $\rho^0-\omega$ mixing 
mechanism~\cite{O'Connell:1997wf,O'Connell:1996ns,O'Connell:1997br} has been developed. Let $A$ be the 
amplitude for the decay $B \rightarrow \rho^{0} ( \omega ) M_1 \; \rightarrow  \pi^{+}  \pi^{-} \; M_1$, then one has,
\begin{equation}\label{eq35}
A=\langle  M_1 \; \pi^{-} \pi^{+}|H^{T}|B \rangle + \langle  M_1 \;  \pi^{-} \pi^{+}|H^{P}|B  \rangle\ ,
\end{equation}
with $H^{T}$ and $H^{P}$ being the Hamiltonians for the tree and penguin operators. Here 
$M_1$ denotes a  pseudo-scalar meson\footnote{The same procedure holds for a vector meson, $M_{1}$, 
see Ref.~\cite{Ajaltouni:2003yt}.}. We can define the relative magnitude and phases between these two contributions  
as follows,
\begin{align}\label{eq36}
A &= \langle  M_1 \;  \pi^{-} \pi^{+}|H^{T}| B \rangle [ 1+re^{i\delta}e^{i\phi}]\ , \nonumber \\   
\bar {A} &= \langle \overline{M_1} \;  \pi^{+}  \pi^{-}|H^{T}|\bar {B} \rangle [ 1+re^{i\delta}e^{-i\phi}]\ ,  
\end{align}
where $\delta$ and $\phi$ are strong and weak phases, respectively. The phase $\phi$ arises 
from the appropriate combination of CKM matrix elements. In case of $b \to d$ or $b \to s$
transitions, $\phi$ is given by $ \phi={\rm arg}[(V_{tb}V_{td}^{\star})/(V_{ub}V_{ud}^{\star})]$
or  ${\rm arg}[(V_{tb}V_{ts}^{\star})/(V_{ub}V_{us}^{\star})]$, respectively. As a result, 
$\sin \phi$ is equal to $\sin \alpha \; (\sin \gamma)$ for $b \to d \; (b \to s)$,  with 
$\alpha \; (\gamma)$ defined in the standard way~\cite{Groom:2000in}. $\sin \phi \; (\cos \phi)$ 
 therefore takes the following form in case of a  $b \to d$ transition,
\begin{align}\label{eq37}
\sin\phi & = \frac{\eta}
{\sqrt {\left(\rho - (\rho^2+\eta^2)(1-\frac{\lambda^2}{2})\right)^2 + \eta^2}}\ , \nonumber \\
\cos\phi & = \frac{\rho- (\rho^2+\eta^2)(1-\frac{\lambda^2}{2})}
{\sqrt {\left(\rho - (\rho^2+\eta^2)(1-\frac{\lambda^2}{2})\right)^2 + \eta^2}}\ ,
\end{align}
\noindent and in case of a $b \to s$ transition,
\begin{align}\label{eq38}
\sin\phi & = \frac{-\eta (1 -\frac{\lambda^2}{2})}
{\sqrt {\left(-\lambda^2 (\rho^2+\eta^2 )- \rho (1-\frac{\lambda^2}{2})\right)^2+ \left( \eta^2 (1-\frac{\lambda^2}{2})\right)^2}}\ , 
\nonumber \\
\cos\phi & = \frac{-\lambda^2 (\rho^2+\eta^2 )- \rho (1-\frac{\lambda^2}{2})}
{\sqrt {\left(-\lambda^2 (\rho^2+\eta^2 )- \rho (1-\frac{\lambda^2}{2})\right)^2+ \left( \eta^2 (1-\frac{\lambda^2}{2})\right)^2}}\ .
\end{align}
\noindent Regarding the parameter, $r$, it represents  the 
absolute value of the ratio of tree and penguin amplitudes:
\begin{equation}\label{eq39}
r \equiv \left| \frac{\langle \rho^{0}(\omega) M_{1}|H^{P}|B \rangle}{\langle\rho^{0}(\omega)
M_{1}|H^{T}|B \rangle} \right|.
\end{equation}

\begin{figure}[hpbt]
\begin{center}
\includegraphics*[width=0.85\columnwidth]{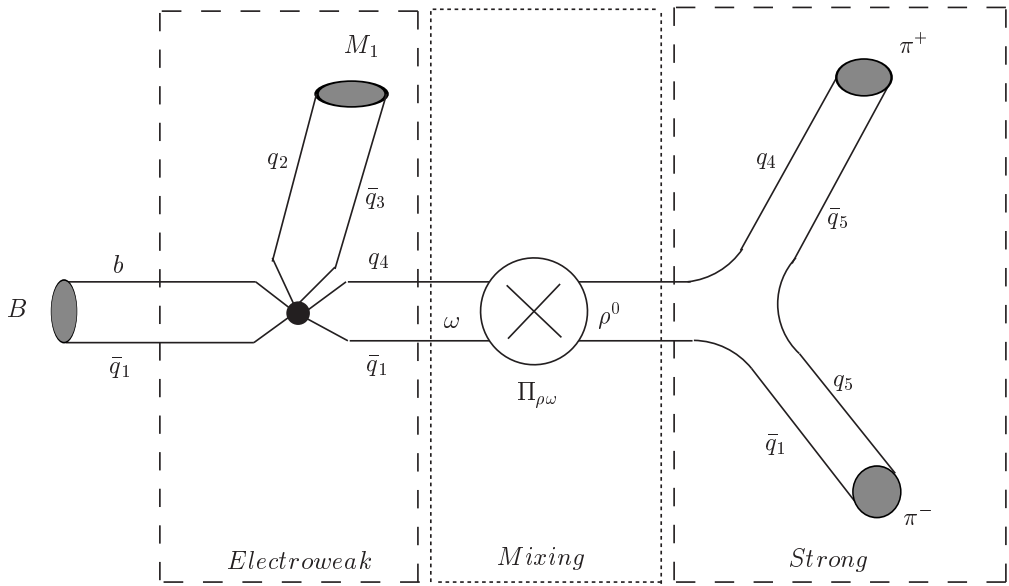}
\caption{Inclusion of the $\rho-\omega$ mixing in $B \to \rho^0 M_1$ decay.}
\label{fig1}
\end{center}
\end{figure}

\noindent With this mechanism (see Fig.~\ref{fig1}), to first  order in  isospin violation, 
we have the following results when the invariant mass of $\pi^{+}\pi^{-}$ is near the 
$\omega$ resonance mass,
\begin{align}\label{eq40}
\langle M_1 \pi^{-} \pi^{+}|H^{T}|B  \rangle & = \frac{g_{\rho}}{s_{\rho}s_{\omega}}
 \tilde{\Pi}_{\rho \omega}(t_{\omega}+t_{\omega}^a) +\frac{g_{\rho}}{s_{\rho}}(t_{\rho}+t_{\rho}^a)\ , \nonumber \\
\langle  M_1 \pi^{-} \pi^{+}|H^{P}|B  \rangle & = \frac{g_{\rho}}{s_{\rho}s_{\omega}} 
\tilde{\Pi}_{\rho \omega}(p_{\omega}+p_{\omega}^a) +\frac{g_{\rho}}{s_{\rho}}(p_{\rho}+p_{\rho}^a)\ .
\end{align}
Here $t_{V}^{(a)} \; (V=\rho \;{\rm  or} \; \omega) $ is the tree (tree annihilation) amplitude
and $p_{V}^{(a)}$ the penguin (penguin annihilation) amplitude for producing a vector meson, 
$V$, $g_{\rho}$ is the coupling for $\rho^{0} \rightarrow \pi^{+}\pi^{-}$, $\tilde{\Pi}_{\rho \omega}$
is the effective $\rho-\omega$ mixing amplitude, and $s_{V}$  is  from the inverse  propagator 
of the vector meson $V$,  $s_{V}=s-m_{V}^{2}+im_{V}\Gamma_{V}$ (with $\sqrt s$  the 
invariant mass of the $\pi^{+}\pi^{-}$ pair). We stress that the direct coupling $ \omega 
\rightarrow \pi^{+} \pi^{-} $ is effectively absorbed into 
$\tilde{\Pi}_{\rho \omega}$~\cite{O'Connell:1994uc,Maltman:1996kj,O'Connell:1997xy,Williams:1998nj,Gardner:1998ta}, 
leading  to the explicit $s$ dependence of $\tilde{\Pi}_{\rho \omega}$. Making the expansion  
$\tilde{\Pi}_{\rho \omega}(s)=\tilde{\Pi}_{\rho \omega}(m_{\omega}^{2})+(s-m_{w}^{2}) 
\tilde{\Pi}_{\rho \omega}^{\prime}(m_{\omega}^{2})$, the  $\rho^{0}-\omega$ mixing parameters
 were determined in the fit of Gardner and O'Connell~\cite{Gardner:1998ie}: $\Re e \; 
\tilde{\Pi}_{\rho \omega}(m_{\omega}^{2})=-3500 \pm 300 \; {\rm MeV}^{2}, \;\;\; \Im m \; 
\tilde{\Pi}_{\rho \omega}(m_{\omega}^{2})= -300 \pm 300 \; {\rm MeV}^{2}$, and  $\tilde{\Pi}_{\rho \omega}^{\prime}
(m_{\omega}^{2})=0.03 \pm 0.04$. In practice, the effect of the derivative term is negligible.
 From Eqs.~(\ref{eq36}, \ref{eq40}) one has,
\vspace{0.8em}
\begin{equation}\label{eq41}
re^{i \delta} e^{i \phi}= \frac{ \tilde {\Pi}_{\rho \omega}(p_{\omega}+p_{\omega}^a)+s_{\omega}(p_{\rho}+p_{\rho}^a)}{\tilde 
{\Pi}_{\rho \omega} (t_{\omega}+t_{\omega}^a) + s_{\omega}(t_{\rho}+t_{\rho}^a)}\ . 
\end{equation}
\vspace{1.0em}
\noindent Defining
\vspace{-1.3em}
\begin{center}
\begin{equation}\label{eq42}
\frac{p_{\omega}+ p_{\omega}^a}{t_{\rho}+t_{\rho}^a} \equiv r^{\prime}e^{i(\delta_{q}+\phi)}\ , \;\;\;\;
\frac{t_{\omega}+t_{\omega}^a}{t_{\rho}+t_{\rho}^a} \equiv \alpha e^{i \delta_{\alpha}}\ , \;\;\;\;
\frac{p_{\rho}+p_{\rho}^a}{p_{\omega}+p_{\omega}^a} \equiv \beta e^{i \delta_{\beta}}\ , 
\end{equation}
\end{center}
where $ \delta_{\alpha}, \delta_{\beta}$ and $ \delta_{q}$ are strong relative phases (absorptive part). 
Substituting Eq.~(\ref{eq42}) into  Eq.~(\ref{eq41})   one finds:
\vspace{0.5em}
\begin{equation}\label{eq43}
re^{i\delta}=r^{\prime}e^{i\delta_{q}} \frac{\tilde{\Pi}_{\rho \omega}+ \beta e^{i \delta_{\beta}} 
s_{\omega}}{s_{\omega}+\tilde{\Pi}_{\rho \omega} \alpha e^{i \delta_{\alpha}}}\ .
\end{equation}
\noindent Defining $\alpha e^{i \delta_{\alpha}}= f + gi, \;\;\; \beta e^{i \delta_{\beta}}= b+ci$ and 
$r^{\prime}e^{i\delta_{q}}=d+ei$, Eq.~(\ref{eq43}) becomes,
\begin{equation}\label{eq44}
re^{i \delta}= \frac{C + i D}{(s-m_{\omega}^{2}+ f \Re e \; \tilde{\Pi}_{\rho \omega} - g  \Im m \;
\tilde{\Pi}_{\rho \omega})^{2}+ (f \Im m \; 
\tilde{\Pi}_{\rho \omega} + g  \Re e \; \tilde{\Pi}_{\rho \omega}  m_{\omega} \Gamma_{\omega})^{2}}\ ,
\end{equation}
%
where $C$ and $D$ are given by:
%
\begin{multline}\label{eq45}
C=\bigl(s-m_{\omega}^{2}+ f \Re e\; \tilde {\Pi}_{\rho \omega} - g  \Im m\; \tilde { \Pi}_{ \rho \omega} \bigr)
 \Biggl\{ d\biggl[ \Re e \;\tilde {\Pi}_{ \rho
 \omega} 
+b(s-m_{ \omega}^{2})-cm_{ \omega} \Gamma_{ \omega}\biggr] \\ 
  -e \biggl[ \Im m \;\tilde { \Pi}_{ \rho \omega} +bm_{ \omega} \Gamma_{ \omega}+c(s-m_{ \omega}^{2})\biggr] 
\Biggr\} \\
   + \bigl(f  \Im m\; \tilde { \Pi}_{ \rho \omega}  +m_{ \omega} \Gamma_{ \omega} + g \Re e\; \tilde {\Pi}_{\rho
 \omega}\bigr) \Biggl\{ e\biggl[
 \Re e \;\tilde 
{\Pi}_{\rho \omega} +b(s-m_{ \omega}^{2})-cm_{ \omega} \Gamma_{ \omega}\biggr]  
 \\
 +d\biggl[ \Im m \;\tilde { \Pi}_{ \rho \omega} 
+bm_{ \omega} \Gamma_{ \omega}+c(s-m_{ \omega}^{2})\biggr] \Biggr\} \ ,
\end{multline}
and
\begin{multline}\label{eq46}
 D=\bigl(s-m_{\omega}^{2}+ f \Re e \;\tilde {\Pi}_{\rho \omega} - g  \Im m\; \tilde { \Pi}_{ \rho \omega}
\bigr) \Biggl\{ e \biggl[ \Re e \;\tilde {\Pi}_{ \rho
 \omega} +
d(s-m_{ \omega}^{2})-cm_{ \omega} \Gamma_{ \omega}\biggr] \\ 
  +d \biggl[ \Im m\; \tilde { \Pi}_{ \rho \omega} +bm_{ \omega} \Gamma_{ \omega}+c(s-m_{ \omega}^{2})\biggr] 
 \Biggr\} \\
   - \bigl(f  \Im m \;\tilde { \Pi}_{ \rho \omega}  +m_{ \omega} \Gamma_{ \omega}+ g \Re e\; \tilde {\Pi}_{\rho
 \omega}\bigr) \Biggl\{ d \biggl[
 \Re e \;\tilde 
{\Pi}_{\rho \omega} +b(s-m_{ \omega}^{2})-cm_{ \omega} \Gamma_{ \omega}\biggr]  
\\
  -e\biggl[ \Im m\; \tilde { \Pi}_{ \rho \omega} 
+bm_{ \omega} \Gamma_{ \omega}+c(s-m_{ \omega}^{2})\biggr] \Biggr\}\ . 
\end{multline}

\noindent Knowing the ratio, $r$, and the strong phase, $\delta$, from Eq.~(\ref{eq43}) 
as well as the weak phase, $\phi$, from the CKM matrix, it is therefore possible to calculate 
the $CP$ violating asymmetry, $a_{CP}$, including the $\rho^{0}-\omega$ 
mixing mechanism:
\vspace{-0.2em}
\begin{equation}\label{eq52}
a_{CP}  \equiv \frac{|A|^{2}-|\bar A|^{2}}{ |A|^{2}+|\bar A|^{2}}=\frac{-2r\sin\delta \sin\phi}{1+2r\cos\delta
\cos\phi+r^2}\ .
\end{equation}
%

\subsection{$\boldsymbol{\eta-\eta^{\prime}}$ mixing scheme}
%
The evaluation of the decay constants for the pseudo-scalar mesons $\eta$ and $\eta^{\prime}$ 
is not trivial. In this section, we recall briefly how the mixing $\eta-\eta^{\prime}$ is taken 
into account, see Ref.~\cite{Beneke:2002jn} for more details. The  $\eta-\eta^{\prime}$ mixing 
scheme requires the assumption that their decay constants follow the pattern of particle state mixing. 
That is known as the Feldmann-Kroll-Stech (FKS) mixing scheme~\cite{Feldmann:1998vh, Feldmann:1999uf}, 
where the axial $U(1)$ anomaly is assumed 
to be the only effect that mixes the two flavor states $| \eta_{q} \rangle$ and $|\eta_{s} \rangle$. 
It is therefore possible to relate the two flavor states to the physical state by the following transformation,
\begin{equation}\label{eq53}
\left(\begin{array}{c}
|\eta \rangle \\
|\eta^{\prime}\rangle
\end{array}  \right)
=
\left(\begin{array}{cc}
\cos \alpha &  -\sin \alpha    \\
\sin \alpha &  \cos \alpha     \\
\end{array}  \right)
\left(\begin{array}{c}
|\eta_{q}\rangle \\
|\eta_{s} \rangle\\
\end{array}  \right)\ ,
\end{equation}
where $|\eta\rangle$ and $|\eta^{\prime}\rangle$ are the physical states and $\alpha$ the mixing 
angle in the flavor basis. Let us now define the matrix elements of the flavor diagonal axial-vector
 and pseudo-scalar densities in terms of decay constants,  $f_{P}^{i}$ and $h_{P}^{i}$, where
 $i$ follows $q$ or $s$:
\begin{align}\label{eq54}
\langle P(q) | \bar{q} \gamma^{\mu} \gamma_{5} q | 0 \rangle = - \frac{i}{\sqrt{2}} f_{P}^{q} q^{\mu}\ , \nonumber \\
\langle P(q) | \bar{s} \gamma^{\mu} \gamma_{5} s | 0 \rangle = - i  f_{P}^{s} q^{\mu}\ , \nonumber \\
2 m_{q}\langle P(q) | \bar{s}  \gamma_{5} s | 0 \rangle = - \frac{i}{\sqrt{2}} h_{P}^{q}\ , \nonumber \\
2 m_{s}\langle P(q) | \bar{s}  \gamma_{5} s | 0 \rangle = - i  h_{P}^{s}\ ,
\end{align}
where $q$ is $u$ or $d$.
Assuming that the same angle, $\alpha$, applies to the decay constants defined previously, $f_{P}^{i}$ and $h_{P}^{i}$, 
this yields:
\begin{equation}\label{eq55}
\left(\begin{array}{cc}
f_{\eta}^{q}(h_{\eta}^{q}) &  f_{\eta}^{s}(h_{\eta}^{s}) \\
f_{\eta^{\prime}}^{q} (h_{\eta^{\prime}}^{q}) &  f_{\eta^{\prime}}^{s} (h_{\eta^{\prime}}^{s}) \\
\end{array}  \right)
=
\left(\begin{array}{cc}
\cos \alpha &  -\sin \alpha    \\
\sin \alpha &  \cos \alpha     \\
\end{array}  \right)
\left(\begin{array}{cc}
f_{q}(h_{q}) & 0\\
0 & f_{s}(h_{s}) \\
\end{array}  \right)\ .
\end{equation}
\noindent The anomaly matrix element, $a_{P}$, which is defined as
\begin{equation}\label{eq56}
\langle P(q) | \frac{\alpha_{s}}{4 \pi} G_{\mu \nu}^{A} \tilde{G}^{A,\mu \nu} | 0 \rangle = a_{P}\ ,
\end{equation}
where $G_{\mu \nu}^{A}$ is  the dual field strength tensor,
can be related to the parameters,  $f_{P}^{i}$ and $h_{i}^{q}$, through the following relation,
\begin{equation}\label{eq57}
a_{P} = \frac{h_{P}^{q}-f^{q}_{P} m^{2}_{P}}{\sqrt{2}}= h_{P}^{s}- f_{P}^{s} m_{P}^{s}\ ,
\end{equation}
by taking the divergence of the axial vector current $\bar{q} \gamma^{\mu} \gamma_{5} q$.
Therefore,  the parameters involved in the FKS scheme can be expressed in terms of
 three independent parameters, $f_{q}$, $f_{s}$ and the mixing angle $\alpha$. They have been determined from a fit 
to experimental data and their values are:
\begin{equation}\label{eq58}
f_{q}= (1.07 \pm 0.02) f_{\pi} \nonumber \ , \;\; f_{s}=  (1.34 \pm 0.06) f_{\pi}
\nonumber \ ,\;\;\;\alpha =  39.3^{o} \pm 1.0^{o} \ .
\end{equation}
%

\section{CKM matrix}\label{part3.1}
%
In most  phenomenological applications, the widely used  CKM matrix  parametrization is 
the {\it Wolfenstein parametrization}~\cite{Wolfenstein:1983yz,Wolfenstein:1964ks}. 
This  has three main advantages in 
comparison with the standard parametrization~\cite{Hocker:2001xe,Chau:1984fp}: 
it allows us to make an explicit hierarchy, in terms of strength couplings,  between quarks; 
it allows us easier analytical derivation and  finally the four independent parameters, $\lambda, A, 
\rho$ and $ \eta$, can be (in)directly measured experimentally. This parametrization 
can  also be described in a geometrical representation, the so-called the unitarity triangle (UT), 
which  offers another way to check  effects of New Physics.

By expanding each element of the CKM matrix  as a power series in the parameter $\lambda =
 \sin \theta_{c} = 0.2224$ ($\theta_{c}$ is the Gell-Mann-Levy-Cabibbo angle), one gets 
(if $O(\lambda^4)$ is neglected)
\begin{equation}\label{eq59}
{\hat V}_{CKM}= \left( \begin{array}{ccc}
1-\frac{1}{2} \lambda^{2} &  \lambda                    & A\lambda^{3}(\rho-i\eta) \\
-\lambda                  & 1-\frac{1}{2}\lambda^{2}    & A\lambda^{2}             \\
A\lambda^{3}(1-\rho-i\eta)& -A\lambda^{2}               &      1                   \\
\end{array}  \right)\ ,
\end{equation}
where $\eta$  plays the well-known role of the  $CP$-violating phase in the Standard Model
 framework. However, it may be more accurate to go beyond the leading order  in terms of $\lambda$ 
in a perturbative expansion of  the CKM matrix.  It was found that the CKM matrix takes
 the following form (up to corrections of $O(\lambda^7)$):
\begin{equation}\label{eq60}
{\hat V}_{CKM}= \left( \begin{array}{ccc}
1-\frac{1}{2} \lambda^{2}- \frac18 \lambda^4 &  \lambda                    & A\lambda^{3}(\rho-i\eta) \\
-\lambda + \frac12 A^2 \lambda^5 (1- 2 (\rho + i \eta))    & 1-\frac{1}{2}\lambda^{2}-\frac18 \lambda^4 (1+4 A^2)    
& A\lambda^{2}             \\
A\lambda^{3}(1-\bar{\rho}-i\bar{\eta})& -A\lambda^{2}+ \frac12 A \lambda^4 (1- 2 (\rho + i \eta)) &      1-\frac12 A^2 \lambda^4    \\
\end{array}  \right)\ ,
\end{equation}
where
\begin{equation}\label{eq61}
\bar{\rho}= \rho \biggl(1-\frac{\lambda^2}{2}\biggr)\;\; {\rm and} \;\;\bar{\eta}= 
\eta \biggl(1-\frac{\lambda^2}{2}\biggr)\ .
\end{equation}
The  corrections to the real and imaginary parts for the other terms can be safely neglected.
In our phenomenological application, we will take into account the above corrections  to $V_{ij}$
because  some of them  appear significant. Finally, due to the unitarity condition of 
the CKM matrix,
\begin{equation}\label{eq62}
{\hat V}_{CKM}^{\dagger} \cdot {\hat V}_{CKM} = {\hat I} = {\hat V}_{CKM} \cdot  {\hat V}_{CKM}^{\dagger}\ ,
\end{equation}
which ensures the absence of flavour-changing neutral-current (FCNC) processes
at the tree level in the Standard Model, the  most useful orthogonality relation 
in charmless  $B$ decays is given by
\begin{equation}\label{eq63}
V_{ud}V_{ub}^*+V_{cd}V_{cb}^*+V_{td}V_{tb}^*=0\ .
\end{equation}
The CKM matrix, expressed in terms of the Wolfenstein parameters,   is constrained with
several experimental data.
The main ones  are  the $b \to u l \bar{\nu}$ and  $b \to c l \bar{\nu}$ decay processes,
$s$ and $d$ mass oscillations, $\Delta m_{s}, \Delta m_{d}$, and  $CP$  violation in the 
kaon system ($\epsilon_{K}$).  In our numerical applications we will take, in case of 
$68\%$ confidence level~\cite{Charles:2004jd},
\begin{equation}\label{eq64}
 0.122 < \rho < 0.232 \;\; {\rm and }\;\; 0.334< \eta <0.414\ , 
\end{equation}
and  in case of $95\%$ confidence level~\cite{Charles:2004jd},
\begin{equation}\label{eq65}
 0.076 < \rho < 0.380  \;\; {\rm and }\;\; 0.280< \eta <0.455\ .
\end{equation}
The values for $A$ and $\lambda$ are assumed to be well determined experimentally~\cite{Charles:2004jd}:
\begin{equation}\label{eq66}
\lambda=0.2265  \;\; {\rm and }\;\; A=0.801\ .
\end{equation}
The angles $\alpha, \beta$ and $\gamma$ (to the Unitarity Triangle)  corresponding to the 
values mentioned previously  for $A, \lambda, \rho$ and $\eta$ are within the following
limits (at $95\%$ confidence level):
\begin{equation}\label{eq67}
70 < \alpha< 130\ , \;\; 20  <  \beta < 30\ , \;\;  50  < \gamma < 70\ .  
\end{equation}
%
%
\section{Input physical parameters}
%
%
\subsection{Quark masses}\label{part3.2}
%
%
The running quark masses are used in order to calculate the 
matrix elements of penguin operators as well as the chiral enhancement factors. The quark mass is  taken 
at the scale $\mu \simeq m_{b}$ in $B$ decays. Therefore one has~\cite{Cheng:2001nj} (in MeV),
\begin{eqnarray}\label{eq68}
m_{u}=  m_{d}= 3.7\ ,\; m_{s}= 90\ ,\; m_{b}= 4200\ ,\; m_{c}= 1300\ ,
\end{eqnarray}
which corresponds to $m_{s}(\mu= 1\;{\rm GeV}) = 140 \;{\rm  MeV}$. For meson  masses,
we shall use the following values~\cite{Groom:2000in} (in  GeV):
\begin{align}
 m_{B^{\pm}}& = 5.279\ ,  & m_{B^{0}}& = 5.279\ , & m_{\eta}   &= 0.515\ , \nonumber\\
m_{K^{\pm}}& = 0.493\ ,   & m_{K^{0}} &  = 0.497\ ,&  m_{\eta^{'}} &= 0.983\ , \nonumber\\
m_{\pi^{\pm}}& = 0.139\ , & m_{\pi^{0}}&  = 0.135\ ,& m_{\phi}     &=  1.019\ , \nonumber\\
m_{\rho^{0}}& = 0.769\ ,   & m_{\omega} &= 0.782\ ,  & m_{K^{*}} &= 0.894\ . \nonumber
\end{align}
%
%
\subsection{Form factors and decay constants}
%
%
The heavy(light)-to-light form factors, $F_{i}(k^{2})$ and $A_{j}(k^{2})$,
depend on 
the inner structure of  the hadrons. 
Here we shall primarily adopt the values of form factors 
for the pseudo-scalar to pseudo-scalar and pseudo-scalar to vector 
transitions 
obtained from QCD sum rule calculations. Moreover, 
we will keep the form factors 
$F_i^{B \to \pi}$ and $F_i^{B \to K}$ unconstrained, 
because of the large uncertainties entering into
their calculation. In this way, the strong 
dependence of the branching ratios,  
for the  decays $B \to X \pi$ and $B \to X K$, 
on the form factors $F_i^{B \to \pi}$ and 
$F_i^{B \to K}$ will be analyzed. For the others, their values\footnote{The uncertainties 
on these values are neglected in our approach.} are the 
following~\cite{Beneke:2002jn, Ball:1998kk,Ball:1998tj,Khodjamirian:1997ub}:
\begin{align}
 A_j^{B \to \rho}& = 0.37 \ ,  & A_j^{B \to \omega}& = 0.33\ , \nonumber\\
A_j^{B \to \phi}& = 0.0\ ,   & A_j^{B \to K^{*}}&  = 0.45\ . \nonumber
\end{align}
For the special case of $F_i^{B \to \eta^{(')}}$, the form factor is parameterized as  follows~\cite{Beneke:2002jn}:  
\begin{eqnarray}\label{eq71}
F_i^{B \to \eta^{(\prime)}}(0)= F_{1} \frac{f_{P}^{q}}{f_{\pi}} + F_{2} \frac{\sqrt{2} f_{P}^{q} 
+ f_{P}^{s}}{\sqrt{3} f_{\pi}}\ ,
\end{eqnarray}
where $F_{1}$ behaves like $F_i^{B \to \pi}(0)$ in the FKS scheme and $F_{2}$ is 
 taken to be around 0.1~\cite{Beneke:2002jn}.

The decay constants for pseudo-scalar, $f_{P}$, and vector, 
$f_{V}$, mesons do not 
suffer from uncertainties as large as those for form factors 
since they are well determined 
experimentally from leptonic and semi-leptonic decays. 
Let us first recall the usual
definition for a pseudo-scalar,
\begin{eqnarray}\label{eq72}
\langle P(q) | \bar{q}_{1} \gamma_{\mu} \gamma_{5} q_{2}| 0 \rangle & = i f_{P} q_{\mu}\ , \nonumber \\
\end{eqnarray}
\vskip -0.5cm
\noindent with $q_{\mu}$ being the momentum of  the pseudo-scalar meson. For a vector meson,
\begin{eqnarray}\label{eq73}
c \langle V(q) | \bar{q}_{1} \gamma_{\mu} q_{2} | 0 \rangle & = f_{V} m_{V} \epsilon_{V}\ ,
\end{eqnarray}
where  $m_{V}$ and $\epsilon_{V}$ are respectively the 
mass and polarization vector of
the vector meson, and $c$ is a constant depending 
on the given meson: $c=\sqrt{2}$ for the $\rho$ 
and $\omega$ and $c=1$ otherwise. Finally, the 
transverse decay constant, appearing in 
the chiral-enhanced factor,  is given by,
\begin{eqnarray}\label{eq74}
\langle V(q) | \bar{q}_{1} \sigma_{\mu \nu}  q_{2}| 0 \rangle & = f_{V}^{\perp}
(q_{\mu}\epsilon^{*}_{\nu}- q_{\nu}\epsilon^{*}_{\mu}) \ . \nonumber \\
\end{eqnarray}
\vskip -0.5cm
\noindent Numerically, in our calculations, for the decay constants we  take
 (in MeV)~\cite{Groom:2000in,Beneke:2002jn},
\begin{eqnarray}\label{eq75}
  f_{\pi}  = 131\ , \;\;   f_B = 200\ ,  \;\;  f_{\rho}  = 209\ ,  \;\;  f_K  = 160 \ ,   \;\;   f_{\phi}  = 221\ ,   \;\; 
 f_{\omega}=187\ ,  \;\; f_K^{*}  = 218\ ,
\end{eqnarray}
and for the transverse vector decay constants (in MeV)~\cite{Beneke:2002jn},
\begin{eqnarray}\label{eq76}
f_{\rho}^{\perp}= 150 \ ,\;\; f_{K^{*}}^{\perp}= 175 \ , \;\; f_{\omega}^{\perp}= 150\ , \;\; f_{\phi}^{\perp}= 175\ . 
\end{eqnarray}
Finally, for the total $B$ decay width, $\Gamma_{B}(= 1/\tau_{B})$, we
use the  world average $B$ life-time values (combined results from ALEPH, CDF, DELPHI, L3, OPAL 
and  SLD)~\cite{Abbaneo:2001bv, Cheng:2000em, Aubert:2003se}:
\begin{align}\label{eq5.15}
\tau_{B^{0}} & = 1.546 \pm 0.021 \; {\rm ps}\ , \nonumber \\
\tau_{B^{+}} & = 1.647 \pm 0.021 \; {\rm ps}\ .
\end{align}
%
%
\subsection{Light cone distribution amplitude}
%
%
QCD factorization involves the light cone distribution amplitude (LCDA) of the mesons where the 
leading twist (twist-2) 
and sub-leading twist (twist-3) distribution amplitudes are taken into account. For a light 
pseudo-scalar meson the LCDA is defined as,
\begin{multline}\label{eq77}
 \langle P(k)| \bar{q}(z_{2})q(z_{1})| 0 \rangle = \\
\frac{if_{P}}{4}{\int}_{0}^{1}dx \ e^{i(xk{\cdot}z_{2}+\bar{x}k{\cdot}z_{1})} \Biggl\{
    k\!\!\!\slash {\gamma}_{5} {\Phi}_{P}(x) - {\mu}_{P} {\gamma}_{5} \biggr[ {\Phi}_{P}^{p}(x)
    - {\sigma}_{{\mu}{\nu}}k^{\mu}z^{\nu}  \frac{{\Phi}_{P}^{\sigma}(x)}{6} \biggr] \Biggr\}\ ,
\end{multline}
where $f_{P}$ is a decay constant, ${\mu}_{P}$ is the chiral enhancement factor and 
$z=z_{2}-z_{1}$. ${\Phi}_{P}(x), {\Phi}_{P}^{p}(x)$ and ${\Phi}_{P}^{\sigma}(x)$ are the 
leading twist and sub-leading twist LCDA's of the mesons, respectively. 
All  distributions are normalized to one. Neglecting
three-particle distributions, such as quark-antiquark-gluon, it
follows from the equations of motion that the asymptotic forms of
the LCDA's must be used for twist-three two particle distribution
amplitudes. They take the forms: 
\begin{eqnarray}\label{eq78}
  {\Phi}_{P}^{p}(x)=1, \;\;\;\;
  {\Phi}_{P}^{\sigma}(x)=6x (1-x)\ .
\end{eqnarray}
However, by taking into account the  higher order terms in the expansion involving 
Gegenbauer polynomials,  the leading-twist light cone amplitude,  $\Phi_P(x)$, becomes:
\begin{eqnarray}\label{eq79}
\Phi_P(x)= 6 x(1-x) \biggl[ 1 + \sum_{n=1}^{\infty} \alpha_{n}^{P}(\mu) C_{n}^{3/2} (2x-1)\biggr]\ ,
\end{eqnarray}
where $\alpha_{n}^{P}(\mu)$ are the 
Gegenbauer moments that depend on the scale $\mu$.
$C_{n}^{3/2}(u)$ are coefficients given by $C_{1}^{3/2}(u)= 3 u$
for $n=1$ and $C_{2}^{3/2}(u)= (3/2) (5 u^{2}-1)$ for $n=2$. 
Regarding the LCDAs of the vector mesons, 
the usual definitions applied here are,
\begin{align}\label{eq80}
 \langle0 | \bar{q}(0){\sigma}_{{\mu}{\nu}}q(z) | V(k,{\lambda}){\rangle}&=
  i ( {\epsilon}_{\mu}^{\lambda}k_{\nu}- {\epsilon}_{\nu}^{\lambda}k_{\mu} )
      f_{V}^{\bot} {\int}_{0}^{1} dx \ e^{-ixk{\cdot}z} {\Phi}_{V}^{\bot}(x)\ , \\
 \langle 0 | \bar{q}(0){\gamma}_{\mu}q(z) | V(k,{\lambda})\rangle &=
    k_{\mu} \frac{{\epsilon}^{\lambda}{\cdot}z}{k{\cdot}z}
    f_{V} m_{V} {\int}_{0}^{1} dx \ e^{-ixk{\cdot}z} {\Phi}_{V}^{\|}(x)\ ,
\end{align}
where  $\epsilon$ is the polarization  vector.  
${\Phi}_{V}^{\bot}(x)$ and ${\Phi}_{V}^{\|}(x)$ are the transverse and longitudinal 
quark distributions of the polarized mesons. Assuming that  the contributions from
${\Phi}_{V}^{\bot}(x)$  are power suppressed,  ${\Phi}_{V}(x)$  takes  the following form,
\begin{eqnarray}\label{eq81}
 {\Phi}_{V}(x)={\Phi}_{V}^{\|}(x)=6x (1-x)\ .
\end{eqnarray}
Similarly to the LCDA for a light pseudo-scalar meson, the leading twist distribution
amplitude for a light vector meson is expanded in terms of Gegenbauer polynomials, 
where $\alpha_{n}^{V}(\mu)$ replaces $\alpha_{n}^{P}(\mu)$. The sub-leading twist 
distribution amplitude, $\Phi_v(x)$, can be written as  follows when the three 
particle distribution is neglected: 
\begin{eqnarray}\label{eq82}
\Phi_v(x)= 3 \sum_{n=1}^{\infty} \alpha_{n,\perp}^{v}(\mu) P_{n+1}(2x-1) \ ,
\end{eqnarray}
where $P_{n+1}(x)$ are the Legendre polynomials  defined such as 
$P_{1}(u)=u$ for $n=0$, $P_{2}(u)=(1/2) (3 u^{2}-1)$ for $n=1$ and  
$P_{3}(u)=(1/2) u (5 u^{2}-3)$ for $n=2$.
Concerning the parameters, $\alpha_{n,\perp}^{v}(\mu)$ and $\alpha_{n}^{M}(\mu)$,
appearing in the expansion of meson distribution amplitudes  in Gegenbauer polynomial forms,
the values used for pseudo-scalar mesons are~\cite{Beneke:2002jn},
\begin{align}\label{eq83}
\alpha_{1}^{\pi} & = 0.0 \ ,  & \alpha_{1}^{K}& = 0.2\ ,  & \alpha_{1}^{\eta}& = 0.0\ ,   
& \alpha_{1}^{\eta^{\prime}}& = 0.0\ , \nonumber \\
\alpha_{2}^{\pi} & = 0.1\ ,   &  \alpha_{2}^{K}&  = 0.1\ ,  & \alpha_{2}^{\eta}& = 0.0\ ,  
& \alpha_{2}^{\eta^{\prime}}& = 0.0\ ,
\end{align}
and for vector mesons~\cite{Beneke:2002jn},
\begin{align}\label{eq84}
\alpha_{1}^{\rho} & = 0.0\ ,  & \alpha_{1}^{\omega}& = 0.0\ ,  & \alpha_{1}^{K^{*}}& = 0.2\ ,   
& \alpha_{1}^{\phi}& = 0.0\ , \nonumber \\
\alpha_{2}^{\rho} & = 0.1\ ,   &  \alpha_{2}^{\omega}&  = 0.0\ ,  & \alpha_{2}^{K^{*}}& = 0.1\ ,  
& \alpha_{2}^{\phi}& = 0.0\ ,\nonumber \\
\alpha_{0,\perp}^{\rho} & = 1.0 \ ,  & \alpha_{0,\perp}^{\omega}& = 1.0\ ,  & \alpha_{0,\perp}^{K^{*}}& = 1.0\ ,   
& \alpha_{0,\perp}^{\phi}& = 1.0\ , \nonumber \\
\alpha_{1,\perp}^{\rho} & = 0.0\ ,   &  \alpha_{1,\perp}^{\omega}&  = 0.0\ ,  & \alpha_{1,\perp}^{K^{*}}& = 0.2\ ,  
& \alpha_{1,\perp}^{\phi}& = 0.0\ ,\nonumber \\
\alpha_{2,\perp}^{\rho} & = 0.1\ ,   &  \alpha_{2,\perp}^{\omega}&  = 0.0\ ,  & \alpha_{2,\perp}^{K^{*}}& = 0.1\ ,  
& \alpha_{2,\perp}^{\phi}& = 0.0\ .
\end{align}
It has to be emphasized that the effects of the above parameters, $\alpha_{n}^{M}$, are 
small enough to support large uncertainties on their given values. 
Finally, $\Phi_v(x)$ and $\Phi_p(x)=(\Phi_P^p(x))$ exhibit unlikely  endpoint divergences (for  $x=0,1$) which
enter into the calculation of the hard spectator scattering kernels and weak 
annihilation contributions. These divergences have been discussed in section 3. 
%
\section{Results and discussions}
%
Assuming that the  parameters such as decay constants, Gegenbauer parameters, quark and meson 
masses, decay widths, involved in QCD factorization have been
constrained by independent studies, we concentrate our efforts on the analysis of the 
form factors, $F^{B\to \pi}_i$ and $F^{B \to K}_i$, the CKM matrix parameters, $\rho$ 
and $\eta$, as well as the hard-scattering (annihilation) phases, $\varphi_{H}^{M_{i}} 
(\varphi_{A}^{M_{i}})$, and parameters, $\varrho_{H}^{M_{i}} (\varrho_{A}^{M_{i}})$, respectively. 
These phases, $\varphi_{H,A}^{M_{i}}$, and  parameters, $\varrho_{H,A}^{M_{i}}$, arise 
in QCD factorization because of divergences coming from the endpoint integrals when the hard scattering 
and annihilation contributions are calculated. We recall that these divergences are parameterized by,
\vspace{0.2cm}
\begin{equation}\label{eq85}
   X^{M_i}_{H,A} = \left( 1 + \varrho^{M_{i}}_{H,A}\,e^{i\varphi^{M_{i}}_{H,A}} \right)
   \ln\frac{m_B}{\Lambda_h} \,, \qquad {\rm with} \;\;
    \qquad \Lambda_h=0.5\,\mbox{GeV} \,,
\end{equation}
where we allow $\varrho^{M_{i}}_{H,A}$ and $\varphi^{M_{i}}_{H,A}$ to vary in the  range of values 
$[-3,+3]$ and $[-180^{0},+180^{0}]$, respectively. The values of  $\varrho^{M_{i}}_{H,A}$ and 
$\varphi^{M_{i}}_{H,A}$ are not universal, since the hard scattering and annihilation 
contributions  depend,  for a given process, on the flavour of  particles in the 
final state. Therefore, each of the pseudo-scalar mesons, $M_{i}$, $(\pi, K, \eta^{(')})$, 
and vector mesons, $M_{i}^*$, $(\rho, \omega, K^*, \phi)$, will have its own set of values for 
$\varrho^{M_{i}}_{H,A}$ and $\varphi^{M_{i}}_{H,A}$.

We have calculated 
the branching ratios (listed below) for $B$ decays into two mesons in the final 
state. We have  focused on  the branching ratios of $B$ decays including a kaon(or a pion) and 
another meson. We also cross checked our results by analyzing  the branching 
ratios including a pion and a kaon in the final state. The determination of our parameters is performed  
by making comparison with experimental and theoretical results in order to obtain the best fit. We take 
into account all the latest data for $b \to s$ and $b \to u$ transitions concentrating on the
CLEO~\cite{Bornheim:2003bv,Eckhart:2002qr,Briere:2001ue,Jessop:2000bv,
Richichi:1999kj,Aubert:2003nm,Aubert:2003wr}, BABAR~\cite{Aubert:2003wr,Aubert:2003fa,Aubert:2003ez,
Aubert:2003tk,Aubert:2003qj,Aubert:2003dn,Aubert:2002jm,Aubert:2002jb,Aubert:2002ng,Aubert:2001zf} 
and BELLE~\cite{Abe:2003rj,unknown:2003jf,Tomura:2003cb,Gordon:2002yt,Huang:2002ev,Abe:2002av} 
branching ratio results. The experimental $CP$ violating measurements are not taken into account
in our analysis. We will  first discuss the  branching ratio results 
for the following  $B$ decay channels without  a pion in the final state:
\begin{align}\label{eq86}
B^- & \to\eta K^-\ , & \bar B^0 & \to\bar K^0 K^0\ , &
\bar B^0 & \to\eta\bar K^0\ , & \bar B^0 & \to K^- K^+\ , \nonumber\\
B^- & \to\eta' K^-\ , & B^- & \to\bar K^0\rho^-\ , &
\bar B^0 & \to\eta'\bar K^0\ , & B^- & \to K^-\rho^0\ , \nonumber\\
B^- & \to K^-\phi\ ,  & \bar B^0 & \to K^-\rho^+\ , &
\bar B^0 & \to\bar K^0\phi\ , &  \bar B^0 & \to\bar K^0\rho^0\ ,\nonumber\\
B^- & \to K^- K^{*0}\ , & B^- & \to K^-\omega\ ,  &
B^- & \to K^- K^0\ , &  \bar B^0 & \to\bar K^0\omega\ ,  \nonumber\\
\bar{B}^0 & \to K^- K^{*+}\ , & \bar{B}^0 & \to \bar{K}^{0} \bar{K}^{*0}\ , &
B^- & \to K^{*-} K^0\ , &  \bar{B}^0 & \to K^{*-} K^+\ , \nonumber\\
\bar{B}^0 & \to \bar{K}^{*0} K^0\ , & & &  & & & & 
\end{align}
then the  $B$ decay channels without  a kaon in the final state:
\begin{align}\label{eq87}
B^- & \to\pi^-\rho^0\ , & \bar B^0 & \to\pi^0\phi\ , &
B^- & \to\pi^0\rho^-\ , &  B^- & \to\pi^-\bar K^{*0}\ , \nonumber\\
\bar B^0 & \to\pi^+\rho^-\ ,&  B^- & \to\pi^0 K^{*-}\ , &
\bar B^0 & \to\pi^-\rho^+\ , & \bar B^0 & \to\pi^+ K^{*-}\ , \nonumber\\
\bar B^0 & \to\pi^\pm\rho^\mp\ , &  \bar B^0 & \to\pi^0\bar K^{*0}\ , &
\bar B^0 & \to\pi^0\rho^0\ , & B^- & \to\pi^-\pi^0\ , \nonumber\\ 
B^- & \to\pi^-\omega\ , & \bar B^0 & \to\pi^+\pi^-\ , &
\bar B^0 & \to\pi^0\omega\ , & \bar B^0 & \to\pi^0\pi^0\ ,  \nonumber\\
B^- & \to\pi^-\phi\ , &  B^- & \to\pi^-\eta\ , &
\bar B^0 & \to\pi^0\eta\ , &  B^- & \to\pi^-\eta'\ , \nonumber\\
\bar B^0 & \to\pi^0\eta'\ , &  & 
\end{align} 
then  the $B$ decay channels with a kaon and a pion in the final state:
\begin{align}
B^- & \to\pi^0 K^-\ , & \bar B^0 & \to\pi^0\bar K^0\ , & \bar B^0 & \to\pi^+ K^-\ , & B^- & \to\pi^-\bar K^0\ ,
\end{align} 
then  ratios of the $B$ branching ratios including either a pion or a kaon in the final state:
\begin{align}\label{eq88}
\frac{\tau^{B^+}}{2 \tau^{B^0}}\Bigl[\frac{\bar{B}^0 \to \pi^+ \pi^-}{B^+ \to \pi^+ \pi^0}\Bigr]\ , & \hspace{1.5cm}
\frac{\tau^{B^0}}{\tau^{B^+}}\Bigl[\frac{B^- \to \pi^- \pi^0}{\bar{B}^0 \to\pi^0 \pi^0}\Bigr]\ ,  &
\Bigl[ \frac{2 B^{\pm} \to \pi^0 K^\pm}{B^{\pm} \to \pi^\pm \bar{K}^0}\Bigr]\ , \nonumber \\
\frac{\tau^{B^+}}{\tau^{B^0}}\Bigl[\frac{\bar{B}^0 \to \pi^\pm K^\mp}{ B^{\pm} \to \pi^\pm \bar{K}^0}\Bigr]\ , & \hspace{1.5cm}
\Bigl[ \frac{\bar{B}^0 \to \pi^\mp K^\pm}{ \bar{B}^0 \to \pi^0 \bar{K}^0}\Bigr]\ , & 
\frac{\tau^{B^0}}{\tau^{B^+}}\Bigl[\frac{B^- \to \pi^- \bar K^{\ast 0}}{ \bar{B}^0 \to \pi^+ K^{\ast -}}\Bigr]\ , \nonumber \\ 
\frac{\tau^{B^0}}{\tau^{B^+}}\Bigl[\frac{B^- \to K^- \phi}{\bar{B}^0 \to \bar{K}^0 \phi}\Bigr]\ ,  & \hspace{1.5cm}
\frac{\tau^{B^0}}{\tau^{B^+}}\Bigl[\frac{B^- \to K^- \eta^{\prime}}{\bar{B}^0 \to \bar{K}^0 \eta^{\prime}}\Bigr]\ , &
\frac{\tau^{B^0}}{\tau^{B^+}}\Bigl[\frac{B^- \to K^- \omega}{\bar{B}^0 \to \bar{K}^0 \omega}\Bigr]\ , \nonumber \\ 
\frac{\tau^{B^+}}{\tau^{B^0}}\Bigl[\frac{\bar{B}^0 \to \pi^\pm \rho^\mp}{B^- \to \pi^- \rho^0}\Bigr]\ . & & 
\end{align}
Finally, based on this previous analysis, we will investigate the  $CP$ violating asymmetries for the 
 $B$ decay channels including  $\rho-\omega$ mixing effects:
\begin{align}\label{eq89}
\bar{B}^0 & \to \rho^{0}(\omega) \pi^0      \to  \pi^+ \pi^- \pi^0\ ,\nonumber \\
\bar{B}^0 & \to \rho^{0}(\omega) \bar{K}^0  \to  \pi^+ \pi^- \bar{K}^0\ ,\nonumber \\
B^- & \to \rho^{0}(\omega) \pi^-            \to  \pi^+ \pi^- \pi^-\ ,\nonumber \\
B^- & \to \rho^{0}(\omega) K^-              \to  \pi^+ \pi^- K^-\ . 
\end{align} 
%
%
\subsection{Branching ratios}
%
The first step of our analysis is to fit the hard-scattering (annihilation) 
phases, $\varphi_{H}^{M_{i}} (\varphi_{A}^{M_{i}})$, and parameters, 
$\varrho_{H}^{M_{i}} (\varrho_{A}^{M_{i}})$, in order to reproduce branching 
ratios for the decay channels $B \to X \pi$ and  $B\to X K$. We keep unconstrained the heavy to light transition 
form factors, $F^{B \to \pi}$ and $F^{B \to K}$, and we use for the other transition form factors, 
the numerical values given in section 6. 
The latter form factors are usually given by the QCD sum rule
calculations. The second step  uses our 
former results in order to show the dependence of the  branching ratios (for 
the $B$ decay channels mentioned previously)  on the CKM 
matrix parameters, $\rho$ and 
$\eta$, as well as on the form factors, 
$F^{B \to \pi}$ and $F^{B \to K}$. A comparison 
analysis is also made between
the naive and QCD factorization approaches. In the case
of naive factorization, we will not include annihilation contributions.
As a reminder, the branching ratio 
definition\footnote{In Eq.~(\ref{eq90}) $\tau_B$ denotes 
either $\tau^{B^0}$ or $\tau^{B^+}$.}
is usually given by:
\begin{equation}\label{eq90}
\mathcal{BR}(B  \rightarrow M_1 M_2)= \frac{ \tau_B}{2 \pi \alpha_k  m_{B}}
\Bigg| V^{T}A^{T}_{M_1,M_2}(a_{1},a_{2})-V^{P}A^{P}_{M_1,M_2}(a_{3}, \cdots, a_{10}) \Bigg|^{2}\ ,
\end{equation}
where the quadratic term in the light meson mass is neglected.
Regarding the branching ratio of  $B \to M_1 M_2$ decays with $M_1\equiv \rho^{0}$, 
to the first order of  isospin violation, it takes the form:
\begin{multline}\label{eq91}
\mathcal{BR}(B  \rightarrow \rho^{0} M_2)=\frac{\tau_B }{2 \pi \alpha_k  m_{B}}\Bigg|
\bigg[V^{T}A^{T}_{\rho^{0},M_2}(a_{1},a_{2})-V^{P}A^{P}_{\rho^{0},M_2}(a_{3}, \cdots, a_{10})\bigg]    \\
+\bigg[V^{T}A^{T}_{\omega,M_2}(a_{1},a_{2})-V^{P}A^{P}_{\omega,M_2}(a_{3}, \cdots, a_{10})\bigg]\frac{
\tilde{\Pi}_{\rho \omega}}{(s_{\rho}-m_{\omega}^{2})+im_{\omega}\Gamma_{\omega}}\Bigg|^{2}\ ,
\end{multline}
with all  usual definitions. $\alpha_k$ holds for 8 or 16 according to  the given decay.

\begin{figure}
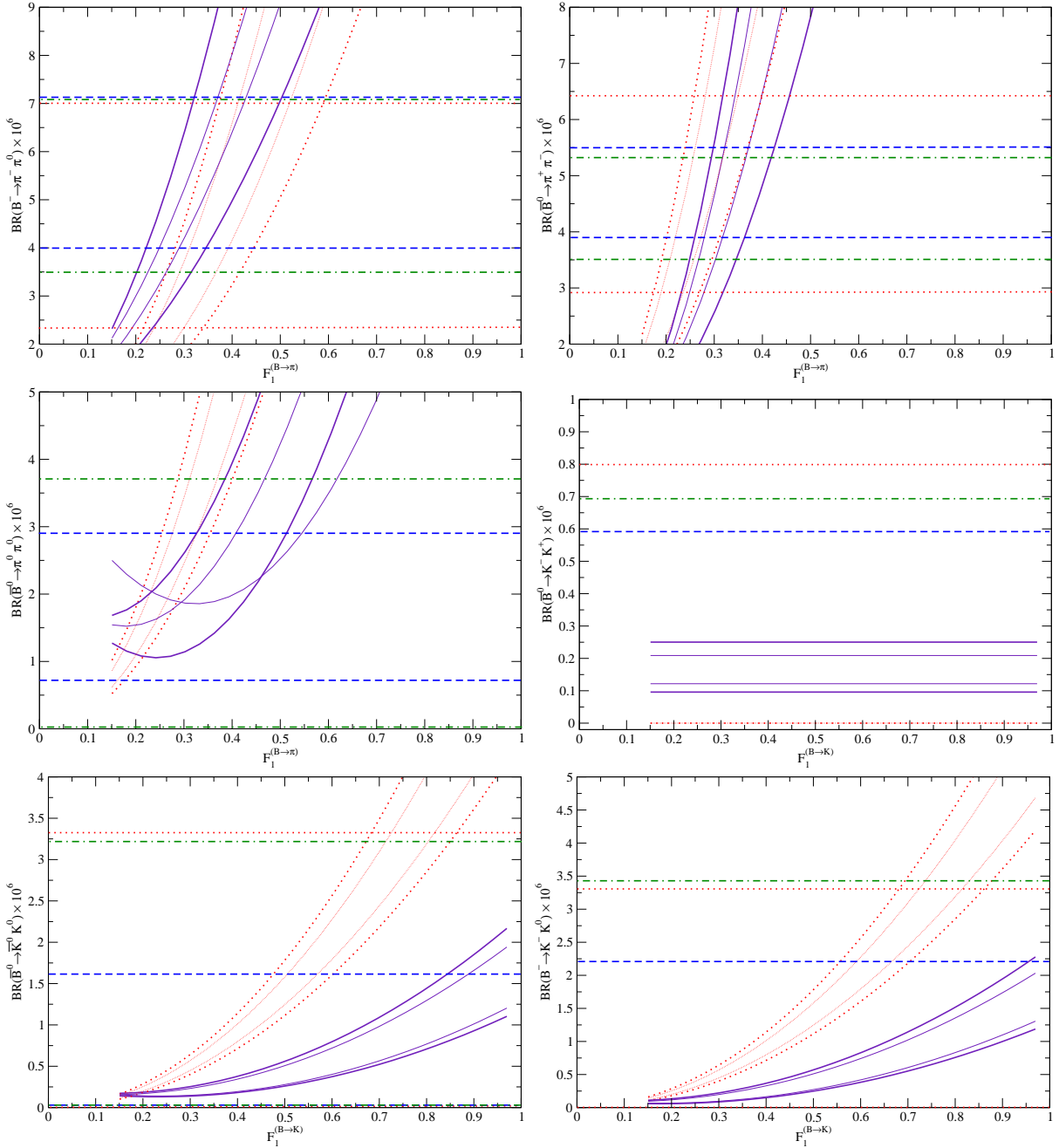

\includegraphics*[width=0.495\columnwidth]{QCDF_frame_BR_pmpz.eps}
\includegraphics*[width=0.495\columnwidth]{QCDF_frame_BR_pppm.eps}
\includegraphics*[width=0.495\columnwidth]{QCDF_frame_BR_pzpz.eps}
\includegraphics*[width=0.495\columnwidth]{QCDF_frame_BR_kmkp.eps}
\includegraphics*[width=0.495\columnwidth]{QCDF_frame_BR_kzbkz.eps}
\includegraphics*[width=0.495\columnwidth]{QCDF_frame_BR_kmkz.eps}
\caption{From top left-handed to bottom right-handed, branching ratios  
$\mathcal{BR}(B^- \to \pi^- \pi^0), \mathcal{BR}(\bar{B}^0 \to \pi^+ \pi^-),  
\mathcal{BR}(\bar{B}^0 \to \pi^0 \pi^0), \mathcal{BR}(\bar{B}^0 \to K^- K^+), 
\mathcal{BR}(\bar{B}^0 \to \bar{K}^0 K^0),$ $\mathcal{BR}(B^- \to K^- K^0)$, as a function of 
CKM matrix elements and form factors $F^{B \to \pi}$ or $F^{B \to K}$. Solid line(dotted line) for 
QCDF(NF) factorization and  for limiting values of the  CKM matrix element parameters
shown  respectively at 68\% (thin line) and 95\% (thick line) of confidence level. 
Notation: horizontal dotted lines: CLEO data; horizontal dashed lines:
BABAR data; horizontal dot-dashed lines: BELLE data.}
\label{fig10}
\end{figure}

\begin{figure}
\includegraphics*[width=0.495\columnwidth]{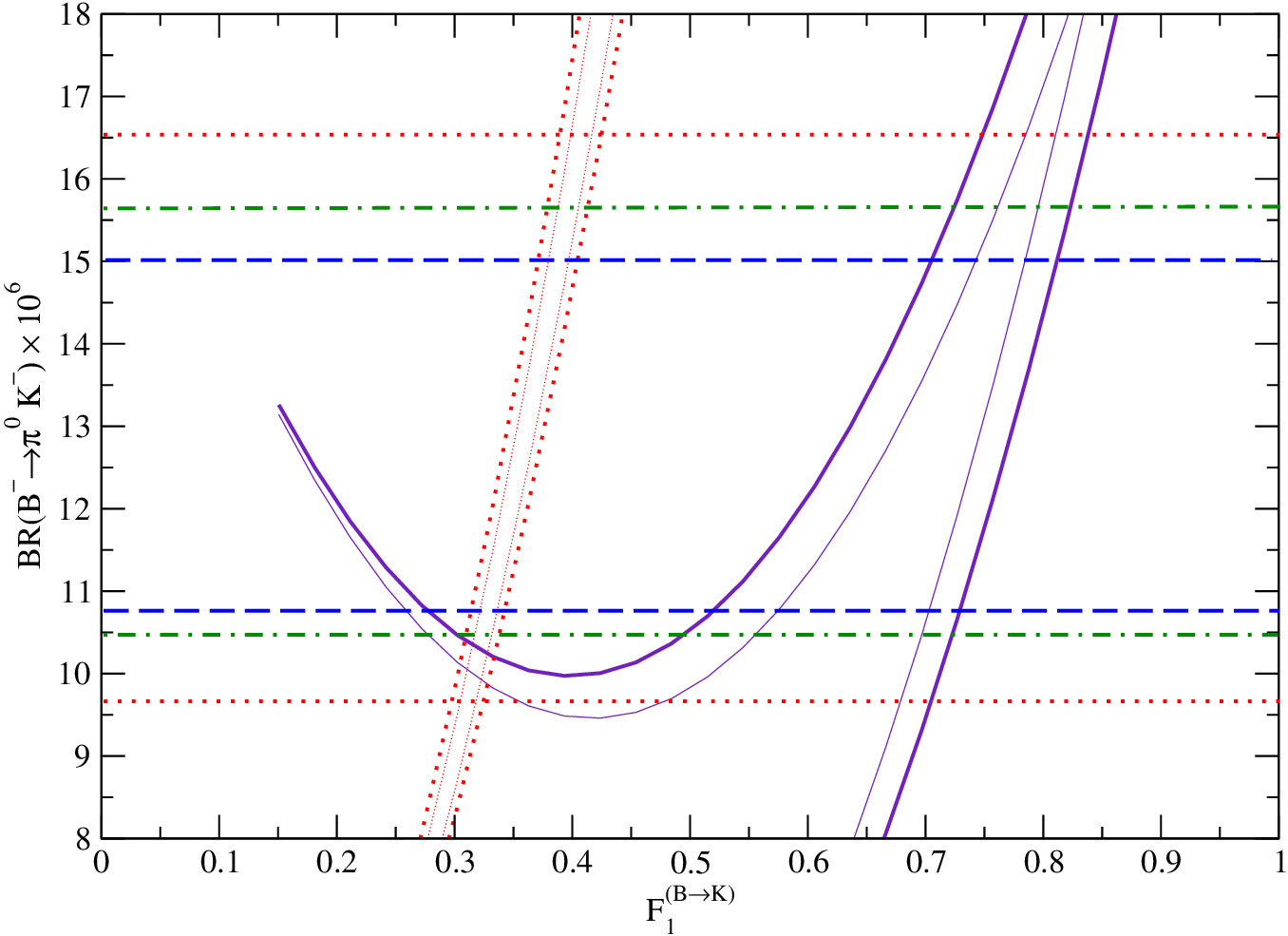}
\includegraphics*[width=0.495\columnwidth]{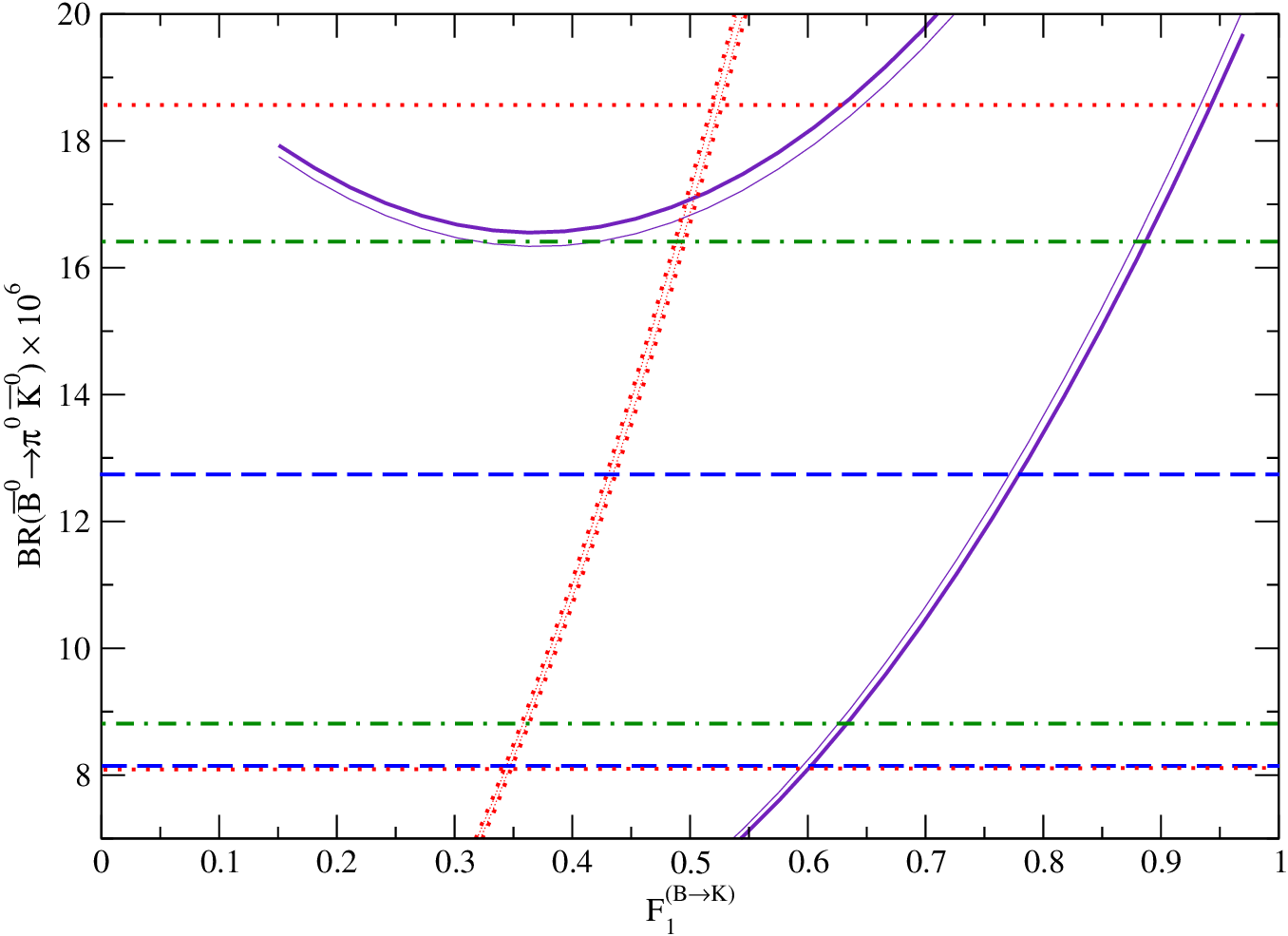}
\includegraphics*[width=0.495\columnwidth]{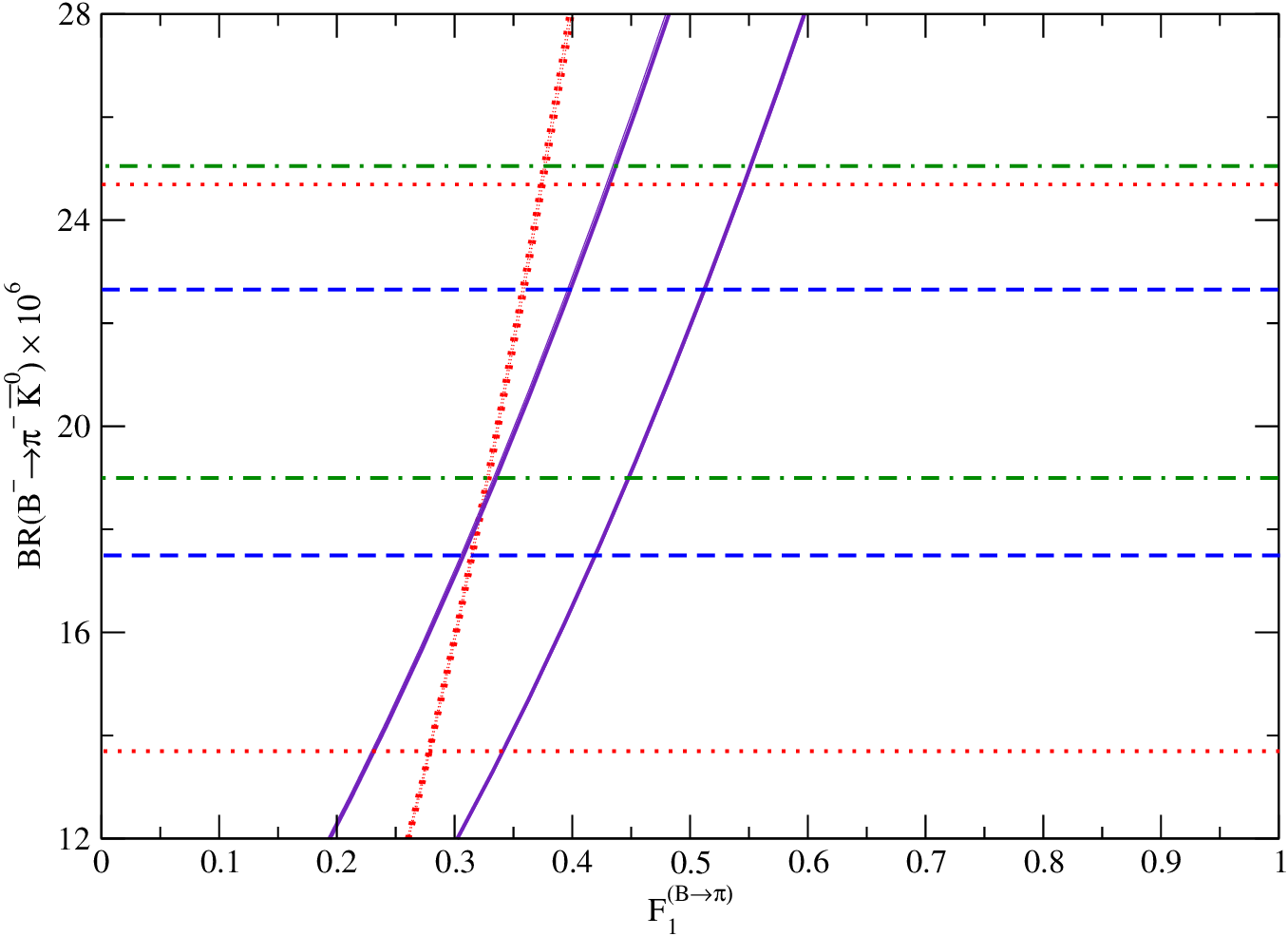}
\includegraphics*[width=0.495\columnwidth]{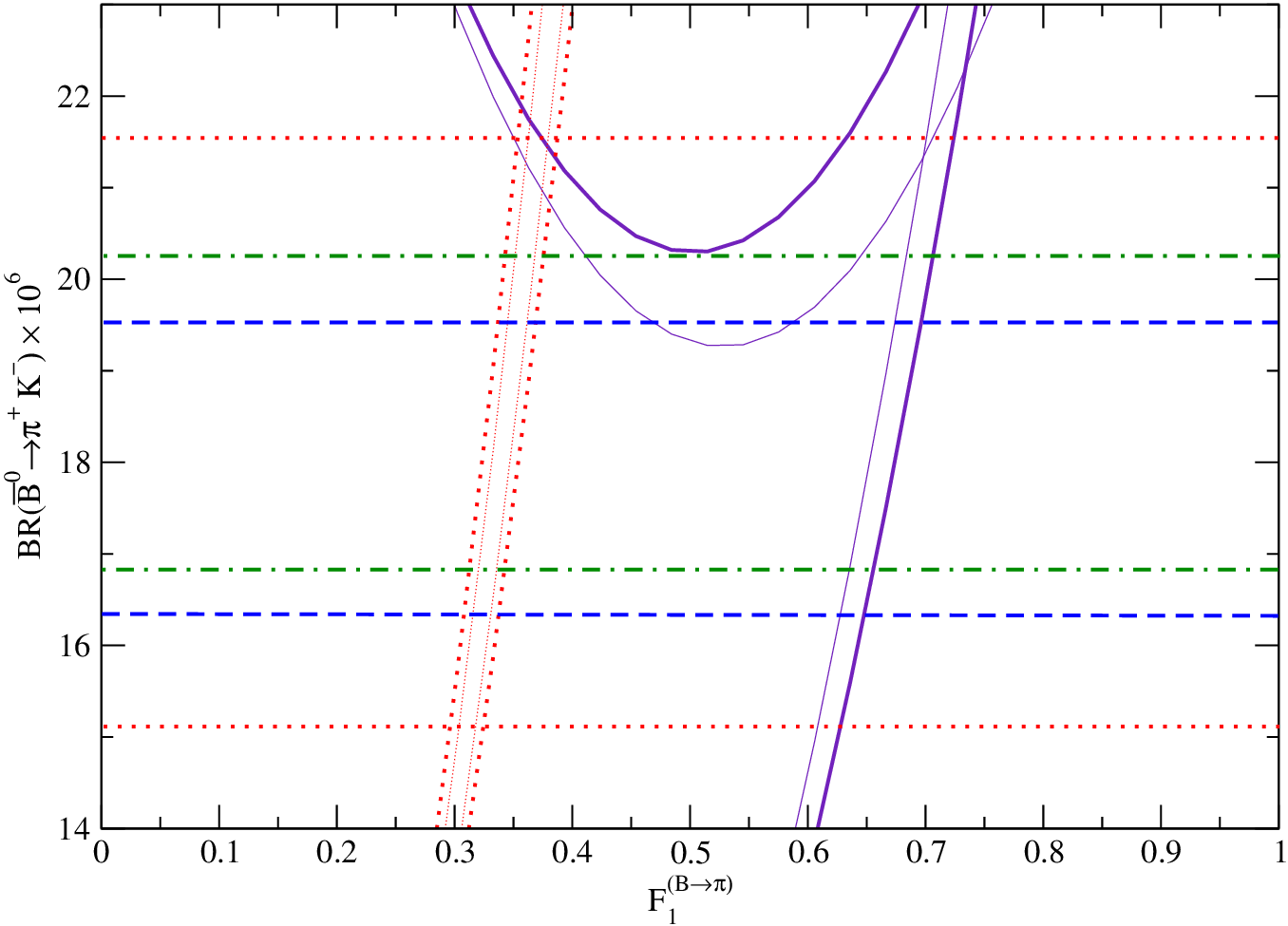}
\includegraphics*[width=0.495\columnwidth]{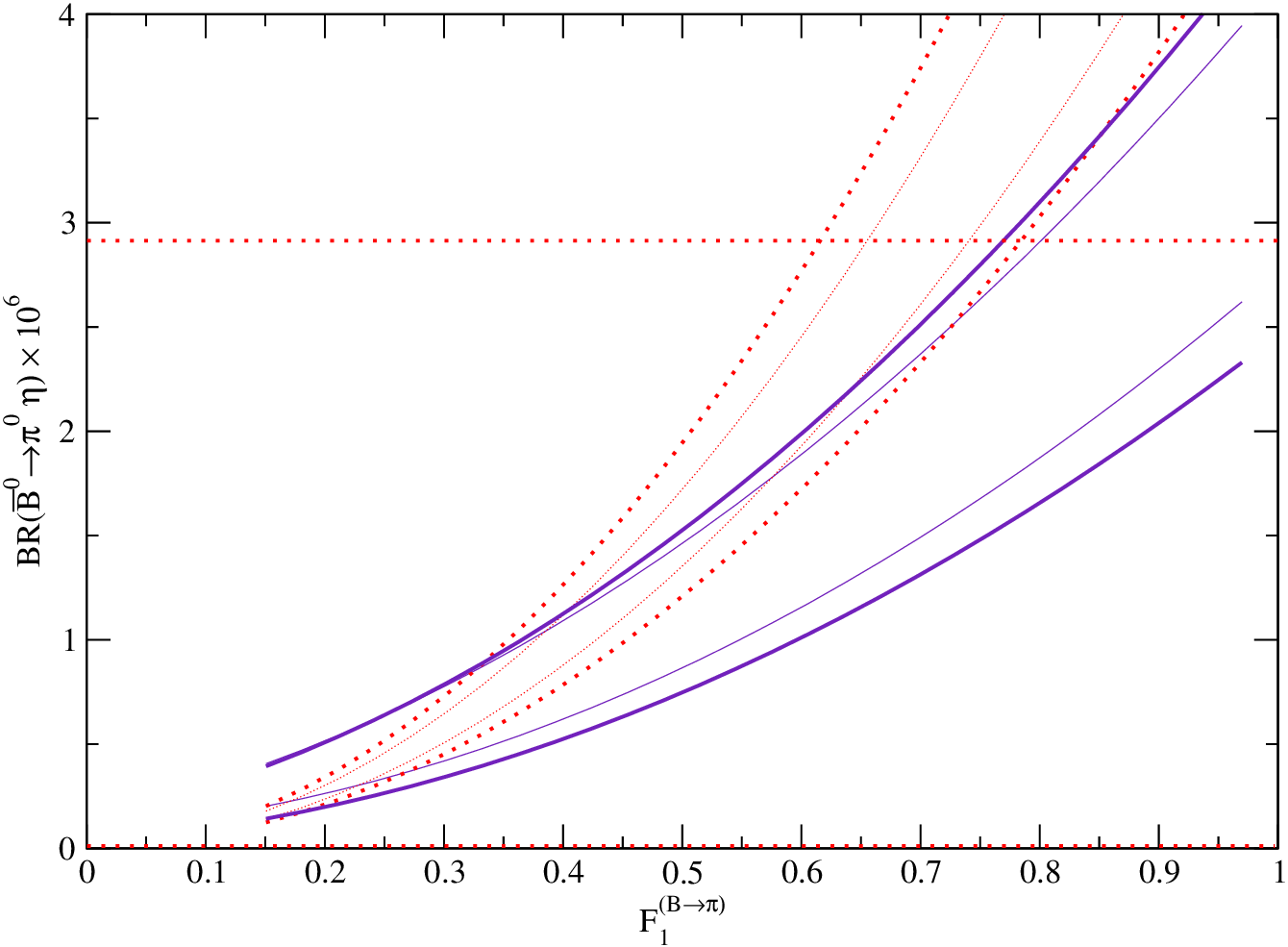}
\includegraphics*[width=0.495\columnwidth]{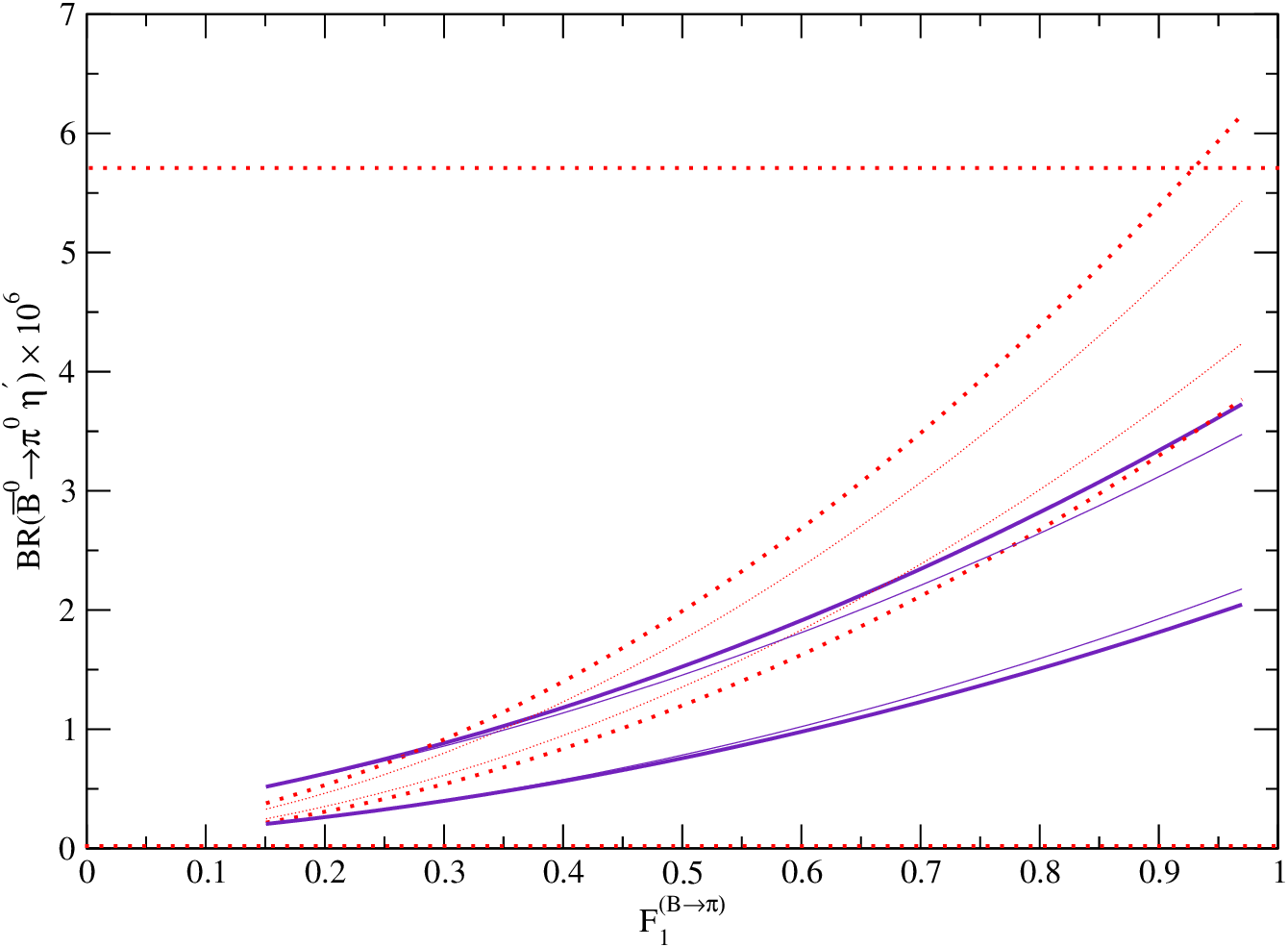}
\caption{From top left-handed to bottom right-handed, branching ratios  
$\mathcal{BR}(B^- \to \pi^0 K^-), \mathcal{BR}(\bar{B}^0 \to \pi^0 \bar{K}^0), 
\mathcal{BR}(B^- \to \pi^- \bar{K}^{0}), \mathcal{BR}(\bar{B}^0 \to \pi^+ K^-), 
\mathcal{BR}(\bar{B}^0 \to \pi^0 \eta), $ $ \mathcal{BR}(\bar{B}^0 \to \pi^0 \eta^{\prime})$, 
as a function of CKM matrix elements and form factors $F^{B \to \pi}$ or $F^{B \to K}$. 
Solid line(dotted line) for QCDF(NF) factorization and  for limiting values of the  
CKM matrix element parameters shown  respectively at 68\% (thin line) and 95\% 
(thick line) of confidence level. Notation: horizontal dotted lines: CLEO data; 
horizontal dashed lines: BABAR data; horizontal dot-dashed lines: BELLE data.}
\label{fig11}
\end{figure}

All the experimental branching ratios for $B$ decays are given in Tables~\ref{tab2},~\ref{tab3},~\ref{tab4} 
and~5 for $B \to X K$, $B \to X \pi$, $B \to \pi K$ and the ratios of 
 $\mathcal{BR}(B \to X K)$/$\mathcal{BR}(B \to X \pi)$, respectively. All the theoretical branching ratios 
are plotted from Fig.~\ref{fig10} to Fig.~\ref{fig20} as a function of the form factors, $F^{B \to \pi}$ or 
$F^{B \to K}$. The variations of the branching ratios with the CKM matrix parameters, $\rho$ and $\eta$, are 
taken into account in our graphs. Finally, in the case of the naive factorization, we use $N_{c}^{eff}=3$ 
for the effective parameter. The reason is to clearly show the differences (the hard scattering spectator and 
annihilations contributions) arising between the two factorization approaches, NF and QCDF.
%
\subsubsection{$\boldsymbol{B\to X \pi}$ with $\boldsymbol{X=\{\rho, \omega, \eta^{(')}, \pi, K^*, \phi\}}$}
%
Let us start by analyzing the branching ratios of  $B$ decay channels  $B\to X \pi$
where $X$ holds for the particles $\rho, \omega, \eta^{(')}, \pi, K^*$ or  $\phi$. In Table~\ref{tab6}, 
are listed the different values for the hard-scattering (annihilation) phases, 
$\varphi_{H}^{M_{i}} (\varphi_{A}^{M_{i}})$, and parameters, $\varrho_{H}^{M_{i}} (\varrho_{A}^{M_{i}})$.
 The values are given  for each of the following  particles $(M_i)$, $\rho, \omega, \eta^{(')}, \pi, K^*$ and  $\phi$, 
and for the minimal (set 2) and maximal (set 1) sets of 
CKM parameters $\rho$ and $\eta$. All the 
experimental results from the BELLE, BABAR and CLEO factories for  $B\to X \pi$ decays are listed
in Tables~\ref{tab3} and~5. The theoretical branching ratios  calculated with the NF and
QCDF factorizations are shown in Tables~7 and~\ref{tab8}. The annihilation and 
hard scattering spectator contributions are explicitly given in these tables.

{}For the decay channel\footnote{This notation 
stands for all of the different channels 
involving two pions. It will be used in a similar way for 
the other decays.}  $B\to \pi \pi$,
our results are shown in Figs.~\ref{fig10} and~\ref{fig19}. 
It appears that there are 
no large discrepancies between the NF and QCDF 
approaches when the form factor,
$F^{B \to \pi}$, is within the range 0.24-0.35. 
Both frameworks yield agreement with 
the experimental results given by the BELLE, BABAR and CLEO (BBC) 
measurements as well as with
the different experimental constraints for the CKM 
matrix  parameters, $\rho$ and $\eta$ when  
the form factor,  $F^{B \to \pi}$, takes  values within 0.24-0.34. 
We underline that the strong
dependence of the branching ratios for $B\to \pi \pi$ 
on the  form factor, $F^{B \to \pi}$, 
provides an excellent test for the QCDF framework if we assume correct 
the value obtained for the  form factor, $F^{B \to \pi}$. The weak contribution of 
the annihilation terms does not
increase the dependence of these branching ratios on the CKM matrix parameters $\rho$ and $\eta$.
The annihilation contribution is even equal to zero in the case of $B^-\to \pi^- \pi^0$.
Regarding the ratios of  $\mathcal{BR}(B^- \to \pi^- \pi^0)/\mathcal{BR}(\bar{B}^0 \to \pi^0 \pi^0)$ and 
$\mathcal{BR}(\bar{B}^0 \to \pi^+ \pi^-)/ \linebreak \mathcal{BR}(B^- \to \pi^- \pi^0)$, plotted in 
Fig.~\ref{fig19}, NF and QCDF do not  show any agreement. The experimental and NF results 
only agree for a very limited set of values of the CKM matrix  parameters, $\rho$ and $\eta$. 
 A  full agreement with the BBC results is found when the QCDF approach  is applied. 
Note as well that the dependence of these ratios on the form factor, 
$F^{B \to \pi}$, vanishes with the 
NF approach but remains when QCDF is used because of the 
inclusion of the annihilation terms.

For the $B\to \pi K^*$ decay channel, our results are shown in Figs.~\ref{fig17} and~\ref{fig20}. 
The agreement between NF and QCDF appears only at   values of the form factor, 
$F^{B \to \pi} \approx 0.2-0.3$,  whereas for higher values of $F^{B \to \pi}$, NF predicts larger 
branching ratios  than QCDF. All the BBC experimental results 
coincide with the theoretical one coming from QCDF when the values of $F^{B \to \pi}$ are about 0.3.
The sensitivity of the branching ratios for  the $B\to \pi K^*$ decays to the CKM matrix
parameters, $\rho$ and $\eta$, is much stronger in the case of  QCDF  than  NF, 
because of the insertion of the annihilation term contribution  in the first approach. 
The strong dependence of the branching ratios on the form factor,  $F^{B \to \pi}$, arises either from  
the only presence of  $F^{B \to \pi}$ in the amplitude or  from the term $\alpha_1 F^{B \to \pi}$ 
(color tree amplitude) when the form factor,
$A_0^{B \to K^{\ast}}$ is involved. Regarding $\mathcal{BR}(\bar{B}^0 \to \pi^0 \bar{K}^{*0})$ and 
$\mathcal{BR}(B^- \to \pi^0 \bar{K}^{*-})$, we estimate their magnitudes at $3.8 \times 10^{-6}$ 
and  $12 \times 10^{-6}$, respectively. These last values remain below the branching ratio upper 
limits only given  by the CLEO factory.
 For the ratio $\mathcal{BR}(B^- \to \pi^- \bar{K}^{*0})/\mathcal{BR}(\bar{B}^0 \to \pi^+ K^{*-})$, an 
agreement is found between NF and QCDF for all values of $F^{B \to \pi}$ but  with  the BELLE data only.  Our
results do not agree with the CLEO ones. The NF approach (in contrast 
with QCDF) provides a ratio which is 
independent of the form factor,  $F^{B \to \pi}$, 
because of the annihilation terms.

Next, we turn to the decay channel  $B\to \pi \rho$, for 
which the results are shown in Figs.~\ref{fig13} and~\ref{fig20}.
First, let us recall that $B \to \pi \rho$ is governed by the tree topology. Therefore it provides a direct
 access to the $|V_{ub}|$ and $|V_{ud}|$ CKM matrix elements.
The QCDF and NF frameworks do agree with each other for the 
branching ratios of $B^- \to \pi^0 \rho^-,
\bar{B}^0 \to \pi^+ \rho^-, \bar{B}^0 \to \pi^{\pm} 
\rho^{\mp}$ and $\bar{B}^0 \to \pi^- \rho^+$, whereas
 no agreement is observed for $\bar{B}^0 \to \pi^0 \rho^0, B^{-} \to \rho^0 \pi^-$.  as well as  for the ratio 
$\mathcal{BR}(\bar{B}^0 \to \rho^{\pm} \pi^{\mp})/\mathcal{BR}(B^- \to \rho^0 \pi^-)$. Experimental results provided by BBC
coincide with the theoretical results if the form factor, $F^{B \to \pi}$, takes values from 0.2 to 0.4, for 
 $B$ decaying into $\pi^0 \rho^-, \pi^+ \rho^-, \pi^{\mp} \rho^{\pm}$. For the remaining decays, i.e.
for  $\pi^0 \rho^0, \pi^- \rho^+$ and $\rho^0 \pi^-$, the agreement between the  BBC results and our results are 
not so clear. Moreover, these branching ratios are weakly 
(or even not at all) sensitive to the form factor,  
$F^{B \to \pi}$: this is either because of the form factor, $A_0^{B \to \rho}$, 
in the amplitude or because the 
main contribution comes from the  color power suppressed tree diagram. 
We note as well a  strong dependence of some  branching 
ratios (i.e. $\pi^- \rho^+, \rho^0 \pi^-$) on the CKM matrix parameters, $\rho$ and $\eta$.
Such $B$ decay channels are  usually required for constraining the CKM matrix because of their high 
sensitivity to the parameters $\rho$ and $\eta$. Finally, the theoretical ratio $\mathcal{BR}(\bar{B}^0 \to \rho^{\pm} \pi^{\mp})
/\mathcal{BR}(B^- \to \rho^0 \pi^-)$ agrees with the BBC data for values of the form factor, $F^{B \to \pi}$, larger than 0.2.

In Fig.~\ref{fig16}, is plotted the decay channel $B\to \pi \omega$. The NF and QCDF frameworks present
similar results only for  values of $F^{B \to \pi}$ larger than 0.35 and 0.7 for the decays $\pi^- \omega$ and 
$\pi^0 \omega$, respectively. Our results for $\pi^0 \omega$ are below the upper experimental limits given by BBC. 
For  $\pi^- \omega$, our results agree with the CLEO data  
in the case of  QCDF and  with BELLE and BABAR if  NF is used.
The similar sensitivity of the branching ratio for $B \to \pi^- \omega$ to the CKM parameters, $\rho$ and $\eta$, when NF 
or QCDF are applied,  comes from the
small effect of the annihilation contribution. However, this does not hold for the  $\pi^0 \omega$ decay. 
Our prediction for  the branching ratio of $\pi^0 \omega$ is around $0.6\; 10^{-6}$.

Regarding  the decay channel $B\to \pi \phi$, our results are shown in Fig.~\ref{fig19}. First, let us note
that this channel gives one of the smallest branching ratio  in $B$ decays because of the penguin dominant 
contribution. In fact, only  the element, $\alpha_3$, (from QCD and electroweak QCD penguin diagrams)  contributes to the amplitude. 
Moreover, the NF and QCDF approaches provide very different results for such decay channel since that 
difference, in terms of magnitude, is of the  order  $10^3$. This difference cannot result from the annihilation 
term (they do not take part of the  amplitude) but from the way of  computing the Wilson coefficients. Our 
theoretical predictions  are $1.0\; 10^{-9}$ and $0.5\; 10^{-9}$ for  $B^-\to \pi^- \phi$ and 
$\bar{B}^0 \to \pi^0 \phi$.

The last decay channel analyzed in this section is $B\to \pi \eta^{(')}$. Our results have been plotted 
in  Figs.~\ref{fig11} and~\ref{fig12}. We observe a global agreement between the NF and QCDF factorizations 
for the four decays, $\pi^0 \eta^{(\prime)}$ and $\pi^- \eta^{(\prime)}$. However, this agreement holds for values
of $F^{B \to \pi}$ about 0.2-0.4 if $\pi^0 \eta^{(\prime)}$ whereas it holds for all  possible values of 
 $F^{B \to \pi}$ if $\pi^- \eta^{(\prime)}$. In both cases, we have a good agreement between the experimental
data given by BBC and our results. For the branching ratios $\bar{B}^0 \to \pi^0 \eta$ and  $\bar{B}^0 \to 
\pi^0 \eta^{\prime}$, our predictions are $0.45\; 10^{-6}$ and $0.5\; 10^{-6}$, respectively. In regards to
the  branching ratio $B^0 \to \pi^- \eta^{\prime}$, we obtained $8.1\; 10^{-6}$. Finally, we note that the 
decay channel $B^- \to \pi^- \eta^{(\prime)}$ is a good candidate for constraining and checking the CKM
parameters, $\rho$ and $\eta$, and the form factor $F^{B \to \pi}$, respectively, because of  its strong sensitivity
to  these parameters.

%
\subsubsection{$\boldsymbol{B\to X K}$  with $\boldsymbol{X=\{\rho, \omega, \eta^{(')}, K, K^*, \phi\}}$}
%
Now, let us focus on  the  branching ratios for  $B$ decay channels  $B\to X K$
where $X$ stands for $\rho, \omega, \eta^{(')}, K, K^*$ or  $\phi$. 
In Table~\ref{tab6}, are listed
the different values for the hard-scattering (annihilation) 
phases, $\varphi_{H}^{M_{i}} 
(\varphi_{A}^{M_{i}})$ and parameters, $\varrho_{H}^{M_{i}} (\varrho_{A}^{M_{i}})$. The values are given  for 
each of the following   particles ($M_i$), $\rho, \omega, \eta^{(')}, K, K^*, \phi$ and 
for the minimal (set 2)  and maximal (set 1) sets of values of the CKM parameters, $\rho$ and $\eta$.
All the experimental results from the BBC factories for the $B\to X K$ decays are listed
in Tables~\ref{tab2} and~5. The theoretical branching ratios  calculated with the 
NF and QCDF factorizations are shown for comparison in Tables~7 and~\ref{tab8} where the annihilation and 
hard scattering spectator contributions are given explicitly.

For the decay channel  $B\to K K$, our results are shown in Fig.~\ref{fig10}. We found no
agreement between the QCDF and NF approaches for all the branching ratios $\bar{K}^0 K^0$, 
$K^- K^+$ and $K^- K^0$. All of our results do agree with the 
upper experimental limits given by the BBC results.
Our predictions for the branching
ratios, $K^- K^0, K^- K^+$ and $\bar{K}^0 K^0$ are 
respectively $0.31\; 10^{-6}, 0.2\; 10^{-6}$ and $0.27\; 10^{-6}$. 
It appears that these branching ratios calculated with QCDF are smaller than those with NF
because of the sizable contribution of the annihilation terms. Moreover, in the case of  $B$ decaying
into $K^- K^+$, the annihilation contribution is the only one which 
participates in the amplitude. As a 
result, NF gives a branching ratio equal to zero in 
the latter case. The same reason explains the independence
of this branching ratio  on the form factor, $F^{B\to K}$. The dependence of 
these branching ratios on the CKM parameters, $\rho$ and $\eta$, as well 
as on the form factor, $F^{B\to K}$, is quite 
similar in both approaches, i.e. NF and QCDF.

Our results for the decay channel $B \to K K^{\ast}$  
are shown in Fig.~\ref{fig18}. Unfortunately, 
there are no available experimental results  for these decays. 
Only  one upper limit is given by CLEO 
for the decay $K^- K^{\ast 0}$. Of the six ways for a $B$ to 
decay into $K K^{\ast}$, only two branching ratios,
$K^- K^{\ast 0}$ and $\bar{K}^0 K^{\ast 0}$, allow one for an agreement between NF and QCDF. 
However, this agreement holds only at low values of the form factor, $F^{B\to K}$, i.e. about 
0.2-0.25. For all the other cases, the NF and QCDF factorization give different branching 
ratio results. We recall that $B \to  K K^{\ast}$ is 
only governed by the penguin topology. 
In the case of the $K^- K^{\ast +}$ and $K^{\ast -} K^+$ decays, the magnitude 
of their branching ratios comes from the annihilation contribution alone. 
That explains 
the independence of these branching ratios on the form factor, $F^{B\to K}$, as well as 
the fact that NF gives magnitudes equal to zero. Regarding the decays $K^{\ast -} K^0$ 
and $K^- K^{\ast 0}$, there is a factor $10^2$ and $10^1$, 
respectively, between the 
results obtained with NF and QCDF. The non sensitivity of these branching ratios
to the form factor, $F^{B\to K}$, arises from 
the only contribution of the form 
factor $F^{B \to K^{\ast}}$ in the amplitude. Because of the annihilation
terms,  a stronger dependence of the branching ratios on the CKM parameters, 
$\rho$ and $\eta$, is observed when QCDF is applied. Finally, we predict the branching 
ratios  for $\bar{K}^0 K^{\ast 0}, \bar{K}^{\ast 0} K^{0}, K^- K^{\ast +}, K^{\ast -} K^{+}, 
K^- K^{\ast 0}$ and $K^{\ast -} K^0$ to be of the order  $0.4\; 10^{-6}, 0.75\; 10^{-6}, 
3.0\; 10^{-6}, 0.6\; 10^{-6}, 0.4\; 10^{-6}$ and $0.68\; 10^{-6}$, respectively.

We have plotted in Fig.~\ref{fig16}  the decay channel $B\to K \rho$. Experimental results
provided by BBC are well reproduced and all of our results are below the upper limits given for  the 
branching ratio data. No strong agreement between NF and QCDF can be confirmed. A weak (or 
even no) dependence of these branching ratios  on the form factor, $F^{B\to K}$,
is observed. This is either due to  the interference of the form factor, $F^{B \to K}$, with 
 the form factor $A_0^{B \to \rho}$ (enhanced by a color-allowed tree diagram, $\alpha_1$) 
or the absence of the $B \to K$ transition in the
amplitude. We also note a stronger sensitivity of the branching ratios to the CKM parameters,
$\rho$ and $\eta$, in the case where the annihilation terms are taken into account. We conclude by
giving our results regarding this channel: $5.8\; 10^{-6}, 5.0\; 10^{-6}$ and $1.7\; 10^{-6}$
are respectively our predicted branching ratios for $\rho^0 \bar{K}^0, \rho^- \bar{K}^0$ and $\rho^0 K^-$.

Next, we consider the decay channel, $B\to K \omega$, for 
which the results are shown in Fig.~\ref{fig17}.
As already the case for $B\to K \rho$, no agreement 
between the NF and QCDF frameworks
is observed. However, the  accurate experimental results given by the BBC factories coincide
with the branching ratios obtained with QCDF when the form factor,  $F^{B\to K}$, is in the   range 
of values  0.2-0.45. The differences between the two frameworks come from the contribution of  the annihilation
terms which strongly enhance the magnitude of these branching ratios. The sensitivity to  the 
 CKM parameters, $\rho$ and $\eta$, arises from the annihilation contribution as well.

For the decay channel, $B\to K \phi$, our results are shown in Fig.~\ref{fig19}. The QCDF 
factorization agrees with the overlap of experimental data for values of the form factor,  
$F^{B\to K}$, equal to 0.2-0.5. We observe a very similar behaviour between the two decays, 
$K^- \phi$ and $\bar{K}^0 \phi$. The small difference only comes  from the annihilation 
contribution. We underline as well a better sensitivity to the 
CKM parameters, $\rho$ and 
$\eta$, in the case where QCDF is applied. The latter behaviour is still due to 
the annihilation term. Finally, we note a 
strong dependence of the branching ratios on the 
form factor, $F^{B\to K}$. It may be a good 
channel for checking the form factor, $F^{B\to K}$, 
if we assume correct the factorization procedure 
or vice-versa. However, the magnitude of 
this decay channel remains very small because it is dominated 
by penguin diagram.

The last decay channel which we investigate is  $B\to K \eta^{(')}$, for 
which our results are shown in Figs.~\ref{fig12} and~\ref{fig20} . 
The decays $K^- \eta$ and $\bar{K}^{0} \eta$ 
show good agreement between the NF and 
QCDF approaches for almost all values of the form factor, $F^{B \to K}$, 
whereas the decays 
$K^- \eta^{\prime}$ and $\bar{K}^{0} \eta^{\prime}$ do not. 
The experimental branching ratios 
and theoretical results agree with each other 
when the form factor, $F^{B \to K}$, is about 0.35 for 
$K \eta^{\prime}$ decays and about 0.55 for $K \eta$ decays. 
We observe a stronger sensitivity of 
the branching ratios to the CKM parameters when the QCDF factorization is used. It has to be noticed
that the decay channel $K \eta^{\prime}$ is an excellent candidate for  the 
determination of both of the CKM 
parameters, $\rho$  and $\eta$, and  the form factor,  $F^{B \to K}$. This statement holds if
the procedure of factorization and  the evaluation of the annihilation contribution are assumed to be correct.
Moreover, the last decay channel is one order of magnitude bigger than those usually observed in $B$ decays. That
makes its determination easier and accurate results are 
expected, thanks to the BBC factories. We also 
predict a branching ratio for the decay $\bar{K}^0 \eta$ to be of the order  $0.9\; 10^{-6}$. Finally, the ratio 
$\mathcal{BR}(B^- \to K^- \eta^{\prime})/\mathcal{BR}(\bar{B}^0 \to \bar{K}^0 \eta^{\prime})$ shows an agreement between NF and QCDF 
at large values of the form factor,  $F^{B \to K}$,  only. However, QCDF agrees with the BELLE and BABAR data
at values of $F^{B \to K}$ about 0.35. No agreement is found with CLEO.
%
\subsubsection{$\boldsymbol{B\to \pi K}$}
%
We conclude our analysis of branching ratios for $B$ decay channels including a pion (or a kaon)
and another meson  in the final state by turning to 
those  where both the kaon and pion are in the final state. 
The experimental results corresponding to  the $B\to \pi K$ decays are given 
in Tables~\ref{tab4} and~5. The theoretical branching ratios including the annihilation and 
hard scattering spectator contributions are given for comparison  in  Table~\ref{tab8}. 

In Figs.~\ref{fig11} and~\ref{fig20} we show the results for 
the decay channel $\pi K$.
We first observe a different behaviour between the NF and QCDF approaches.  This is due to the 
annihilation contribution which strongly enhances these branching ratios. As a result, NF and
QCDF only coincide to each other for a small set of values of the form factors, $F^{B \to K}$, and 
$F^{B \to \pi}$, which are of the order   0.3-0.4. We note as well an excellent agreement between 
our results and the branching ratio data provided by the BBC factories. The strong dependence of the 
branching ratios for  the 
decays $\pi K$ on the CKM parameters, $\rho$ and $\eta$, when QCDF is applied, gives one the opportunity
to efficiently constrain  these latter parameters. At the same time, it 
allows us to accurately model 
the annihilation contribution, since it arises in the 
amplitude times  a CKM matrix element 
$V_{pb} V_{pD}^{\ast}$ (in the case of $b \to d$ transition).  We finally underline that the sensitivity
of this decay channel to the forms factors, $F^{B \to K}$, and $F^{B \to \pi}$, contributes effectively to
the tests of the procedure of factorization of $B$ decaying into two light mesons. 

Regarding the ratios
$\mathcal{BR}(\bar{B}^0 \to \pi^+ K^-)/\mathcal{BR}(\bar{B}^0 \to \pi^0 \bar{K}^0), 
\mathcal{BR}(B^- \to \pi^0 K^-)/\mathcal{BR}(B^- \linebreak \to \pi^- \bar{K}^0)$
and  $\mathcal{BR}(\bar{B}^0 \to \pi^+ K^-)/\mathcal{BR}(B^- \to \pi^- \bar{K}^0)$, no agreement between 
NF and QCDF is found. This is not true  for the latter ratio if the values of the form factor,  
$F^{B \to \pi}$, are about 0.15-0.45. Our predictions for these ratios and those given by BBC do reasonably
 agree. Finally, the ratios calculated within the NF framework do not depend on form factor because of 
the omission of the annihilation terms, whereas they do when QCDF is used.

Before going further in our analysis, let us draw the main conclusions of this analysis of  the branching 
ratios for $B$ decay channels  $B \to \pi X$ and $B \to K X$ where $X=\rho, \omega, \eta^{(')}, \pi, K, K^*$ 
and $\phi$. We have calculated the branching ratios for $B$ decays into two mesons in the final state where we  have 
made comparison with the NF and QCDF frameworks. The results show that the form factors, $F^{B \to \pi}$ and 
$F^{B \to K}$, are equal to $0.31 \pm 0.12$ and $0.37 \pm 0.13$, respectively. At the same time, we have 
determined the four unknown parameters used in QCDF by performing a fit 
to all of the branching ratio data provided by 
the BBC facilities. We note that we did not take into account any $CP$ violating
asymmetry experimental values, $a_{CP}^{exp}$, in our fit. 
These phases, $\varphi_{H,A}^{M_i}$, and parameters, 
$\varrho_{H,A}^{M_i}$ are assumed, in our analysis, to be non  universal regarding  the mesons. It appears that
their determination is very sensitive to the values of the CKM parameters, $\rho$ and $\eta$, and unfortunately, 
at the present time, it is not possible to draw any firm conclusions concerning this point. Let us note that
the NF factorization gives, in first approximation, the right order of magnitude of  branching ratios 
in most of the cases. We have used $N_c^{eff}=3$ for the effective parameter. This emphasizes the differences
between NF and QCDF since, in that case, the differences mainly come from the hard scattering spectator 
as well as the annihilation terms. In most of the investigated cases, we observed a quite good agreement 
between NF and QCDF if we restrict the transition form factors, 
$F^{B \to \pi}$ or $F^{B \to K}$, to the range 
of values 0.2-0.4. This common behaviour was expected 
because the non factorizable terms are usually 
power suppressed corrections (see Eq. (19)). If we use  $N_c^{eff}=2$ for the effective parameter in NF,
our results remain coherent with the previous observation.
However, the theoretical branching ratio results obtained by 
applying the QCDF framework do provide better agreement with the BBC data. This is also due to  the annihilation 
term effects included only in the QCDF approach. It has to be emphasized as well  that the latter contribution takes
importance if we analyze the dependence of  branching ratios on the CKM parameters $\rho$ and $\eta$. 
Finally, over all the analyzed branching ratios, some of them are more suitable for the analysis of the CKM 
matrix elements and others for checking the factorization procedure if we assume correct  the 
transition form factors. Let us mention, $\bar{B}^0 \to \pi^{\pm} \rho^{\mp}, B^- \to \pi^0 \rho^-$, and 
$B^- \to \pi^- \pi^0, \bar{B}^0 \to \pi^+ \pi^-, \bar{K}^0 \to \bar{K}^0 \eta^{\prime}$, respectively. 
Regarding the effect of the annihilation contribution, the following $B$ decay channels are the most 
interesting since the amplitude of their branching ratios is only proportional to the annihilation term:
$\bar{B}^0 \to K^- K^+, \bar{B}^0 \to K^- K^{\ast +}$ and $\bar{B}^0 \to K^{\ast -} K^+$.

%
\subsection{$\boldsymbol{CP}$ violating asymmetry}

Now, let us focus on the violation of  $CP$ symmetry. 
We concentrate our analysis on
$B$ decays such as $B \to \rho^0(\omega) M_2 \to \pi^+ \pi^- M_2$, 
where $M_2$ is either
a pion or a kaon. It has been shown in previous studies that the 
CP-violating asymmetry 
parameter, $a_{CP}$, may be large when the 
invariant mass $\pi^+ \pi^-$ is in the 
vicinity of the $\omega$ resonance. This enhancement is known as $\rho-\omega$ 
mixing effect and we refer interested 
readers to Refs.~\cite{Guo:2000uc, Leitner:2002br, Leitner:2002xh} 
for more details. In this last section, 
we are using all the parameters collected thanks to
the analysis of branching ratios of $B$ decays.

In the application of the QCD factorization, we define,
for practical reason, three amplitudes, $t^{u}, p^{u}$ and $p^{c}$. We set the 
amplitudes as  follows:
\begin{align}\label{eq47}
A_{Bf} &= |A_{1}| e^{i \delta_{1} + i \phi_{1}} + |A_{2}| e^{i \delta_{2} + i \phi_{2}}+
|A_{3}| e^{i \delta_{3} + i \phi_{3}}\ , \nonumber \\
\bar{A}_{Bf} &= |\bar{A}_{1}| e^{i \delta_{1} - i \phi_{1}} + |\bar{A}_{2}| e^{i \delta_{2} - i \phi_{2}}+
|\bar{A}_{3}| e^{i \delta_{3} - i \phi_{3}}\ , 
\end{align}
where the first term refers to the ``tree'' contribution, $A_1 \equiv t^{u}$, instead of the 
remaining terms define the ``penguin'' contributions with $A_2 \equiv p^{u}$, and  $A_3 \equiv p^{c}$.
In that case, the  $CP$ violating asymmetry, $a_{CP}$, takes  the following form:
\begin{multline}\label{eq48}
a_{CP}  \equiv \frac{|A|^{2}-|\bar A|^{2}}{ |A|^{2}+|\bar A|^{2}}= \\
\frac{-2 ( r_1 \sin \phi_{12} \sin \delta_{12} + r_2 \sin \phi_{13} \sin \delta_{13}+r_1 r_2\sin \phi_{23} \sin \delta_{23})}
{1+ r_1^2 + r_2^2+ 2 (r_1 \cos \phi_{12} \cos \delta_{12} + r_2 \cos \phi_{13} \cos \delta_{13}+ 
r_1 r_2  \cos \phi_{23} \cos \delta_{23})}\ ,
\end{multline}
where the convention for the used parameters is:
\begin{equation}\label{eq49}
r_1 = \left| \frac{A_2}{A_1}\right| \ , \;\;\;\; r_2 =  \left|\frac{A_3}{A_1}\right|\ , \;\;{\rm and}\;\; 
\phi_{ij}=\phi_i-\phi_j\ ,  \;\;\;\; \delta_{ij}=\delta_i-\delta_j\ . 
\end{equation}
The parameters, $r_i$, and the phases, $\delta_{ij}$, are given by applying 
Eq.~(\ref{eq43}) whereas $\delta_{23}$ is obtained from the ratio $\lambda_u p^u/ \lambda_c p^c$ 
where $\lambda_p= V_{pb}V^{*}_{pD}$ with $D=d,s$.  
Finally, the phases $\phi_{13}= \phi_{23}
={\rm arg}[(V_{cb}V_{cd}^{\star})/(V_{ub}V_{ud}^{\star})]$ for $b \to d$ transition, 
$\phi_{13}= \phi_{23}={\rm arg}[(V_{cb}V_{cs}^{\star})/(V_{ub}V_{us}^{\star})]$ for $b \to s$ 
transition and $\phi_{12}=0$ in both transitions.
$\sin \phi_{13} \; (\cos \phi_{13})$ takes  the following form in case of  $b \to d$ transition,
\begin{align}\label{eq50}
\sin\phi_{13} = \sin\phi_{23}& = \frac{-(\lambda \eta - \frac12 A^2 \lambda^5)}
{\sqrt {\left(\eta \lambda- \frac{A^{2} \lambda^{5}}{2}\right)^{2}+\left(\lambda \rho + \frac12 A^{2} \lambda^{5}
(2 \eta^{2} + \rho (-1+2 \rho))\right)^{2}}}\ , \nonumber \\
\cos\phi_{13}= \cos\phi_{23} & = \frac{\frac12 A^2 \lambda^5 (\rho -2 \rho^2 -2 \eta^2) - \lambda \rho}
{\sqrt {\left(\eta \lambda- \frac{A^{2} \lambda^{5}}{2}\right)^{2}+\left(\lambda \rho + \frac12 A^{2} \lambda^{5}
(2 \eta^{2} + \rho (-1+2 \rho))\right)^{2}}}\ ,
\end{align}
\noindent and in case of $b \to s$ transition, $\sin \phi_{13} \; (\cos \phi_{13})$ is given by:
\begin{align}\label{eq51}
\sin\phi_{13} = \sin\phi_{23} & = \frac{\eta}{\sqrt {\rho^2+\eta^{2}}}\ , \nonumber \\
\cos\phi_{13}= \cos\phi_{23} & = \frac{\rho}{\sqrt {\rho^2+\eta^{2}}}\ .
\end{align}
%
%
%
\subsubsection{$\boldsymbol{B \rightarrow \rho^{0}(\omega) \pi \to \pi^+ \pi^- \pi}$}
%
We have investigated the $CP$ violating asymmetry, $a_{CP}$, for the  $B$ decays such as
$B \rightarrow \rho^{0}(\omega) \pi \to \pi^+ \pi^- \pi$. In Fig.~\ref{fig14},  we  show the 
$CP$ violating asymmetry for $B^{-} \to \rho^{0}(\omega) \pi^{-}\to \pi^{+} \pi^{-} \pi^{-}$ 
and  $\bar{B}^{0} \to \rho^{0}(\omega) \pi^{0} \to \pi^{+} \pi^{-} \pi^{0}$ respectively,
as a function of the energy, $\sqrt{S}$, of the two pions coming from 
$\rho^0$ decay, the form factor, $F_{1}^{B \to \pi}$, and the 
CKM matrix element parameters $\rho$ and $\eta$. For 
comparison, on the same plot we show 
the $CP$ violating asymmetries, $a_{CP}$, when NF is 
applied as well as QCDF where default values for the 
phases, $\varphi_{H,A}^{M_i}$, and parameters, $\varrho_{H,A}^{M_i}$ are 
used. In the latter case, we take
 $\varphi_{H,A}^{M_i}=0$ and  $\varrho_{H,A}^{M_i}=1$ for all the particles.

Focusing first  on Fig.~\ref{fig14}, where the asymmetry for  $B^{-} \to \rho^{0}(\omega) 
\pi^{-}\to \pi^{+} \pi^{-} \pi^{-}$ is plotted, we observe that the $CP$ violating 
asymmetry parameter, $a_{CP}$, can be large  outside the region where the invariant mass of the 
$\pi^{+} \pi^{-}$ pair is in   the vicinity of the $\omega$ resonance. 
This is the first consequence of  QCD factorization, since within this framework, the strong phase can 
be  generated not only by the $\rho-\omega$ mechanism but also  by the  Wilson coefficients. We recall that the 
Wilson coefficients include all of the final state interactions at  order $\alpha_{s}$. This shows as well that the 
non factorizable contribution effects  are important and can modify the strong interaction phase. 
Because of the strong phase\footnote{In comparison with QCDF, pQCD predicts large strong phases and direct $CP$ 
asymmetries.} that is either at the order of $\alpha_{s}$ or power suppressed by $\Lambda_{QCD}/m_{b}$,
the $CP$ violating asymmetry, $a_{CP}$, may be small but a large asymmetry cannot be excluded.

At the $\omega$ resonance, the asymmetry parameter, $a_{CP}$, for $B^- \to \pi^+ \pi^- \pi^-$, is around $0\%$ in our case. 
In comparison, the asymmetry  parameter,  $a_{CP}$,  (still at the $\omega$ resonance) obtained by applying 
the naive factorization gives  $-10\%$ whereas it gives $-2\%$ in case of QCDF with default values for
 $\varphi_{H,A}^{M_i}$ and $\varrho_{H,A}^{M_i}$. The results 
are quite different between  these  approaches because of the  strong phase  mentioned previously.
On the same figure, the asymmetry violating parameter, $a_{CP}$, 
is shown for the decay
$\bar{B}^0 \to \pi^+ \pi^- \pi^0$ as a function 
of $\sqrt{S}$, the form factor, $F^{B \to \pi}$, and 
for one set of CKM parameters, $\rho$ and $\eta$. In the vicinity of the $\omega$ resonance, the QCDF
approach gives an asymmetry of the order $-8\%$. We obtain $-20\%$ and $+5\%$ in the case of NF and QCDF
with the default values for   $\varphi_{H,A}^{M_i}$ and $\varrho_{H,A}^{M_i}$.

It appears as well that the
asymmetry depends strongly on the CKM matrix parameters $\rho$ and $\eta$, as expected.  When QCDF is applied, 
the asymmetry for the decay $B^- \to \pi^+ \pi^- \pi^-$, varies from $12\%$ to $+5\%$ outside the region of the
$\omega$ resonance whereas for the decay $\bar{B}^0 \to \pi^+ \pi^- \pi^0$, the asymmetry varies from $10\%$ down to
$-20\%$, depending on the CKM matrix element parameters, $\rho$ and $\eta$. 
In the vicinity of the $\omega$ resonance, the asymmetry, 
$a_{CP}$, takes values from $-2\%$ to
$5\%$ for $B^- \to \pi^+ \pi^- \pi^-$ 
and from $5\%$ to $-30\%$ for $\bar{B}^0 \to \pi^+ \pi^- \pi^0$ 
when $\rho$ and $\eta$ vary. In both decays, we note as well a dependence of the asymmetry on the form factor, 
$F^{B \to \pi}$. This dependence reaches usually its maximum when  the asymmetry is given at the $\omega$ 
resonance. However, this dependence remains under control because of 
the constraints obtained for their values 
by the analysis of $B$ branching ratios.

In Fig.~15, the ratio, $r_i$, of penguin to tree amplitude for $B \to \pi^+ \pi^- \pi$ is given as 
a function of $\sqrt{S}$, the form factor $F^{B \to \pi}$ and for one set of CKM parameter, $\rho$ and $\eta$.
 For both decays, i.e. $B^- \to \pi^+ \pi^- \pi^-$ and $\bar{B}^0 \to \pi^+ \pi^- \pi^0$ we observe similar results 
for $r_{1}=p^{u}/t^{u}$ and  $r_{2}=p^{c}/t^{u}$. As  is 
expected for dominant tree decays, the contribution coming from the tree diagram, $t^{u}$, is bigger than that one coming from the 
penguins, $p^{u}$ or $p^{c}$. Finally, as mentioned, one of the main reasons for the interest in $\rho-\omega$ mixing
is to provide an opportunity to remove the phase uncertainty mod$(\pi)$ in the determination of the CKM
angle $\alpha$ in the case of $b \to u$ transition. Knowing the sign of the $CP$ violating asymmetry at the 
$\omega$ resonance gives us the angle $\alpha$ without any ambiguity. In Fig.~15, we present the evolution of 
$\sin \delta_{12}, \sin \delta_{13}$ and $\sin \delta_{23}$ as a function of $\sqrt{S}$, the form factor $F^{B \to \pi}$
and for one set of CKM parameters $\rho$ and $\eta$. In both decays,  $\sin \delta_{ij}$ goes to a maximum
or a minimum when $\sqrt{S}$ is in the vicinity of the $\omega$ resonance.

%
\subsubsection{$\boldsymbol{B \rightarrow \rho^{0}(\omega) K \to \pi^+ \pi^- K}$}
%
%

After the analysis of the $CP$ asymmetry in 
$B^{\pm,0} \to \rho^{0}(\omega) \pi^{\pm,0} 
\rightarrow \pi^+ \pi^- \pi^{\pm,0}$, we 
conclude our work by focusing on the 
asymmetry in $B^{\pm,0} \rightarrow \pi^+ \pi^- K^{\pm,0}$.
Plotted in Fig.~14   is the direct $CP$ violating 
asymmetry, $a_{CP}$, for $B^{-} 
\to  \rho^{0}(\omega) K^{-} \to \pi^{+} \pi^{-} K^{-}$ and for $\bar{B}^{0}\to  \rho^{0}(\omega)  \bar{K}^{0}
\to \pi^{+} \pi^{-} \bar{K}^{0}$, as a function of $\sqrt{S}$, the form factor, $F^{B \to K}$, and for 
one set of CKM parameters $\rho$ and $\eta$. 

For the decay $B^- \to \pi^+ \pi^- K^-$, the asymmetry, $a_{CP}$, in the vicinity of the $\omega$ resonance, is 
about $+60\%$ with QCDF, $-40\%$ with NF and $-45\%$ with QCDF and default values 
for $\varphi_{H,A}^{M_i}$ and $\varrho_{H,A}^{M_i}$. For the decay $\bar{B}^0 \to \pi^+ \pi^- \bar{K}^0$, when 
$\sqrt{S}$ is near the $\omega$ resonance, the asymmetry, $a_{CP}$ is about $+70\%$ with QCDF, $-60\%$ with 
NF and $-15\%$ with QCDF and usual default values for  $\varphi_{H,A}^{M_i}$ and $\varrho_{H,A}^{M_i}$.
There is no agreement, for the value of the asymmetry between the naive  and QCD factorization at the 
$\omega$ resonance except that, in both cases,  the $CP$ violating asymmetry, $a_{CP}$ reaches its maximum 
in the vicinity of $\omega$. 
Similar conclusions can be drawn to that of previous case  regarding the sensitivity of the 
asymmetry parameter, $a_{CP}$, on the form factor, $F^{B \to K}$, as well as  the CKM matrix element 
parameters, $\rho$ and $\eta$. In Fig.~14, the ratio, $r_i$, of penguin to tree amplitudes, for 
$\bar{B}^0 \to \pi \pi \bar{K}^{0}$ and $B^- \to \pi \pi K^-$ is plotted. We observe that the ratio 
$r_1=p^u/t^u$ is very small. This underlines the contribution of the tree diagram, $t^u$, in 
comparison with the penguin one, $p^u$. We observe as well that the ratio $r_2=p^c/t^u$ is much bigger 
than $r_1=p^u/t^u$. This is due to an additional contribution of the $c$ quark when the amplitude for 
$\bar{B}^0 \to \omega \bar{K}^0$ and $B^- \to \omega K^-$ are calculated (see Eqs. 138 and~139 in Appendix B). 
As usual, we note that $\rho-\omega$ mixing strongly enhances  the ratio, $r_{i}$, at the $\omega$ resonance.
As we did for $B$ decaying into $\pi \pi \pi$, we can remove the ambiguity for the determination of the 
angle $\gamma$ that arises from the conventional determination of $\sin 2 \gamma$ in indirect $CP$ violation.
In Fig.~15, $\sin \delta_{ij}$, as a function of $\sqrt{S}$, the form factor, $F^{B \to K}$, and for one set 
of CKM parameters, $\rho$ and $\eta$, for $B \to \pi \pi K$ is shown. For both decays, $\sin \delta_{ij}$, is large
in the vicinity of the $\omega$ resonance i.e. around $775-785$ MeV. As for $B \to \pi \pi \pi$, we note 
that the strong phase, $\delta_{ij}$, can remain large outside the 
region where the mass of the $\pi^+\pi^-$ pair 
is in the vicinity of the $\omega$ resonance. 
This underlines the dynamical mechanism of creating a strong phase
not only at the $\omega$ resonance but for all values of $\sqrt{S}$.

From this analysis, it is clear that to take into account 
$\rho-\omega$ mixing in $B$ decays such as $B \to \rho(\omega) M_i$ allows us to ``amplify'' the hadronic interaction
near the $\omega$ resonance and it provides an excellent 
test of the Standard Model through direct $CP$ 
violation.

%
\section{Summary and discussion}

The calculation of the hadronic matrix elements that appear in the $B$ decay 
amplitude is non trivial.  The main  difficulty is  to express the hadronic matrix elements
which represent the transition between the meson $B$ and the final state. 
Non-leptonic $B$ decay amplitudes involve hadronic matrix elements
$\langle M_1 M_2 | O_i| B \rangle$,  built on  four quark operators. In a first 
approximation, this yields a product of two quark currents translated
in terms of form factor and decay constant: that gives the naive factorization. 
Radiative, non-factorizable corrections 
coming from the light quark spectator of the 
$B$ meson are included in QCD factorization. 
In that case the main uncertainty comes 
from the $O(\Lambda_{QCD}/m_B)$ terms. 
In this paper, we first investigated  in a phenomenological way, the dependence on the form
factors, $F^{B \to \pi}$ and $F^{B \to K}$, of all the branching ratios for $B$ decaying
into $B \to \pi X$ or $B \to K X$, where $X$ is either a pseudo-scalar ($\pi, K, \eta^{(\prime)}$), 
or a vector ($\rho, \omega, K^{*}, \phi$) mesons.

We have investigated the branching ratios for $B$ decays with 
two different methods: the NF and QCDF frameworks have been applied in order to underline 
the differences occurring between these two kinds of factorization. We observe 
that the NF factorization gives, in first approximation, the right order of magnitude 
of  branching ratios in most of the cases.  However, the theoretical branching ratio results
obtained with QCDF  do provide better agreement with the BELLE, BABAR and CLEO 
experimental data. From our analysis, it appears that the hard scattering spectator 
contribution is quite small in regards to  the annihilation effect. The annihilation
contributions in $B$ decays play an important role since they contribute significantly 
to the magnitude of the amplitude. The annihilation diagram contribution to the total decay  amplitude 
may strongly modify (in a positive or negative way) the total amplitude. Let us mention some decays such
as $\bar{B}^0 \to K^- K^+, \bar{B}^0 \to K^- K^{\ast +}$ and $\bar{B}^0 \to K^{\ast -} K^+$. 
We emphasize as well  that the annihilation contribution cannot be neglected if 
we analyze  the dependence of branching ratios on the CKM parameters, $\rho$ and $\eta$. 

An analysis  of more than 50 $B$ decays shows that the transition form factors, $F^{B \to \pi}$ and 
$F^{B \to K}$, are respectively equal to $0.31 \pm 0.12$ and $0.37 \pm 0.13$, if one wants
to reproduce the experimental results. This statement assumes that the procedure of factorization
is accurate enough and it confirms the values of 
the form factors calculated within the QCD sum-rule and 
light-cone frameworks.

We have determined the four unknown parameters, $\varphi_{H,A}^{M_i}$ and  $\varrho_{H,A}^{M_i}$, 
used in QCDF by performing a fit of all the branching ratios with all the experimental data provided 
by the B-factories. These phases, $\varphi_{H,A}^{M_i}$, and parameters, $\varrho_{H,A}^{M_i}$, are 
assumed, in our analysis, to be non  universal for mesons. It is obvious that their determination 
is very sensitive to  two quantities, at least: the experimental branching ratio data and the values of the CKM 
parameters, $\rho$ and $\eta$.  At the time being,  it is not possible to draw any firm conclusions 
concerning the values for the phases, $\varphi_{H,A}^{M_i}$, and parameters, $\varrho_{H,A}^{M_i}$. 
A fine tuning requires more accurate  experimental data for branching ratios as well as more  accurate
CKM matrix parameters $\rho$ and $\eta$. 

Finally, by analyzing the branching ratios, we find that 
some of them are more suitable for 
the analysis of the CKM matrix elements, $V_{ub}, V_{ud}$, 
whereas others can be 
used for checking the factorization procedure if we assume correct the values of the transition form factors 
(or vice-versa): let us mention, $\bar{B}^0 \to \pi^{\pm} \rho^{\mp}, B^- \to \pi^0 \rho^-$, and 
$B^- \to \pi^- \pi^0, \bar{B}^0 \to \pi^+ \pi^-, \bar{K}^0 \to \bar{K}^0 \eta^{\prime}$, respectively. 

Next, we analysed the $CP$ violating asymmetry 
parameter, $a_{CP}$, for the $B$ decays
$B\rightarrow \rho^{0}(\omega) \pi \to \pi^+ \pi^- \pi$ and $B \rightarrow \rho^{0}(\omega) \pi \to \pi^+ \pi^- K$.
This analysis was perfomed with the QCD factorization and comparisons with 
the so-called naive factorization
were also made. We included $\rho-\omega$ mixing in order 
to investigate its effect on this $CP$ violating asymmetry.
The mixing through isospin violation of an
$\omega$ to $\rho$, which then decays into two pions, 
allows us to obtain a difference of the strong phase reaching
its maximum at the $\omega$ resonance. $\rho-\omega$ mixing  
provides an opportunity
to remove  the phase uncertainty mod$(\pi)$ in the determination of two  CKM angles, $\alpha$ in the case of 
$B \to \rho \pi$ and $\gamma$ in the case of $B \to \rho K$. This phase uncertainty usually arises from the 
conventional determination of $\sin 2 \alpha$ or $\sin 2 \gamma$~\cite{Buchalla:2004tw, Buchalla:2003jr} in indirect $CP$ violation.
We have observed large discrepancies in our results regarding the asymmetry when NF, QCDF and QCDF with defauts
values are applied. In the naive factorization, the large strong phase only comes from $\rho-\omega$ mixing that
yields a large asymmetry at the $\omega$ resonance, 
as well as a very small asymmetry far away from the
$\omega$ resonance. In QCDF, the strong phase can be generated dynamically. However, the mechanism suffers
from end-point singularities which are not well controlled. 
The determination of the parameters, 
$\varphi_{H,A}^{M_i}$, and $\varrho_{H,A}^{M_i}$, for the hard scattering spectator and annihilation
contributions can only be achieved thanks to  an  analysis of $B$ branching ratios. However,
this analysis requires accurate experimental data from the $B$-factories. Unfortunately, too many $B$ decay
channels are still uncertain and therefore they do not allow us to draw final conclusions about these
crucial parameters $\varphi_{H,A}^{M_i}$, and $\varrho_{H,A}^{M_i}$. It is the reason why the $CP$ violating 
asymmetry, calculated in QCDF, can vary so much according to the values used for the four unknown parameters.

It is now apparent that the Cabibbo-Kobayashi-Maskawa 
matrix is the dominant source of 
$CP$ violation in flavour changing processes 
in $B$ decays. The corrections to this dominant 
source coming from beyond the Standard Model are not expected to be large. 
In fact, the main  
remaining uncertainty is  to deal with the procedure of factorization. 
In many cases, 
naive factorization allows us to obtain the right 
order of magnitude for the branching ratios 
in $B$ decays,  but fails in predicting 
large $CP$ violating asymmetries if no 
particular mechanism (i.e. $\rho-\omega$ mixing) is adding to reproduce the strong phase.
The QCDF gives us an explicit picture of factorization in
the heavy quark limit. It takes into account all the leading contributions as well as 
subleading corrections to the naive factorization. However, the end-point singularities 
arising in the treatment of the hard scattering spectator and annihilation contributions do 
not make QCDF as predictive as was expected. 
The soft collinear effective theory (SCET) has 
been proposed as a new procedure for factorization. For 
more details see Refs.~\cite{ Bauer:2001cu, 
Bauer:2001yt, Bauer:2000yr, Bauer:2001ct, Bauer:2002uv}. 
In the last case, it allows one to formulate 
a collinear factorization theorem in terms of 
effective operators where new effective degrees 
of freedom are involved,
in order to take into account the collinear, soft and ultrasoft quarks
and gluons. All of these investigations allow us 
to increase our knowledge of $B$ physics and to 
look for new physics beyond the Standard Model.

\subsubsection*{Acknowledgments}
%
This work was supported in part by DOE contract DE-AC05-84ER40150,
under which SURA operates Jefferson Lab and by the special Grants for 
``Jing Shi Scholar'' of Beijing Normal University. One of us (O.L.) would 
like to thank Z.~J.~Ajaltouni for 
correspondence. A new graphical interface, 
so-called JaxoDraw~\cite{Binosi:2003yf},  has been 
used for drawing all the  Feynman diagrams.
%
%
\section*{Appendix}\label{sec8}
%
%

 \begin{appendix}
 \section{The annihilation amplitudes}
 \subsection{The annihilation amplitudes for $\boldsymbol{B{\to}PV}$}
 \begin{multline}
 {\cal A}^{a}({\overline{B}}^{0}{\to}{\rho}^{+} {\pi}^{-}) =
 -i \frac{G_{F}}{\sqrt{2}} f_{B} f_{\pi} f_{\rho} \sum_{p=u}^{c} V_{pb}V_{pd}^{\ast}
  \Bigg\{ \delta_{pu} b_{1}^{p}(\pi,\rho)+ b_{3}^{p}(\rho,\pi)+b_{4}^{p}(\rho,\pi)+b_{4}^{p}(\pi,\rho)
\\
- \frac{1}{2}b_{3}^{p,ew}(\rho,\pi) -\frac{1}{2}b_{4}^{p,ew}(\rho,\pi)+b_{4}^{p,ew}(\pi,\rho)
 \Bigg\}\ ,
 \end{multline}
 \begin{multline}
 {\cal A}^{a}({\overline{B}}^{0}{\to}{\pi}^{+}{\rho}^{-}) =
  -i \frac{G_{F}}{\sqrt{2}} f_{B} f_{\pi} f_{\rho} \sum_{p=u}^{c} V_{pb}V_{pd}^{\ast}
 \Bigg\{ \delta_{pu} b_{1}^{p}(\rho,\pi)+ b_{3}^{p}(\pi,\rho)+b_{4}^{p}(\pi,\rho)+b_{4}^{p}(\rho,\pi)
\\
- \frac{1}{2}b_{3}^{p,ew}(\pi,\rho) -\frac{1}{2}b_{4}^{p,ew}(\pi,\rho)+b_{4}^{p,ew}(\rho,\pi)
 \Bigg\}\ ,
 \end{multline}
  \begin{multline}
  {\cal A}^{a}({\overline{B}}^{0}{\to}{\pi}^{0}{\rho}^{0})= 
  i \frac{G_{F}}{2 \sqrt{2}} f_{B} f_{\pi} f_{\rho} \sum_{p=u}^{c} V_{pb}V_{pd}^{\ast}
  \Bigg\{ -\delta_{pu} (b_{1}^{p}(\pi,\rho)+ b_{1}^{p}(\rho,\pi)) 
 -b_{3}^{p}(\pi,\rho)-2b_{4}^{p}(\pi,\rho)\\
+\frac{1}{2}b_{3}^{p,ew}(\pi,\rho)-\frac{1}{2}b_{4}^{p,ew}(\pi,\rho)
-b_{3}^{p}(\rho,\pi)-2b_{4}^{p}(\rho,\pi)+\frac{1}{2}b_{3}^{p,ew}(\rho,\pi)-\frac{1}{2}b_{4}^{p,ew}(\rho,\pi)
 \Bigg\}\ ,
 \end{multline}
 \begin{multline}
  {\cal A}^{a}(B^{-}{\to}{\pi}^{-}{\rho}^{0})=  
  -i \frac{G_{F}}{2} f_{B} f_{\pi} f_{\rho} \sum_{p=u}^{c} V_{pb}V_{pd}^{\ast}
  \Bigg\{ \delta_{pu} (b_{2}^{p}(\rho,\pi)-b_{2}^{p}(\pi,\rho))-b_{3}^{p}(\pi,\rho)\\-b_{3}^{p,ew}(\pi,\rho)
+b_{3}^{p}(\rho,\pi)+b_{3}^{p,ew}(\rho,\pi) \Bigg\}\ ,
 \end{multline}
 \begin{multline}
  {\cal A}^{a}(B^{-}{\to}{\rho}^{-}{\pi}^{0})
  =
  -i \frac{G_{F}}{2} f_{B} f_{\pi} f_{\rho} \sum_{p=u}^{c} V_{pb}V_{pd}^{\ast}
  \Bigg\{ \delta_{pu} (b_{2}^{p}(\pi,\rho)-b_{2}^{p}(\rho,\pi))-b_{3}^{p}(\rho,\pi)\\-b_{3}^{p,ew}(\rho,\pi)
+b_{3}^{p}(\pi,\rho)+b_{3}^{p,ew}(\pi,\rho) \Bigg\}\ ,
 \end{multline}
 \begin{multline}
  {\cal A}^{a}({\overline{B}}^{0}{\to}{\pi}^{0}{\omega})= 
  i \frac{G_{F}}{2 \sqrt{2}} f_{B} f_{\pi} f_{\omega} \sum_{p=u}^{c} V_{pb}V_{pd}^{\ast}
\Bigg\{ -\delta_{pu} (b_{1}^{p}(\pi,\omega)+ b_{1}^{p}(\omega,\pi)) +b_{3}^{p}(\pi,\omega)\\
-\frac{1}{2}b_{3}^{p,ew}(\pi,\omega)-\frac{3}{2}b_{4}^{p,ew}(\pi,\omega)+
b_{3}^{p}(\omega,\pi)
-\frac{1}{2}b_{3}^{p,ew}(\omega,\pi)-\frac{3}{2}b_{4}^{p,ew}(\omega,\pi) \Bigg\}\ ,
 \end{multline}
 \begin{multline}
  {\cal A}^{a}(B^{-}{\to}{\pi}^{-}{\omega})
  = 
  -i \frac{G_{F}}{2} f_{B} f_{\pi} f_{\omega} \sum_{p=u}^{c} V_{pb}V_{pd}^{\ast}
  \Bigg\{\delta_{pu} (b_{2}^{p}(\pi,\omega) + b_{2}^{p}(\omega,\pi))
  + b_{3}^{p}(\pi,\omega)   \\
+b_{3}^{p,ew}(\pi,\omega) +  b_{3}^{p}(\omega,\pi) +  b_{3}^{p,ew}(\omega,\pi) \Bigg\}\ , 
 \end{multline}
  \begin{multline}
  {\cal A}^{a}({\overline{B}}^{0}{\to}{\rho}^{+} K^{-})
  = 
  - i \frac{G_{F}}{\sqrt{2}} f_{B} f_{K} f_{\rho} \sum_{p=u}^{c} V_{pb}V_{ps}^{\ast}
  \Bigg\{ b_{3}^{p}(\rho,K)- \frac{1}{2}b_{3}^{p,ew}(\rho,K) \Bigg\}\ , \hspace{2.0cm}
 \end{multline}
 \begin{multline}
  {\cal A}^{a}({\overline{B}}^{0}{\to}{\rho}^{0}{\overline{K}}^{0})
   = 
   - i \frac{G_{F}}{2} f_{B} f_{K} f_{\rho}\sum_{p=u}^{c} V_{pb}V_{ps}^{\ast}
   \Bigg\{ -b_{3}^{p}(\rho,K)+\frac{1}{2}b_{3}^{p,ew}(\rho,K) \Bigg\}\ ,\hspace{1.3cm}
 \end{multline}
 \begin{multline}
  {\cal A}^{a}(B^{-}{\to}{\rho}^{0} K^{-})
  = 
  -i \frac{G_{F}}{2} f_{B} f_{K} f_{\rho} \sum_{p=u}^{c} V_{pb}V_{ps}^{\ast}    
  \Bigg\{\delta_{pu} b_{2}^{p}(\rho,K)
  + b_{3}^{p}(\rho,K) + b_{3}^{p,ew}(\rho,K) \Bigg\}\ ,
 \end{multline}
 \begin{multline}
  {\cal A}^{a}(B^{-}{\to}{\rho}^{-} {\overline{K}}^{0})
  =
  -i \frac{G_{F}}{\sqrt{2}} f_{B} f_{K} f_{\rho}\sum_{p=u}^{c} V_{pb}V_{ps}^{\ast}
  \Bigg\{\delta_{pu} b_{2}^{p}(\rho,K)+
  b_{3}^{p}(\rho,K)+ b_{3}^{p,ew}(\rho,K) \Bigg\}\ ,
 \end{multline}
 \begin{multline}
  {\cal A}^{a}(B^{-}{\to}K^{-}{\omega})
   =
   -i \frac{G_{F}}{2} f_{B} f_{K} f_{\omega}  \sum_{p=u}^{c} V_{pb}V_{ps}^{\ast} 
  \Bigg\{b_{3}^{p}(\omega,K) + b_{3}^{p,ew}(\omega,K) \Bigg\}\ , \hspace{1.6cm} 
 \end{multline}
 \begin{multline}
  {\cal A}^{a}({\overline{B}}^{0}{\to}{\overline{K}}^{0}{\omega})
   = 
   -i \frac{G_{F}}{2} f_{B} f_{K} f_{\omega}\sum_{p=u}^{c} V_{pb}V_{ps}^{\ast}
  \Bigg\{   b_{3}^{p}(\omega,K)- \frac{1}{2}b_{3}^{ew}(\omega,K) \Bigg\}\ ,\hspace{2.1cm}
 \end{multline}
 \begin{multline}
 {\cal A}^{a}({\overline{B}}^{0}{\to}{\pi}^{+}K^{{\ast}-})
  = -i \frac{G_{F}}{\sqrt{2}} f_{B} f_{\pi} f_{K^{\ast}}\sum_{p=u}^{c} V_{pb}V_{ps}^{\ast}
 \Bigg\{  b_{3}^{p}(\pi,K^{\ast})
 - \frac{1}{2}b_{3}^{p,ew}(\pi,K^{\ast}) \Bigg\}\ ,\hspace{1.2cm}
 \end{multline}
 \begin{multline}
  {\cal A}^{a}({\overline{B}}^{0}{\to}{\pi}^{0}{\overline{K}}^{{\ast}0})
  =   
  - i \frac{G_{F}}{2} f_{B} f_{\pi} f_{K^{\ast}}\sum_{p=u}^{c} V_{pb}V_{ps}^{\ast}
  \Bigg\{ -b_{3}^{p}(\pi,K^{\ast})+\frac{1}{2}b_{3}^{p,ew}(\pi,K^{\ast}) \Bigg\}\ ,
 \end{multline}
 \begin{multline}
  {\cal A}^{a}(B^{-}{\to}{\pi}^{-}{\overline{K}}^{{\ast}0})
  = 
  -i \frac{G_{F}}{\sqrt{2}} f_{B} f_{\pi} f_{K^{\ast}}\sum_{p=u}^{c} V_{pb}V_{ps}^{\ast}
\Bigg\{ \delta_{pu} b_{2}^{p}(\pi,K^{\ast})  +b_{3}^{p}(\pi,K^{\ast})  \\ +
 b_{3}^{p,ew}(\pi,K^{\ast}) \Bigg\}\ ,
 \end{multline}
 \begin{multline}
  {\cal A}^{a}(B^{-}{\to}{\pi}^{0}K^{{\ast}-})
  = 
  - i \frac{G_{F}}{2} f_{B} f_{\pi} f_{K^{\ast}} \sum_{p=u}^{c} V_{pb}V_{ps}^{\ast}
   \Bigg\{ \delta_{pu}  b_{2}^{p}(\pi,K^{\ast})+b_{3}^{p}(\pi,K^{\ast})\\
+b_{3}^{p,ew}(\pi,K^{\ast})\Bigg\}\ ,
 \end{multline}
 \begin{multline}
  {\cal A}^{a}(B^{-}{\to}K^{-}K^{{\ast}0})
   = 
   -i \frac{G_{F}}{\sqrt{2}} f_{B} f_{K} f_{K^{\ast}} \sum_{p=u}^{c} V_{pb}V_{ps}^{\ast}
  \Bigg\{ \delta_{pu} b_{2}^{p}(K,K^{\ast})+b_{3}^{p}(K,K^{\ast})\\ +
b_{3}^{p,ew}(K,K^{\ast}) \Bigg\}\ ,
 \end{multline}
 \begin{multline}
  {\cal A}^{a}({\overline{B}}^{0}{\to}K^{-}K^{\ast +})
   = 
  -i\frac{G_{F}}{\sqrt{2}}f_{B}f_{K}f_{K^{\ast}} \sum_{p=u}^{c} V_{pb}V_{ps}^{\ast}
   \Bigg\{\delta_{pu} b_{1}^{p}(K,K^{\ast})+b_{4}^{p}(K,K^{\ast})\\+b_{4}^{p,ew}(K,K^{\ast})
   +b_{4}^{p}(K^{\ast},K)-\frac12 b_{4}^{p,ew}(K^{\ast},K)\Bigg\}\ ,
  \end{multline}
   \begin{multline}
  {\cal A}^{a}({\overline{B}}^{0}{\to}{\overline{K}}^{0}K^{\ast 0})
   = 
  i\frac{G_{F}}{\sqrt{2}}f_{B}f_{K} f_{K^{\ast}} \sum_{p=u}^{c} V_{pb}V_{ps}^{\ast}
  \Bigg\{ b_{3}^{p}(K,K^{\ast})+b_{4}^{p}(K,K^{\ast})-\frac12 b_{3}^{p,ew}(K,K^{\ast})\\ 
-\frac12 b_{4}^{p,ew}(K,K^{\ast})
  +b_{4}^{p}(K^{\ast},K)-\frac12 b_{4}^{p,ew}(K^{\ast},K)\Bigg\}\ ,
   \end{multline}
 \begin{multline}
  {\cal A}^{a}(B^{-}{\to}K^{\ast -}K^{0})
   = 
   -i \frac{G_{F}}{\sqrt{2}} f_{B} f_{K} f_{K^{\ast}} \sum_{p=u}^{c} V_{pb}V_{ps}^{\ast}
  \Bigg\{ \delta_{pu} b_{2}^{p}(K^{\ast},K)+b_{3}^{p}(K^{\ast},K)\\
  +b_{3}^{p,ew}(K^{\ast},K) \Bigg\}\ ,
 \end{multline}
 \begin{multline}
  {\cal A}^{a}({\overline{B}}^{0}{\to}K^{\ast -}K^{+})
   = 
  -i\frac{G_{F}}{\sqrt{2}}f_{B}f_{K}f_{K^{\ast}} \sum_{p=u}^{c} V_{pb}V_{ps}^{\ast}
   \Bigg\{\delta_{pu} b_{1}^{p}(K^{\ast},K)+b_{4}^{p}(K^{\ast},K)\\ +b_{4}^{p,ew}(K^{\ast},K)
   +b_{4}^{p}(K,K^{\ast})-\frac12 b_{4}^{p,ew}(K,K^{\ast})\Bigg\}\ ,
  \end{multline}
   \begin{multline}
  {\cal A}^{a}({\overline{B}}^{0}{\to}{\overline{K}}^{\ast 0}K^{0})
   = 
  -i\frac{G_{F}}{\sqrt{2}}f_{B}f_{K} f_{K^{\ast}}\sum_{p=u}^{c} V_{pb}V_{ps}^{\ast}
  \Bigg\{ b_{3}^{p}(K^{\ast},K)+b_{4}^{p}(K^{\ast},K)\\-\frac12 b_{3}^{p,ew}(K^{\ast},K)
-\frac12 b_{4}^{p,ew}(K^{\ast},K)
  +b_{4}^{p}(K,K^{\ast})-\frac12 b_{4}^{p,ew}(K,K^{\ast})\Bigg\}\ ,
   \end{multline}
 \begin{multline}
  {\cal A}^{a}(B^{-}{\to}K^{-}{\phi})
   = 
   -i \frac{G_{F}}{\sqrt{2}} f_{B} f_{K} f_{\phi}\sum_{p=u}^{c} V_{pb}V_{ps}^{\ast}
  \Bigg\{ b_{3}^{p}(K,\phi)+ b_{3}^{p,ew}(K,\phi) \Bigg\}\ ,\hspace{1.6cm} 
 \end{multline}
  \begin{multline}
  {\cal A}^{a}({\overline{B}}^{0}{\to}{\overline{K}}^{0}{\phi})
   = 
   - i \frac{G_{F}}{\sqrt{2}} f_{B} f_{K} f_{\phi}\sum_{p=u}^{c} V_{pb}V_{ps}^{\ast}
   \Bigg\{  b_{3}^{p}(K,{\phi})-\frac{1}{2}b_{3}^{p,ew}(K,{\phi}) \Bigg\}\ , \hspace{2.0cm}
 \end{multline}
 \begin{multline}
  {\cal A}^{a}(B^{-}{\to}{\pi}^{-}{\phi})
={\cal A}^{a}({\overline{B}}^{0}{\to}{\pi}^{0}{\phi}) =0\ . \hspace{7.8cm}
 \end{multline}
%
  \subsection{The annihilation amplitudes for $\boldsymbol{B{\to}PP}$}
%
 \begin{multline}
  {\cal A}^{a}({\overline{B}}^{0}{\to}{\pi}^{+}{\pi}^{-})
   = 
  i\frac{G_{F}}{\sqrt{2}}f_{B}f^{2}_{\pi}\sum_{p=u}^{c} V_{pb}V_{pd}^{\ast}
   \Bigg\{ \delta_{pu} b_{1}^{p}(\pi,\pi)+
   b_{3}^{p}(\pi,\pi)+2b_{4}^{p}(\pi,\pi)\\
  -\frac{1}{2}b_{3}^{p,ew}(\pi,\pi)
  +\frac{1}{2}b_{4}^{p,ew}(\pi,\pi)\Bigg\}\ ,
  \end{multline}
  \begin{multline}
  {\cal A}^{a}({\overline{B}}^{0}{\to}{\pi}^{0}{\pi}^{0})
  =
 -i\frac{G_{F}}{\sqrt{2}}f_{B}f^{2}_{\pi}\sum_{p=u}^{c} V_{pb}V_{pd}^{\ast}
   \Bigg\{ -\delta_{pu} b_{1}^{p}(\pi,\pi)-
   b_{3}^{p}(\pi,\pi)-2b_{4}^{p}(\pi,\pi)\\
  +\frac{1}{2}b_{3}^{p,ew}(\pi,\pi)
  -\frac{1}{2}b_{4}^{p,ew}(\pi,\pi)\Bigg\}\ ,
  \end{multline}
  \begin{multline}
  {\cal A}^{a}(B^{-}{\to}{\pi}^{0}{\pi}^{-})
   =
   0\ ,\hspace{10.5cm}
  \end{multline}
   \begin{multline}
  {\cal A}^{a}({\overline{B}}^{0}{\to}{\overline{K}}^{0}K^{0})
   = 
  i\frac{G_{F}}{\sqrt{2}}f_{B}f_{K}^{2}\sum_{p=u}^{c} V_{pb}V_{ps}^{\ast}
  \Bigg\{ b_{3}^{p}(\overline{K},K)+b_{4}^{p}(\overline{K},K)-\frac12 b_{3}^{p,ew}(\overline{K},K)\\ 
-\frac12 b_{4}^{p,ew}(\overline{K},K)
  +b_{4}^{p}(K,\overline{K})-\frac12 b_{4}^{p,ew}(K,\overline{K})\Bigg\}\ ,
   \end{multline}
   \begin{multline}
  {\cal A}^{a}(B^{-}{\to}K^{-}K^{0})
   = 
  i\frac{G_{F}}{\sqrt{2}}f_{B}f_{K}^{2}\sum_{p=u}^{c} V_{pb}V_{ps}^{\ast}
   \Bigg\{\delta_{pu} b_{2}^{p}(\overline{K},K)+b_{3}^{p}(\overline{K},K)
+b_{3}^{p,ew}(\overline{K},K)\Bigg\}\ ,
   \end{multline}
   \begin{multline}
  {\cal A}^{a}({\overline{B}}^{0}{\to}K^{-}K^{+})
   = 
  i\frac{G_{F}}{\sqrt{2}}f_{B}f_{K}^{2}\sum_{p=u}^{c} V_{pb}V_{ps}^{\ast}
   \Bigg\{\delta_{pu} b_{1}^{p}(\overline{K},K)+b_{4}^{p}(\overline{K},K)+b_{4}^{p,ew}(\overline{K},K)\\
+b_{4}^{p}(K,\overline{K})-\frac12 b_{4}^{p,ew}(K,\overline{K})\Bigg\}\ ,
  \end{multline}
   \begin{multline}
  {\cal A}^{a}({\overline{B}}^{0}{\to}{\pi}^{+}K^{-})
  =
   i \frac{G_{F}}{\sqrt{2}} f_{B} f_{\pi} f_{K}\sum_{p=u}^{c} V_{pb}V_{ps}^{\ast}
 \Bigg\{b_{3}^{p}(\pi,K)- \frac{1}{2}b_{3}^{p,ew}(\pi,K) \Bigg\}\ ,\hspace{1.5cm}
  \end{multline}
   \begin{multline}
  {\cal A}^{a}({\overline{B}}^{0}{\to}{\pi}^{0}{\overline{K}}^{0})
  =
  i\frac{G_{F}}{2}f_{B}f_{\pi}f_{K}\sum_{p=u}^{c} V_{pb}V_{ps}^{\ast}
  \Bigg\{
  -b_{3}^{p}(\pi,K)+\frac{1}{2}b_{3}^{p,ew}(\pi,K)\Bigg\}\ , \hspace{1.5cm}
  \end{multline}
   \begin{multline}
  {\cal A}^{a}(B^{-}{\to}{\pi}^{0}K^{-})
   = 
  i\frac{G_{F}}{2}f_{B}f_{\pi}f_{K} \sum_{p=u}^{c} V_{pb}V_{ps}^{\ast}
   \Bigg\{ \delta_{pu} b_{2}^{p}(\pi,K)+b_{3}^{p}(\pi,K)+b_{3}^{p,ew}(\pi,K)\Bigg\}\ ,
  \end{multline}
   \begin{multline}
  {\cal A}^{a}(B^{-}{\to}{\pi}^{-}{\overline{K}}^{0})
  =i \frac{G_{F}}{\sqrt{2}} f_{B} f_{\pi} f_{K}\sum_{p=u}^{c} V_{pb}V_{ps}^{\ast}
  \Bigg\{ \delta_{pu} b_{2}^{p}(\pi,K)+b_{3}^{p}(\pi,K)+b_{3}^{p,ew}(\pi,K)\Bigg\}\ , 
  \end{multline}
   \begin{multline}
  {\cal A}^{a}({\overline{B}}^{0}{\to}{\pi}^{0}{{\eta}^{({\prime})}})
  = 
  -i\frac{G_{F}}{2\sqrt{2}}f_{B}f_{\pi}f_{{\eta}^{({\prime})}}^{q}\sum_{p=u}^{c} V_{pb}V_{ps}^{\ast}
  \Bigg(
\Bigg\{
\delta_{pu}(-b_{1}^{p}(\pi,\eta_q^{(\prime)})-2b_{S1}^{p}(\pi,\eta_q^{(\prime)})\\
-b_{1}^{p}(\eta_q^{(\prime)},\pi))+b_{3}^{p}(\pi,\eta_q^{(\prime)})-\frac12 b_{3}^{p,ew}(\pi,\eta_q^{(\prime)})
-\frac32 b_{4}^{p}(\pi,\eta_q^{(\prime)})+2 b_{S3}^{p}(\pi,\eta_q^{(\prime)})\\
-b_{S3}^{p,ew}(\pi,\eta_q^{(\prime)})
-3 b_{S4}^{p,ew}(\pi,\eta_q^{(\prime)})+b_{3}^{p}(\eta_q^{(\prime)},\pi)-\frac12 b_{3}^{p,ew}(\eta_q^{(\prime)},\pi)
-\frac32 b_{4}^{p}(\eta_q^{(\prime)},\pi)
\Bigg\}
\\
+\Bigg(\sqrt{2}\frac{f_{\eta^{({\prime})}}^{s}}{f_{\eta^{({\prime})}}^{q}}\Bigg)
\Bigg\{ -\delta_{pu} b_{S1}^{p}(\pi,\eta_s^{(\prime)})+b_{S3}^{p}(\pi,\eta_s^{(\prime)})
-\frac12 b_{S3}^{p,ew}(\pi,\eta_s^{(\prime)})-\frac32  b_{S4}^{p,ew}(\pi,\eta_s^{(\prime)})
\Bigg\}
\Bigg)\ ,
  \end{multline}
   \begin{multline}
  {\cal A}^{a}(B^{-}{\to}{\pi}^{-}{\eta^{({\prime})}})
   = 
  i\frac{G_{F}}{2}f_{B}f_{\pi}f_{\eta^{({\prime})}}^{q}\sum_{p=u}^{c} V_{pb}V_{ps}^{\ast}
  \Bigg(
\Bigg\{\delta_{pu}(b_{2}^{p}(\pi,\eta_q^{(\prime)})+b_{2}^{p}(\eta_q^{(\prime)},\pi) 
+ 2b_{S2}^{p}(\pi,\eta_q^{(\prime)}))\\
 + b_{3}^{p}(\pi,\eta_q^{(\prime)})+b_{3}^{p,ew}(\pi,\eta_q^{(\prime)})
 +2 b_{S3}^{p}(\pi,\eta_q^{(\prime)}) +2 b_{S3}^{p,ew}(\pi,\eta_q^{(\prime)}) +
 b_{3}^{p}(\eta_q^{(\prime)},\pi)+b_{3}^{p,ew}(\eta_q^{(\prime)},\pi)
\Bigg\} \\
+ \Bigg(\frac{2}{\sqrt{2}}\frac{f_{\eta^{({\prime})}}^{s}}{f_{\eta^{({\prime})}}^{q}}\Bigg)
\Bigg\{ \delta_{pu}  b_{S2}^{p}(\pi,\eta_s^{(\prime)})
+b_{S3}^{p}(\pi,\eta_s^{(\prime)})+b_{S3}^{p,ew}(\pi,\eta_s^{(\prime)})
\Bigg\}
\Bigg)\ ,   
  \end{multline}
   \begin{multline}
  {\cal A}^{a}({\overline{B}}^{0}{\to}{\overline{K}}^{0}{\eta}^{({\prime})})
   =
  i\frac{G_{F}}{2}f_{B}f_{K}f_{{\eta}^{({\prime})}}^{q}\sum_{p=u}^{c} V_{pb}V_{ps}^{\ast}
\Bigg(
\Bigg \{ 2 b_{S3}^{p}(K,\eta_q^{(\prime)})-b_{S3}^{p,ew}(K,\eta_q^{(\prime)})\\+b_{3}^{p}(\eta_q^{(\prime)},K)-
\frac12 b_{3}^{p,ew}(\eta_q^{(\prime)},K)
\Bigg\}
+ \Bigg(\frac{2}{\sqrt{2}}\frac{f_{\eta^{({\prime})}}^{s}}{f_{\eta^{({\prime})}}^{q}}\Bigg)
\Bigg \{
b_{3}^{p}(K,\eta_s^{(\prime)})-\frac12 b_{3}^{p,ew}(K,\eta_s^{(\prime)})\\+b_{S3}^{p}(K,\eta_s^{(\prime)}) 
-\frac12 b_{S3}^{p,ew}(K,\eta_s^{(\prime)})
\Bigg \}
\Bigg)\ ,
   \end{multline}
  \begin{multline}
    {\cal A}^{a}(B^{-}{\to}K^{-}{\eta}^{({\prime})})
   = 
   i\frac{G_{F}}{2}f_{B}f_{K}f_{{\eta}^{({\prime})}}^{q}\sum_{p=u}^{c} V_{pb}V_{ps}^{\ast}
  \Bigg(
\Bigg\{\delta_{pu}(2 b_{S2}^{p}(K,\eta_q^{(\prime)})+ b_{2}^{p}(\eta_q^{(\prime)},K))\\
+2 b_{S3}^{p}(K,\eta_q^{(\prime)})+2 b_{S3}^{p,ew}(K,\eta_q^{(\prime)})+b_{3}^{p}(\eta_q^{(\prime)},K)+
b_{3}^{p,ew}(\eta_q^{(\prime)},K)
\Bigg\}
+ \Bigg(\frac{2}{\sqrt{2}}\frac{f_{\eta^{({\prime})}}^{s}}{f_{\eta^{({\prime})}}^{q}}\Bigg)
\Bigg\{ \delta_{pu} (b_{2}^{p}(K,\eta_s^{(\prime)})\\
+b_{S2}^{p}(K,\eta_s^{(\prime)}))
+ b_{3}^{p}(K,\eta_s^{(\prime)})+ b_{3}^{p,ew}(K,\eta_s^{(\prime)})+b_{S3}^{p}(K,\eta_s^{(\prime)})
+b_{S3}^{p,ew}(K,\eta_s^{(\prime)})
\Bigg\}
\Bigg)\ .
   \end{multline}
All the  singlet weak annihilation coefficients, $b_{Si}$, appearing into the decay amplitude of 
$B \to X \eta^{(\prime)}$ (with $X=\{\pi, K\})$ can be neglected in first approximation. Their 
expressions can be found in Ref.~\cite{Beneke:2003zv}.
%
\section{Amplitudes}
%
%
\subsection{The decay amplitudes for $\boldsymbol{B \to PV}$}
We use the following definition for the $\alpha_{i}^{p}(M_1,M_2)$ parameters:
\begin{align*}
   \alpha_1^p(M_1,M_2) & = a_1^p(M_1,M_2) \,; \\
   \alpha_2^p(M_1,M_2) & = a_2^p(M_1,M_2) \,; \\
   \alpha_3^p(M_1,M_2) & = \left\{
    \begin{array}{cl} 
     a_3^p(M_1,M_2) - a_5^p(M_1,M_2) \,;
      & \quad \mbox{if~} M_1 M_2=PP, \,VP \,, \\
     a_3^p(M_1,M_2) + a_5^p(M_1,M_2) \,;
      & \quad \mbox{if~} M_1 M_2=PV  \,, 
    \end{array}\right. \\
   \alpha_4^p(M_1,M_2) &= \left\{
    \begin{array}{cl} 
     a_4^p(M_1,M_2) + r_{\chi}^{M_2}\,a_6^p(M_1,M_2) \,;
      & \quad \mbox{if~} M_1 M_2=PP, \,PV \,, \\
     a_4^p(M_1,M_2) - r_{\chi}^{M_2}\,a_6^p(M_1,M_2) \,;
      & \quad \mbox{if~} M_1 M_2=VP\,,   
    \end{array}\right. \\
   \alpha_{3,\rm EW}^p(M_1,M_2) &= \left\{
    \begin{array}{cl} 
     a_9^p(M_1,M_2) - a_7^p(M_1,M_2) \,;
      & \quad \mbox{if~} M_1 M_2=PP, \,VP \,, \\
     a_9^p(M_1,M_2) + a_7^p(M_1,M_2) \,;
      & \quad \mbox{if~} M_1 M_2=PV  \,, 
    \end{array}\right. \\
   \alpha_{4,\rm EW}^p(M_1,M_2) &= \left\{
    \begin{array}{cl} 
     a_{10}^p(M_1,M_2) + r_{\chi}^{M_2}\,a_8^p(M_1,M_2) \,;
      & \quad \mbox{if~} M_1 M_2=PP, \,PV \,, \\
     a_{10}^p(M_1,M_2) - r_{\chi}^{M_2}\,a_8^p(M_1,M_2) \,;
      & \quad \mbox{if~} M_1 M_2=VP\,,
     \end{array}\right.
\end{align*}
with $r_{\chi}^{M_2}$ defined in section 3.2.
%
\begin{multline}
   {\cal A} ({\overline{B}}^{0} \to \rho^+  \pi^-)  
   =
    -i \frac{G_F}{\sqrt{2}}m_B^2 f_\pi A_0^{B\to \rho} (m_\pi^2) \sum_{p=u}^{c} V_{pb}V_{pd}^{\ast}
 \Bigg \{\delta_{pu} \alpha_1^{p}(\rho,\pi) + \alpha_4^{p}(\rho,\pi) \\+\alpha_4^{p,ew}(\rho,\pi)\Bigg \}\ ,
\end{multline}
\begin{multline}
   {\cal A} ({\overline{B}}^{0} \to \pi^+ \rho^-)  
   =  
   -i \frac{G_F}{\sqrt{2}}m_B^2 f_\rho F_1^{B\to \pi } (m_\rho^2) \sum_{p=u}^{c} V_{pb}V_{pd}^{\ast}
 \Bigg \{\delta_{pu} \alpha_1^{p}(\pi,\rho) + \alpha_4^{p}(\pi,\rho) \\+\alpha_4^{p,ew}(\pi,\rho)\Bigg \}\ ,
\end{multline}
\begin{multline}
  {\cal A} ({\overline{B}}^{0} \to \pi^0 \rho^0 )
  =
  i\frac{G_F}{2\sqrt{2}} m_B^2 \sum_{p=u}^{c} V_{pb}V_{pd}^{\ast}
  \Bigg(f_\pi A_0^{B\to \rho }(m_\pi^2)\Bigg\{\delta_{pu} \alpha_2^{p}(\rho,\pi)
  - \alpha_4^{p}(\rho,\pi)\\+\frac32 \alpha_3^{p,ew}(\rho,\pi)
+\frac12 \alpha_4^{p,ew}(\rho,\pi)\Biggr\}  +f_\rho F_1^{B\to\pi } (m_\rho^2)
   \Bigg\{ \delta_{pu} \alpha_2^{p}(\pi,\rho)
  - \alpha_4^{p}(\pi,\rho)+\frac32 \alpha_3^{p,ew}(\pi,\rho)\\+\frac12 \alpha_4^{p,ew}(\pi,\rho)\Bigg\} \Bigg)\ , 
\end{multline}
\begin{multline}
  {\cal A} ( B^- \to \pi^- \rho^0 )
  = 
  -i \frac{G_F}{2 } m_B^2  \sum_{p=u}^{c} V_{pb}V_{pd}^{\ast}
    \Bigg(f_\pi A_0^{B\to \rho }(m_\pi^2)
  \Bigg\{\delta_{pu} \alpha_1^{p}(\rho,\pi)  +\alpha_4^{p}(\rho,\pi)\\
  +\alpha_4^{p,ew}(\rho,\pi)
    \Bigg\} 
  +  f_\rho F_1^{B\to \pi } (m_\rho^2)
  \Bigg \{\delta_{pu} \alpha_2^{p}(\pi,\rho)-\alpha_4^{p}(\pi,\rho)+ \frac23 \alpha_3^{p,ew}(\pi,\rho)
   +  \frac12 \alpha_4^{p,ew}(\pi,\rho)\Bigg \}\Bigg) \ ,
\end{multline}
\begin{multline}
    {\cal A} ( B^- \to \pi^0  \rho^-)
    =
    -i\frac{G_F}{2 } m_B^2 \sum_{p=u}^{c} V_{pb}V_{pd}^{\ast}
   \Bigg( f_\pi A_0^{B\to \rho }(m_\pi^2) 
\Bigg \{\delta_{pu} \alpha_2^{p}(\rho,\pi)- \alpha_4^{p}(\rho,\pi)+ \\ \frac32  \alpha_3^{p,ew}(\rho,\pi) 
+\frac12  \alpha_4^{p,ew}(\rho,\pi)\Bigg\} 
    +   f_\rho F_1^{B\to \pi } (m_\rho^2)
   \Bigg \{ \delta_{pu} \alpha_1^{p}(\pi,\rho) + \alpha_4^{p}(\pi,\rho)+\alpha_4^{p,ew}(\pi,\rho)\Bigg\} \Bigg)\ ,
\end{multline}
\begin{multline}
     {\cal A} ( B^0 \to \pi^0 \omega)  
   =  
   i\frac{G_F}{2 \sqrt{2}} m_B^2\sum_{p=u}^{c} V_{pb}V_{pd}^{\ast}
   \Bigg(f_\pi A_0^{B\to \omega }(m_\pi^2)   
   \Bigg \{-\delta_{pu} \alpha_2^{p}(\omega,\pi) 
  +\alpha_4^{p}(\omega,\pi)  \\+ \alpha_4^{p,ew}(\omega,\pi) \Bigg \} 
   +    f_\omega F_1^{B\to \pi } (m_\omega^2)
   \Bigg \{  \delta_{pu}\alpha_2^{p}(\pi,\omega) + 
  2 \alpha_3^{p}(\pi,\omega) +  \alpha_4^{p}(\pi,\omega) + \frac12 \alpha_3^{p}(\pi,\omega)  \Bigg\} \Bigg)\ ,
\end{multline}
\begin{multline}
     {\cal A} ( B^- \to \pi^- \omega)  
   =  
   -i\frac{G_F}{2 } m_B^2\sum_{p=u}^{c} V_{pb}V_{pd}^{\ast}
   \Bigg(f_\pi A_0^{B\to \omega }(m_\pi^2)   
   \Bigg \{\delta_{pu} \alpha_1^{p}(\omega,\pi) 
  +\alpha_4^{p}(\omega,\pi)  \\+ \alpha_4^{p,ew}(\omega,\pi)\Bigg \} 
   +    f_\omega F_1^{B\to \pi } (m_\omega^2)
   \Bigg \{  \delta_{pu}\alpha_2^{p}(\pi,\omega) + 
  2 \alpha_3^{p}(\pi,\omega) +  \alpha_4^{p}(\pi,\omega) + \frac12 \alpha_3^{p,ew}(\pi,\omega) \Bigg\} \Bigg)\ ,
\end{multline}
\begin{multline}
  {\cal A} ({\overline{B}}^{0} \to \rho^+ K^{-}) 
  = 
  -i\frac{G_F}{\sqrt{2}} m_B^2 f_K A_0^{B\to \rho }(m_K^2)\sum_{p=u}^{c} V_{pb}V_{ps}^{\ast}
\Bigg \{\delta_{pu} \alpha_1^{p}(\rho,K) + \alpha_4^{p}(\rho,K)\\
+\alpha_4^{p,ew}(\rho,K)\Bigg \}\ ,
\end{multline}
\begin{multline}
   {\cal A} ({\overline{B}}^{0} \to  \rho^0 {\overline K}^0)   
   = 
   -i\frac{G_F}{2} m_B^2 \sum_{p=u}^{c} V_{pb}V_{ps}^{\ast}
   \Bigg(f_{K} A_0^{B\to \rho } (m_{K}^2)
    \Bigg \{-\alpha_4^{p}(\rho,K)+\frac12 \alpha_4^{p,ew}(\rho,K)\Bigg\} \\
   +  f_\rho F_1^{B\to K }(m_\rho^2)\Bigg \{\delta_{pu} \alpha_2^{p}(K,\rho)+ \frac32 \alpha_3^{p,ew}(K,\rho)
\Bigg\}\Bigg)\ ,
\end{multline}
\begin{multline}
  {\cal A} ( B^- \to \rho^0 K^-)  
  =  
   -i\frac{G_F}{2} m_B^2 \sum_{p=u}^{c} V_{pb}V_{ps}^{\ast}
   \Bigg(
     f_{K} A_0^{B\to \rho } (m_{K}^2)
    \Bigg\{ \delta_{pu} \alpha_1(\rho,K) + \alpha_4^{p}(\rho,K)  \\  +\alpha_4^{p,ew}(\rho,K)\Bigg \} 
    +   f_\rho F_1^{B\to K }(m_\rho^2)
    \Bigg \{\delta_{pu} \alpha_2(K,\rho) +
   \frac32 \alpha_{3}^{p,ew}(K,\rho) \Bigg \} \Bigg)\ ,
\end{multline}
\begin{multline}
      {\cal A} ( B^- \to \rho^- {\overline K}^0) 
     =
    -i  \frac{G_F}{\sqrt{2}}m_B^2 
     f_{K} A_0^{B\to \rho } (m_{K}^2)  \sum_{p=u}^{c} V_{pb}V_{ps}^{\ast} \Bigg \{
    \alpha_4^{p}(\rho,K)-\frac12 \alpha_4^{p,ew}(\rho,K) \Bigg \}\ ,
\end{multline}
\begin{multline}
   {\cal A} ( B^- \to K^- \omega) 
   =  
   -i\frac{G_F}{2} m_B^2 \sum_{p=u}^{c} V_{pb}V_{ps}^{\ast}
   \Bigg(
    f_{K} A_0^{B\to \omega } (m_{K}^2)
   \Bigg \{\delta_{pu} \alpha_1^{p}(\omega,K) +\alpha_4^{p}(\omega,K)\\+ \alpha_4^{p,ew}(\omega,K)
    \Bigg \} +  
     f_\omega F_1^{B\to K }(m_\omega^2)
    \Bigg \{\delta_{pu} \alpha_2^{p}(K,\omega) + 2 \alpha_3^{p}(K,\omega)+ \delta_{pc} \frac12 \alpha_3^{p,ew}(K,\omega)  
     \Bigg \}\Bigg)\ ,
\end{multline}
\begin{multline}
  {\cal A} ({\overline{B}}^{0} \to {\overline K}^0 \omega)   
  =
  -i \frac{G_F}{2} m_B^2 \sum_{p=u}^{c} V_{pb}V_{ps}^{\ast}
  \Bigg( f_{K} A_0^{B\to \omega } (m_{K}^2)
  \Bigg\{\alpha_4^{p}(\omega,K) -\frac12 \alpha_4^{p,ew}(\omega,K)\Bigg \} \\
  +    f_\omega F_1^{B\to K }(m_\omega^2)
  \Bigg \{ \delta_{pu} \alpha_2^{p}(K,\omega) +2 \alpha_3^{p}(K,\omega)
   +  \delta_{pc} \frac12 \alpha_3^{p,ew}(K,\omega)
  \Bigg \} \Bigg)\ ,
\end{multline}
\begin{multline}
  {\cal A} ({\overline{B}}^{0} \to \pi^+ K^{*-})
  =
  -i\frac{G_F}{\sqrt{2}} m_B^2 f_{K^*} F_1^{B\to \pi }(m_{K^*}^2)\sum_{p=u}^{c} V_{pb}V_{ps}^{\ast}
  \Bigg \{ \delta_{pu} \alpha_1^{p}(\pi,K^{\ast})+\alpha_4^{p}(\pi,K^{\ast})\\ +
\alpha_4^{p,ew}(\pi,K^{\ast})
   \Bigg \} ,
\end{multline}
\begin{multline}
  {\cal A} ({\overline{B}}^{0} \to \pi^0  \bar{K}^{\ast 0})
  =
  -i\frac{G_F}{2} m_B^2 \sum_{p=u}^{c} V_{pb}V_{ps}^{\ast} \Bigg(
  f_{K^\ast} F_1^{B\to \pi }(m_{K^\ast}^2) \Bigg \{ -\alpha_4^{p}(\pi,K^{\ast})\\+ 
   \frac12 \alpha_4^{p,ew}(\pi,K^{\ast})\Bigg \} 
  + f_{\pi} F_1^{B\to K^{\ast} }(m_{\pi}^2)
    \Bigg \{ \delta_{pu} \alpha_2^{p}(K^{\ast},\pi)+ \frac32 \alpha_3^{p,ew}(K^{\ast},\pi)
   \Bigg \} \Bigg) \ ,
\end{multline}
\begin{multline}
        {\cal A} ( B^- \to \pi^- {\overline K}^{*0}) 
       =
      -i \frac{G_F}{\sqrt{2}}m_B^2 
       f_{K^*} F_1^{B\to \pi } (m_{K^{*}}^2)  \sum_{p=u}^{c} V_{pb}V_{ps}^{\ast} \Bigg \{
 \alpha_4^{p}(\pi,K^{\ast}) \\
-\frac12 \alpha_4^{p,ew}(\pi,K^{\ast}) \Bigg \} \ ,
\end{multline}
\begin{multline}
   {\cal A} ( B^- \to \pi^0 K^{\ast -})  
   = 
   -i\frac{G_F}{2} m_B^2 \sum_{p=u}^{c} V_{pb}V_{ps}^{\ast}
    \Bigg (
    f_\pi A_0^{B\to K^{\ast}}(m_\pi^2)
    \Bigg \{\delta_{pu} \alpha_2^{p}(K^{\ast},\pi) + \\
\frac32 \alpha_3^{p,ew}(K^{\ast},\pi) \Bigg \}
   +   f_{K^{\ast}} F_1^{B\to \pi } (m_{K^{*}}^2) \Bigg \{
    \delta_{pu} \alpha_1^{p}(\pi,K^{\ast}) + \alpha_4^{p}(\pi,K^{\ast}) +\alpha_4^{p,ew}(\pi,K^{\ast})\Bigg \} \Bigg)\ ,
\end{multline}
\begin{multline}
      {\cal A} (B^- \to K^-  K^{\ast 0}) 
     = 
     {\cal A} ({\overline{B}}^{0} \to {\overline K}^0  K^{\ast 0}) =
     \\
     -i\frac{G_F}{\sqrt{2}}m_B^2 f_{K^{\ast}}
         F_1^{B\to K } (m_{K^{\ast}}^2)  \sum_{p=u}^{c} V_{pb}V_{ps}^{\ast} \Bigg \{
\alpha_4^{p}(K,K^{\ast})-\frac12 \alpha_4^{p,ew}(K,K^{\ast}) \Bigg \}\ ,
\end{multline}
\begin{multline}
   {\cal A}( \bar  B^0 \to K^-  K^{\ast +})= {\cal A}( \bar  B^0 \to K^{\ast -}  K^{+})
   = 0\ , \hspace{6.2cm}
\end{multline}
\begin{multline}
      {\cal A} ( B^- \to K^{\ast-}  K^0)
     ={\cal A} ( {\overline{B}}^{0} \to {\overline K}^{*0}  K^0) =
     \\
    -i\frac{G_F}{\sqrt{2}} m_B^2 
    f_{K} A_0^{B\to K^{\ast} } (m_{K}^2) \sum_{p=u}^{c} V_{pb}V_{ps}^{\ast} \Bigg \{
      \alpha_4^{p}(K^{\ast},K)-\frac12 \alpha_4^{p,ew}(K^{\ast},K) \Bigg \}\ ,
\end{multline}
\begin{multline}
      {\cal A} ( B^- \to K^- \phi) 
    = 
    {\cal A} (\bar{B}^{0} \to \bar{K}^0 \phi)
    = \\
    -i \frac{G_F}{\sqrt{2}} m_B^2 
    f_\phi F_1^{B\to K}(m_\phi^2) \sum_{p=u}^{c} V_{pb}V_{ps}^{\ast} \Bigg \{
  \alpha_3^{p}(K,\phi)+ \alpha_4^{p}(K,\phi)- \delta_{pc} \frac12 \alpha_3^{p,ew}(K,\phi)
\Bigg \}\ ,
\end{multline}
\begin{multline}
    {\cal A} ( B^{-} \to \pi^- \phi) 
    = 
     -\sqrt{2} {\cal A} ({\overline{B}}^{0} \to \pi^0 \phi)= \\
     -i\frac{G_F}{\sqrt{2}} m_B^2 f_\phi F_1^{B\to \pi }(m_\phi^2) \sum_{p=u}^{c} V_{pb}V_{ps}^{\ast} \Bigg \{
   \alpha_3^{p}(\pi,\phi) - \frac12 \alpha_3^{p,ew}(\pi,\phi)\Bigg \}\ ,
\end{multline}
%
  \subsection{The decay amplitudes for $\boldsymbol{B{\to}PP}$}
%
\begin{multline}
    {\cal A}(\bar B^0 \to \pi^+  \pi^-) 
   = 
   i\frac{G_F}{\sqrt{2}} f_\pi F_0^{B\to \pi } (m_\pi^2) m_B^2 \sum_{p=u}^{c} V_{pb}V_{pd}^{\ast} 
   \Bigg\{\delta_{pu} \alpha_1^{p}(\pi,\pi) + \alpha_4^{p}(\pi,\pi)\\+ \alpha_4^{p,ew}(\pi,\pi)\Bigg\}\ , 
\end{multline}
\begin{multline}
   {\cal A}(\bar B^0 \to \pi^0 \pi^0 ) 
  = 
   -i\frac{G_F}{\sqrt{2}}f_\pi F_0^{B\to \pi } (m_\pi^2) m_B^2 \sum_{p=u}^{c} V_{pb}V_{pd}^{\ast} 
   \Bigg\{ \delta_{pu} \alpha_2^{p}(\pi,\pi)-\alpha_4^{p}(\pi,\pi)  \\+\frac32 \alpha_3^{p,ew}(\pi,\pi) + 
\frac12 \alpha_4^{p,ew}(\pi,\pi)\Bigg \}\ ,
\end{multline}
\begin{multline}
   {\cal A}(B^- \to \pi^- \pi^0 )  
   = 
   i\frac{G_F}{{2}}f_\pi F_0^{B\to \pi } (m_\pi^2) m_B^2 \sum_{p=u}^{c} V_{pb}V_{pd}^{\ast}   
   \Biggl\{ \delta_{pu}(\alpha_1^{p}(\pi,\pi)+\alpha_2^{p}(\pi,\pi)) \\+ \frac32 \alpha_4^{p,ew}(\pi,\pi)
+\frac32 \alpha_3^{p,ew}(\pi,\pi)  \Bigg \}\ , 
\end{multline}
\begin{multline}
   {\cal A}( \bar  B^0 \to \overline{K}^0 K^0) 
   = 
  i\frac{G_F}{\sqrt{2}}f_K F_0^{B\to K } (m_{K}^2) m_B^2 \sum_{p=u}^{c} V_{pb}V_{ps}^{\ast}
\Bigg \{ \alpha_4^{p}(\overline{K},K)- \frac12 \alpha_4^{p,ew}(\overline{K},K)\Bigg \} \ ,
\end{multline}
\begin{multline}
   {\cal A}( B^- \to K^- K^0) 
  = 
  i\frac{G_F}{\sqrt{2}}f_K F_0^{B\to K } (m_{K}^2) m_B^2 \sum_{p=u}^{c} V_{pb}V_{ps}^{\ast} 
\Bigg \{ \alpha_4^{p}(K,K)- \frac12 \alpha_4^{p,ew}(K,K)\Bigg \} \ ,
\end{multline}
\begin{multline}
   {\cal A}( \bar  B^0 \to K^+ \ K^-) 
   = 0\ , \hspace{10.2cm}
\end{multline}
\begin{multline}
       {\cal A}(\bar B^0 \to \pi^+ K^- ) 
    = 
    i\frac{G_F}{\sqrt{2}} f_K F_0^{B\to \pi } (m_K^2) m_B^2  \sum_{p=u}^{c} V_{pb}V_{ps}^{\ast} 
     \Bigg \{ \delta_{pu} \alpha_1^{p}(\pi,K)\\  +
\alpha_4^{p}(\pi,K)+\alpha_4^{p,ew}(\pi,K)
\Bigg \}\ ,
\end{multline}
\begin{multline}
    {\cal A}(\bar B^0 \to \bar \pi^0  K^0) 
   = 
   i \frac{G_F}{2} m_B^2 \sum_{p=u}^{c} V_{pb}V_{ps}^{\ast} 
\Bigg( f_K F_0^{B\to \pi}(m_K^2) \Bigg \{ -\alpha_4^{p}(\pi,K)+ \frac12 \alpha_4^{p,ew}(\pi,K) \Bigg \} \\
+f_\pi F_0^{B\to K} (m_\pi^2) 
\Bigg \{ \delta_{pu} \alpha_2^{p}(K,\pi)+ \frac32 \alpha_3^{p,ew}(K,\pi)
\Bigg \} \Bigg) \ ,
\end{multline}
\begin{multline}
    {\cal A}( B^- \to \pi^0 K^-)   
   = 
   i\frac{G_F}{{2}} m_B^2 \sum_{p=u}^{c} V_{pb}V_{ps}^{\ast}  
\Bigg( f_K F_0^{B\to \pi } (m_K^2) \Bigg \{  \delta_{pu} \alpha_1^{p}(\pi,K)\\
+\alpha_4^{p}(\pi,K)+\alpha_4^{p,ew}(\pi,K) \Bigg \} 
 + f_\pi F_0^{B\to K } (m_\pi^2)
  \Bigg\{\delta_{pu}  \alpha_2^{p}(K,\pi)+ \frac32   \alpha_3^{p,ew}((K,\pi)
   \Bigg \} \Bigg) \ ,
\end{multline}
\begin{multline}
    {\cal A}(B^- \to \pi^- \bar K^0)  
    =
   \hspace{0.11cm} i\frac{G_F}{\sqrt{2}}f_K F_0^{B\to \pi } (m_K^2) m_B^2 \sum_{p=u}^{c} V_{pb}V_{ps}^{\ast}  
\Bigg \{ \alpha_4^{p}(\pi,K)- \frac12 \alpha_4^{p,ew}(\pi,K)\Bigg \}\ , 
\end{multline}
\begin{multline}
         {\cal A}( \bar B^0 \to \pi^0 \eta ^{(\prime)} )   
    =
    -i\frac{G_F}{2\sqrt{2}}m_B^2 \sum_{p=u}^{c} V_{pb}V_{pd}^{\ast} 
   \Bigg(  f_{\eta^{(\prime)}_q}  F_0^{B\to \pi}(m_{\eta^{(\prime)}_q}^{2})  
\Bigg \{ 
\delta_{pu} \alpha_2^{p}(\pi,\eta ^{(\prime)}_q)+ 2 \alpha_3^{p}(\pi,\eta ^{(\prime)}_q) \\ +
\alpha_4^{p}(\pi,\eta ^{(\prime)}_q)+  \frac12 \alpha_3^{p,ew}(\pi,\eta ^{(\prime)}_q)
- \frac12 \alpha_4^{p,ew}(\pi,\eta ^{(\prime)}_q)
\Bigg \}
+    f_\pi F_0^{B\to \eta ^{(\prime)}_q }(m_\pi^2) 
\Bigg \{ 
-\delta_{pu} \alpha_{2}^{p}(\eta ^{(\prime)}_q,\pi) \\ +\alpha_4^{p}(\eta ^{(\prime)}_q,\pi)
 -\frac12 \alpha_4^{p,ew}(\eta ^{(\prime)}_q,\pi) - \frac32 \alpha_3^{p,ew}(\eta ^{(\prime)}_q,\pi)
\Bigg \}
+ \sqrt{2} f_{\eta^{(\prime)}_s}  F_0^{B\to \pi}(m_{\eta^{(\prime)}_s}^{2}) 
\Bigg \{ 
\alpha_3^{p}(\pi,\eta ^{(\prime)}_s)  \\ -\frac12 \alpha_3^{p,ew}(\pi,\eta ^{(\prime)}_s)
\Bigg \}
+\sqrt{2} f_{\eta^{(\prime)}_c}  F_0^{B\to \pi}(m_{\eta^{(\prime)}_c}^{2}) 
\Bigg \{
\delta_{pc} \alpha_2^{p}(\pi,\eta ^{(\prime)}_c)+ \alpha_3^{p}(\pi,\eta ^{(\prime)}_c)
\Bigg \} \Bigg) \ ,   
\end{multline}
\begin{multline}
   {\cal A}( B^- \to \pi^- \eta ^{(\prime)} )   
   =
   i\frac{G_F}{2}m_B^2 \sum_{p=u}^{c} V_{pb}V_{pd}^{\ast} 
   \Bigg(  f_{\eta^{(\prime)}_q}  F_0^{B\to \pi}(m_{\eta^{(\prime)}_q}^{2})   
\Bigg \{ 
\delta_{pu} \alpha_2^{p}(\pi,\eta ^{(\prime)}_q)+ 2 \alpha_3^{p}(\pi,\eta ^{(\prime)}_q) \\ +
\alpha_4^{p}(\pi,\eta ^{(\prime)}_q)+  \frac12 \alpha_3^{p,ew}(\pi,\eta ^{(\prime)}_q) 
- \frac12 \alpha_4^{p,ew}(\pi,\eta ^{(\prime)}_q)
\Bigg \}
+  f_\pi F_0^{B\to \eta ^{(\prime)}_q }(m_\pi^2) 
\Bigg \{ 
\delta_{pu} \alpha_{1}^{p}(\eta ^{(\prime)}_q,\pi)+ \\ \alpha_4^{p}(\eta ^{(\prime)}_q,\pi)
 + \alpha_4^{p,ew}(\eta ^{(\prime)}_q,\pi)
\Bigg \}
+ \frac{2}{\sqrt{2}}  f_{\eta^{(\prime)}_s}  F_0^{B\to \pi}(m_{\eta^{(\prime)}_s}^{2}) 
\Bigg \{ 
\alpha_3^{p}(\pi,\eta ^{(\prime)}_s)- \frac12 \alpha_3^{p,ew}(\pi,\eta ^{(\prime)}_s)
\Bigg \}
\\+\frac{2}{\sqrt{2}} f_{\eta^{(\prime)}_c}  F_0^{B\to \pi}(m_{\eta^{(\prime)}_c}^{2})
\Bigg \{
\delta_{pc} \alpha_2^{p}(\pi,\eta ^{(\prime)}_c)+ \alpha_3^{p}(\pi,\eta ^{(\prime)}_c)
\Bigg \} \Bigg) \ ,
\end{multline}
\begin{multline}
   {\cal A}( \bar B^0 \to \bar K^0 \eta ^{(\prime)} )  
  = 
  i\frac{G_F}{2}  m_B^2 \sum_{p=u}^{c} V_{pb}V_{ps}^{\ast}  
\Bigg(  f_{\eta ^{(\prime)}_q} F_0^{B\to K }(m_{\eta ^{(\prime)}_q}^2)
\Bigg\{ \delta_{pu} \alpha_2^{p}(K,\eta ^{(\prime)}_q) + 2 \alpha_3^{p}(K,\eta ^{(\prime)}_q) \\
+ \frac12 \alpha_3^{p,ew}(K,\eta ^{(\prime)}_q) \Bigg\} +
 f_K   F_0^{B\to \eta ^{(\prime)}_q}(m_K^2)
\Bigg\{
 \alpha_4^{p}(\eta ^{(\prime)}_q,K)
-\frac12 \alpha_4^{p,ew}(\eta ^{(\prime)}_q,K)
\Bigg\} \\
+ \frac{2}{\sqrt{2}} f_{\eta ^{(\prime)}_s}   F_0^{B\to K}(m_{\eta ^{(\prime)}_s}^2)
   \Bigg \{ \alpha_3^{p}(K,\eta ^{(\prime)}_s)+ \alpha_4^{p}(K,\eta ^{(\prime)}_s) 
- \frac12  \alpha_3^{p,ew}(K,\eta ^{(\prime)}_s)- \frac12  \alpha_4^{p,ew}(K,\eta ^{(\prime)}_s)
\Bigg\}
 \\ +
\frac{2}{\sqrt{2}} f_{\eta ^{(\prime)}_c}  F_0^{B\to K}(m_{\eta ^{(\prime)}_c}^2)
 \Bigg \{ 
\delta_{pc} \alpha_2^{p}(K,\eta ^{(\prime)}_c)+\alpha_3^{p}(K,\eta ^{(\prime)}_c) 
\Bigg\}
 \Bigg) \ , 
\end{multline}
\begin{multline}
   {\cal A}( B^- \to K^- \eta ^{(\prime)} ) 
   = 
  i\frac{G_F}{2}  m_B^2 \sum_{p=u}^{c} V_{pb}V_{ps}^{\ast}  
\Bigg(   f_{\eta ^{(\prime)}_q} F_0^{B\to K}(m_{\eta ^{(\prime)}_q}^2)   
\Bigg\{ \delta_{pu} \alpha_2^{p}(K,\eta ^{(\prime)}_q) + 2 \alpha_3^{p}(K,\eta ^{(\prime)}_q) \\
+ \frac12 \alpha_3^{p,ew}(K,\eta ^{(\prime)}_q) \Bigg\} +
    f_K F_0^{B\to \eta ^{(\prime)}_q }(m_K^2)
\Bigg\{
\delta_{pu} \alpha_1^{p}(\eta ^{(\prime)}_q,K)+ \alpha_4^{p}(\eta ^{(\prime)}_q,K)
+ \alpha_4^{p,ew}(\eta ^{(\prime)}_q,K)
\Bigg\} \\
+ \frac{2}{\sqrt{2}}  f_{\eta ^{(\prime)}_s} F_0^{B\to K}(m_{\eta ^{(\prime)}_s}^2)
   \Bigg \{ \alpha_3^{p}(K,\eta ^{(\prime)}_s)+ \alpha_4^{p}(K,\eta ^{(\prime)}_s) 
- \frac12  \alpha_3^{p,ew}(K,\eta ^{(\prime)}_s)- \frac12  \alpha_4^{p,ew}(K,\eta ^{(\prime)}_s)
\Bigg\}
 \\ +
\frac{2}{\sqrt{2}} f_{\eta ^{(\prime)}_c} F_0^{B\to K}(m_{\eta ^{(\prime)}_c}^2)
 \Bigg \{ 
\delta_{pc} \alpha_2^{p}(K,\eta ^{(\prime)}_c)+\alpha_3^{p}(K,\eta ^{(\prime)}_c) 
\Bigg\}
 \Bigg) \ . 
\end{multline}
\end{appendix}

\newpage
%
%
\section*{Figure captions} 
%
\begin{itemize}
\item{Fig.~\ref{fig2} Graphical representation of the QCD factorization formula.}
\item{Fig.~\ref{fig3} Vertex corrections, (first row) and 
hard spectator scattering and penguin corrections (second row) at the order $\alpha_s$.}
\item{Fig.~\ref{fig4} Order $\alpha_s$ corrections to the weak annihilation.}
\item{Fig.~\ref{fig1} Inclusion of the $\rho-\omega$ mixing in $B \to \rho^0 M_1$ decay.}
\item{Fig.~\ref{fig10} Branching ratios  
$\mathcal{BR}(B^- \to \pi^- \pi^0), \mathcal{BR}(\bar{B}^0 \to \pi^+ \pi^-),  
\mathcal{BR}(\bar{B}^0 \to \pi^0 \pi^0), \linebreak \mathcal{BR}(\bar{B}^0 \to K^- K^+), 
\mathcal{BR}(\bar{B}^0 \to \bar{K}^0 K^0), \mathcal{BR}(B^- \to K^- K^0)$, as a function of 
CKM matrix elements and form factors $F^{B \to \pi}$ or $F^{B \to K}$.}
\item{Fig.~\ref{fig11} Branching ratios  
$\mathcal{BR}(B^- \to \pi^0 K^-), \mathcal{BR}(\bar{B}^0 \to \pi^0 \bar{K}^0), 
\mathcal{BR}(B^- \to \pi^- \bar{K}^{0}), \linebreak \mathcal{BR}(\bar{B}^0 \to \pi^+ K^-), 
\mathcal{BR}(\bar{B}^0 \to \pi^0 \eta), \mathcal{BR}(\bar{B}^0 \to \pi^0 \eta^{\prime})$, 
as a function of CKM matrix elements and form factors $F^{B \to \pi}$ or $F^{B \to K}$.}
\item{Fig.~\ref{fig12} Branching ratios  
$\mathcal{BR}(B^- \to \pi^- \eta), \mathcal{BR}(B^- \to \pi^-  \eta^{\prime}), 
\mathcal{BR}(\bar{B}^0 \to  \bar{K}^0 \eta), \linebreak \mathcal{BR}(\bar{B}^0 \to \bar{K}^{0} \eta^{\prime}), 
\mathcal{BR}(B^- \to K^- \eta), \mathcal{BR}(B^-  \to K^- \eta^{\prime})$, as a function of 
CKM matrix elements and form factors $F^{B \to \pi}$ or $F^{B \to K}$.}
\item{Fig.~\ref{fig13} Branching ratios  
$\mathcal{BR}(\bar{B}^0 \to \pi^- \rho^+), \mathcal{BR}(\bar{B}^0 \to \pi^+ \rho^-), 
\mathcal{BR}(\bar{B}^0 \to \pi^0 \rho^0), \linebreak \mathcal{BR}(B^- \to \pi^- \rho^0), 
\mathcal{BR}(B^- \to \pi^0 \rho^-), \mathcal{BR}(\bar{B}^0 \to \rho^{\pm} \pi^{\mp})$,  
as a function of CKM matrix elements and form factor $F^{B \to \pi}$. }
\item{Fig.~\ref{fig16} Branching ratios  $\mathcal{BR}(\bar{B}^0 \to \pi^0 \omega),   
\mathcal{BR}(B^- \to \pi^- \omega), \mathcal{BR}(\bar{B}^0 \to \rho^+ K^-), 
\linebreak  \mathcal{BR}(\bar{B}^0 \to \rho^0 \bar{K}^0), 
\mathcal{BR}(B^- \to \rho^0 K^-), \mathcal{BR}(B^- \to \rho^- \bar{K}^0)$,  as a function of 
CKM matrix elements and form factors $F^{B \to \pi}$ or $F^{B \to K}$.}
\item{Fig.~\ref{fig17} Branching ratios  $\mathcal{BR}(B^- \to K^- \omega), \mathcal{BR}(\bar{B}^0 \to \bar{K}^0 \omega), 
\mathcal{BR}(\bar{B}^0 \to \pi^+ K^{\ast -}), \linebreak \mathcal{BR}(\bar{B}^0 \to \pi^0 \bar{K}^{\ast 0}),
 \mathcal{BR}(B^- \to \pi^- \bar{K}^{\ast 0}), \mathcal{BR}(B^- \to \pi^0  K^{\ast -})$, as a function of 
CKM matrix elements and form factors $F^{B \to \pi}$ or  $F^{B \to K}$.}
\item{Fig.~\ref{fig18} Branching ratios  $\mathcal{BR}(\bar{B}^0 \to K^{-} K^{\ast +}), \mathcal{BR}(B^- \to K^-  K^{\ast 0}),
\mathcal{BR}(\bar{B}^0 \to \linebreak \bar{K}^0 K^{\ast 0}),  \mathcal{BR}(B^- \to K^{\ast -} K^{0}), 
\mathcal{BR}(\bar{B}^{0} \to K^{\ast -} K^+), \mathcal{BR}(\bar{B}^{0} \to \bar{K}^{\ast 0} K^0)$,  
as a function of CKM matrix elements and form factor  $F^{B \to K}$.}
\item{Fig.~\ref{fig19}  Branching ratios  $\mathcal{BR}(B^- \to K^- \phi), \mathcal{BR}(\bar{B}^0 \to \bar{K}^0 \phi), 
\mathcal{BR}(B^- \to \pi^- \phi), \linebreak \mathcal{BR}(\bar{B}^0 \to \pi^0 \phi), 
\mathcal{BR}(B^- \to \pi^- \pi^0)/\mathcal{BR}(\bar{B}^0 \to \pi^0 \pi^0)  , 
\mathcal{BR}(\bar{B}^0 \to \pi^+ \pi^-)/ 2\mathcal{BR}(B^- \to \pi^- \pi^0)$,  as a function of 
CKM matrix elements and form factors $F^{B \to \pi}$ or  $F^{B \to K}$.}
\item{Fig.~\ref{fig20}  Branching ratios for 
$\mathcal{BR}(\bar{B}^0 \to \pi^+ K^- )/\mathcal{BR}(\bar{B}^0 \to \pi^0 \bar{K}^0), 
\mathcal{BR}(\bar{B}^0 \to \pi^+ K^- ) \linebreak  /\mathcal{BR}(B^- \to \pi^- \bar{K}^0), 
2 \mathcal{BR}(B^- \to \pi^0 K^-)/\mathcal{BR}(B^- \to \pi^- \bar{K}^0), 
\mathcal{BR}(B^- \to \pi^- \bar{K}^{\ast 0})/\linebreak \mathcal{BR}(\bar{B}^0 \to \pi^+ K^{\ast -}), 
\mathcal{BR}(\bar{B}^0 \to \rho^{\pm} \pi^{\mp})/\mathcal{BR}(B^- \to \rho^0 \pi^-), 
\mathcal{BR}(B^- \to K^- \eta^{\prime})/ \linebreak \mathcal{BR}(\bar{B}^0 \to \bar{K}^0 \eta^{\prime}),$
as a function of CKM matrix elements and form factors  $F^{B \to \pi}$ or  $F^{B \to K}$. }
\item{Fig.~\ref{fig14} $CP$ violating asymmetry, $a_{CP}$,  as a function of $\sqrt{S}$, for  $B^{-} \to \pi^{+} 
\pi^{-} K^{-}, \linebreak \bar{B}^{0} \to \pi^{+} \pi^{-} \bar{K}^0$ for limiting values of the CKM matrix elements and for different 
values of the form factor $F_{1}^{B \to K}(m_{\rho}^{2})$.   $CP$ violating asymmetry, $a_{CP}$,  as a function of $\sqrt{S}$, 
for $B^{-} \to \pi^{+} \pi^{-} \pi^-, \bar{B}^{0} \to \pi^{+} \pi^{-} \pi^0$ for limiting values of the CKM matrix elements and for 
different values of the form factor $F_{1}^{B \to \pi}$. The ratio of penguin to tree amplitudes, 
$r_i$, as a function of $\sqrt{S}$, for  $B^{-} \to \pi^{+} 
\pi^{-} K^{-}, \bar{B}^{0} \to \pi^{+} \pi^{-} \bar{K}^0$ for limiting values of the CKM matrix elements and for different 
values of the form factor $F_{1}^{B \to K}$. }
\item{Fig.~\ref{fig15} The ratio of 
penguin to tree amplitudes, $r_i$, as a function of $\sqrt{S}$ for  $B^{-} \to \pi^{+} 
\pi^{-} \pi^{-}, \bar{B}^{0} \to \pi^{+} \pi^{-} \pi^0$ for limiting values of the CKM matrix elements and for different 
values of the form factor $F_{1}^{B \to K}$.  $\sin \delta_{ij}$, as a function of  $\sqrt{S}$ for  $B^{-} \to \pi^{+} 
\pi^{-} K^{-}, \bar{B}^{0} \to \pi^{+} \pi^{-} \bar{K}^0$ for limiting values of the CKM matrix elements and for different 
values of the form factor $F_{1}^{B \to K}$.  $\sin \delta_{ij}$, 
as a function of  $\sqrt{S}$ for  $B^{-} \to \pi^{+} 
\pi^{-} \pi^{-}, \bar{B}^{0} \to \pi^{+} \pi^{-} \pi^0$ for limiting values of the CKM matrix elements and for different 
values of the form factor $F_{1}^{B \to \pi}$. }
\end{itemize}
\newpage
%
\section*{Table  captions}
%
%
\begin{itemize}
\item{Tab.~\ref{tab1} Upper table: Wilson coefficients $C_i$ in the NDR scheme. 
Input parameters are $\Lambda^{(5)}_{\overline{\rm
MS}}=0.225$\,$\mathrm{GeV}$,  $m_t(m_t)=167$\,$\mathrm{GeV}$, $m_b(m_b)=4.2$\,
$\mathrm{GeV}$, $M_W=80.4$\, $\mathrm{GeV}$, $\alpha=1/129$, and $\sin^2\!
\theta_W=0.23$. Lower table: Wilson coefficients $C_i$ in naive factorization.}
\item{Tab.~\ref{tab2} Experimental branching ratio data from the BELLE, CLEO and BABAR 
$B$-factories for the $B \to K  X$ channel where $X$ stands for $\eta^{(\prime)}, 
\omega, \phi, \rho$ and $K^{(*)}$.}
\item{Tab.~\ref{tab3} Experimental branching ratio data from the BELLE, CLEO and BABAR 
$B$-factories for the $B \to \pi  X$ channel where $X$ stands for $\eta^{(\prime)}, \omega, 
\phi, \rho, \pi$ and $K^{*}$.}
\item{Tab.~\ref{tab4} Experimental branching ratio data from the BELLE, CLEO and BABAR 
$B$-factories for the $B \to \pi K$ channel.}
\item{Tab.~5 Experimental data from the BELLE, CLEO and BABAR $B$-factories 
for the ratios between branching ratios involving pions (first case), pion and kaon 
(second case), kaon and $\phi$ or $\eta$ (third case) and kaon $\omega$ or  pion 
$\rho$ (fourth case).}
\item{Tab.~\ref{tab6} Phases $\varphi_{A,H}^{M_i}$ and parameters $\varrho_{A,H}^{M_i}$ for the 
annihilation and hard-spectator scattering contributions, respectively, for $K, K^*, 
\pi, \rho, \omega, \eta^{(\prime)}, \phi$ and  determined for  the $B \to K X$ 
and $B \to \pi X$  channels.}
\item{Tab.~7 Theoretical branching ratios for the $B \to \pi  X$ and $B \to K  X$ channels
where $X$ stands for $\eta^{(\prime)}, \omega, \phi, \rho, \pi$ and $K^{*}$. Values are given for naive
factorization, QCD factorization, and annihilation and hard scattering spectator contributions.}
\item{Tab.~\ref{tab8} Theoretical branching ratios for the $B \to K  X$ channel and  ratios between
$B \to K  X$ and $B \to \pi  X$ channels. Values are given for naive
factorization, QCD factorization, and annihilation and hard scattering spectator contributions.}
\end{itemize}
\newpage
%
\newpage
%
%
\begin{figure}
\includegraphics*[width=0.495\columnwidth]{QCDF_frame_BR_pmet.eps}
\includegraphics*[width=0.495\columnwidth]{QCDF_frame_BR_pmetp.eps}
\includegraphics*[width=0.495\columnwidth]{QCDF_frame_BR_kzbet.eps}
\includegraphics*[width=0.499\columnwidth]{QCDF_frame_BR_kzbetp.eps}
\includegraphics*[width=0.495\columnwidth]{QCDF_frame_BR_kmet.eps}
\includegraphics*[height=5.75cm,width=7.85cm,clip=true]{QCDF_frame_BR_kmetp.eps}
\caption{From top left-handed to bottom right-handed, branching ratios  
$\mathcal{BR}(B^- \to \pi^- \eta), \mathcal{BR}(B^- \to \pi^-  \eta^{\prime}), 
\mathcal{BR}(\bar{B}^0 \to  \bar{K}^0 \eta), \mathcal{BR}(\bar{B}^0 \to \bar{K}^{0} \eta^{\prime}), 
\mathcal{BR}(B^- \to K^- \eta), \mathcal{BR}(B^-  \to K^- \eta^{\prime})$, as a function of 
CKM matrix elements and form factors $F^{B \to \pi}$ or $F^{B \to K}$. Solid line(dotted line) for 
QCDF(NF) factorization and  for limiting values of the  CKM matrix element parameters
shown  respectively at 68\% (thin line)  and 95\% (thick line) of confidence level. 
Notation: horizontal dotted lines: CLEO data; horizontal dashed lines:
BABAR data; horizontal dot-dashed lines: BELLE data.}
\label{fig12}
\end{figure}
\begin{figure}
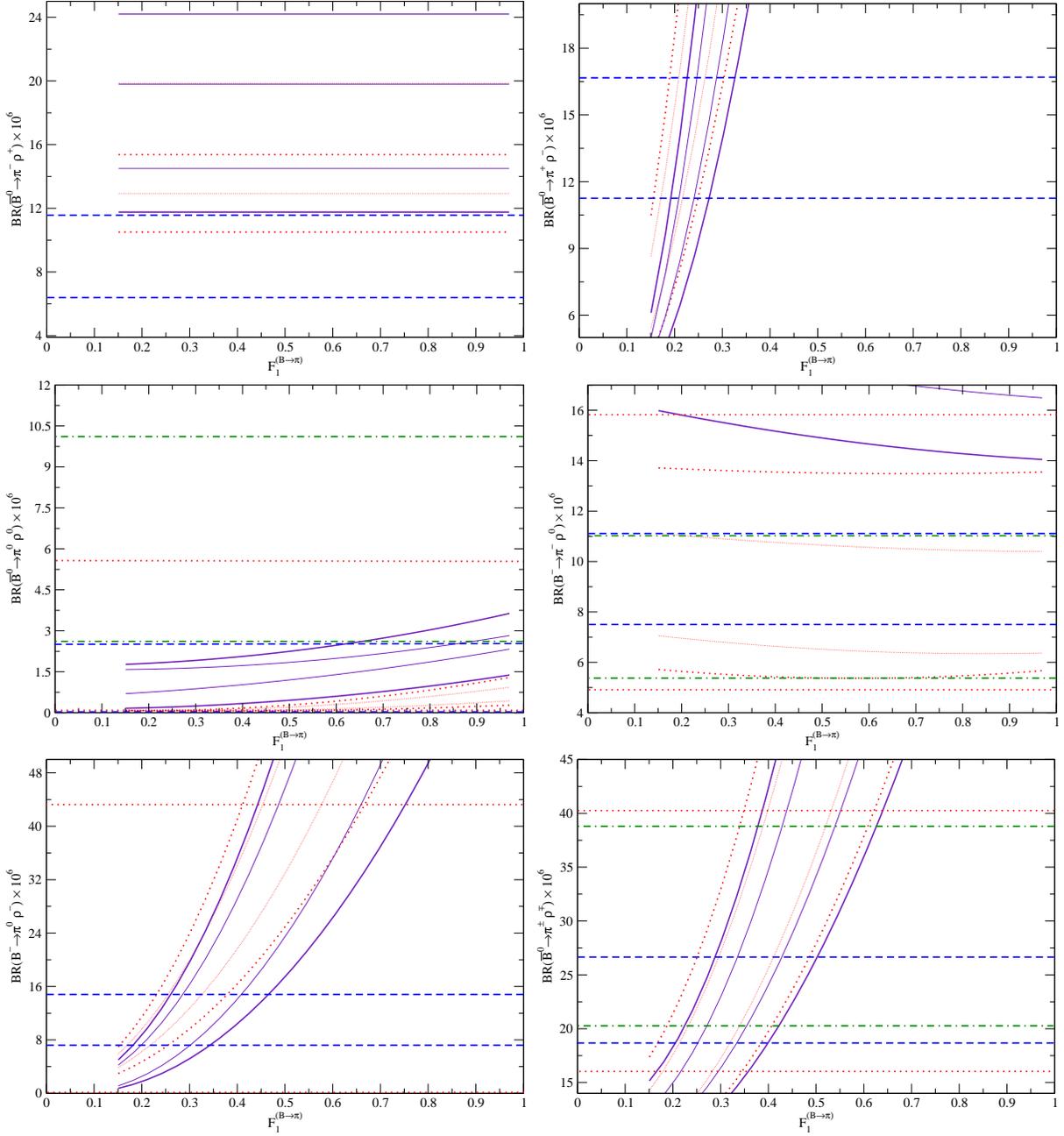

\includegraphics*[width=0.495\columnwidth]{QCDF_frame_BR_pmrp.eps}
\includegraphics*[width=0.495\columnwidth]{QCDF_frame_BR_pprm.eps}
\includegraphics*[width=0.495\columnwidth]{QCDF_frame_BR_pzrz.eps}
\includegraphics*[width=0.486\columnwidth]{QCDF_frame_BR_pmrz.eps}
\includegraphics*[width=0.495\columnwidth]{QCDF_frame_BR_pzrm.eps}
\includegraphics*[width=0.495\columnwidth]{QCDF_frame_BR_ppmrmp.eps}
\caption{From top left-handed to bottom right-handed, branching ratios  
$\mathcal{BR}(\bar{B}^0 \to \pi^- \rho^+), \mathcal{BR}(\bar{B}^0 \to \pi^+ \rho^-), 
\mathcal{BR}(\bar{B}^0 \to \pi^0 \rho^0), \mathcal{BR}(B^- \to \pi^- \rho^0), 
\mathcal{BR}(B^- \to \pi^0 \rho^-), $ $ \mathcal{BR}(\bar{B}^0 \to \rho^{\pm} \pi^{\mp})$,  
as a function of CKM matrix elements and form factors $F^{B \to \pi}$ or $F^{B \to K}$. Solid 
line(dotted line) for QCDF(NF) factorization and  for limiting values of the  
CKM matrix element parameters shown  respectively at 68\% (thin line) and 95\% (thick line) 
of confidence level. Notation: horizontal dotted lines: CLEO data;  horizontal dashed lines:
BABAR data;  horizontal dot-dashed lines: BELLE data.}
\label{fig13}
\end{figure}

\newpage
\begin{figure}
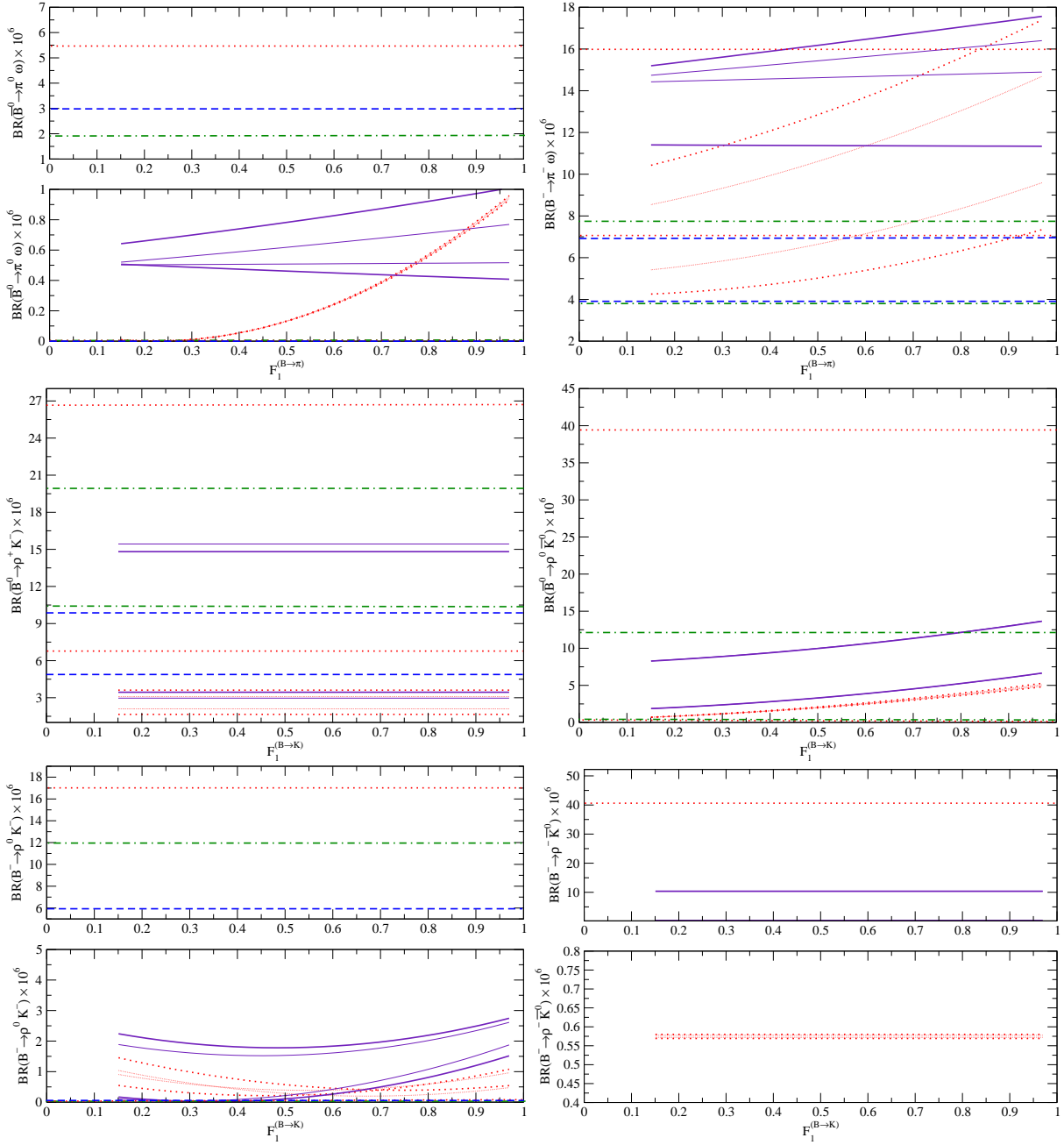

\includegraphics*[width=0.495\columnwidth]{QCDF_frame_BR_pzom.eps}
\includegraphics*[width=0.495\columnwidth]{QCDF_frame_BR_pmom.eps}
\includegraphics*[width=0.495\columnwidth]{QCDF_frame_BR_rpkm.eps}
\includegraphics*[width=0.495\columnwidth]{QCDF_frame_BR_rzkzb.eps}
\includegraphics*[width=0.495\columnwidth]{QCDF_frame_BR_rzkm.eps}
\includegraphics*[width=0.500\columnwidth]{QCDF_frame_BR_rmkzb.eps}
\caption{From top left-handed to bottom right-handed, 
branching ratios  $ \mathcal{BR}(\bar{B}^0 \to \pi^0 \omega),  \mathcal{BR}(B^- \to \pi^- \omega), 
\mathcal{BR}(\bar{B}^0 \to \rho^+ K^-),  \mathcal{BR}(\bar{B}^0 \to \rho^0 \bar{K}^0), 
\mathcal{BR}(B^- \to \rho^0 K^-), $ $ \mathcal{BR}(B^- \to \rho^- \bar{K}^0)$,  as a function of 
CKM matrix elements and form factors $F^{B \to \pi}$ or $F^{B \to K}$. Solid line(dotted line) for 
QCDF(NF) factorization and  for limiting values of the  CKM matrix element parameters
shown  respectively at 68\%  (thin line) and 95\% (thick line) of confidence level. 
Notation: horizontal dotted lines: CLEO data; horizontal  dashed lines:
BABAR data; horizontal  dot-dashed lines: BELLE data.
}
\label{fig16}
\end{figure}
\newpage
\begin{figure}
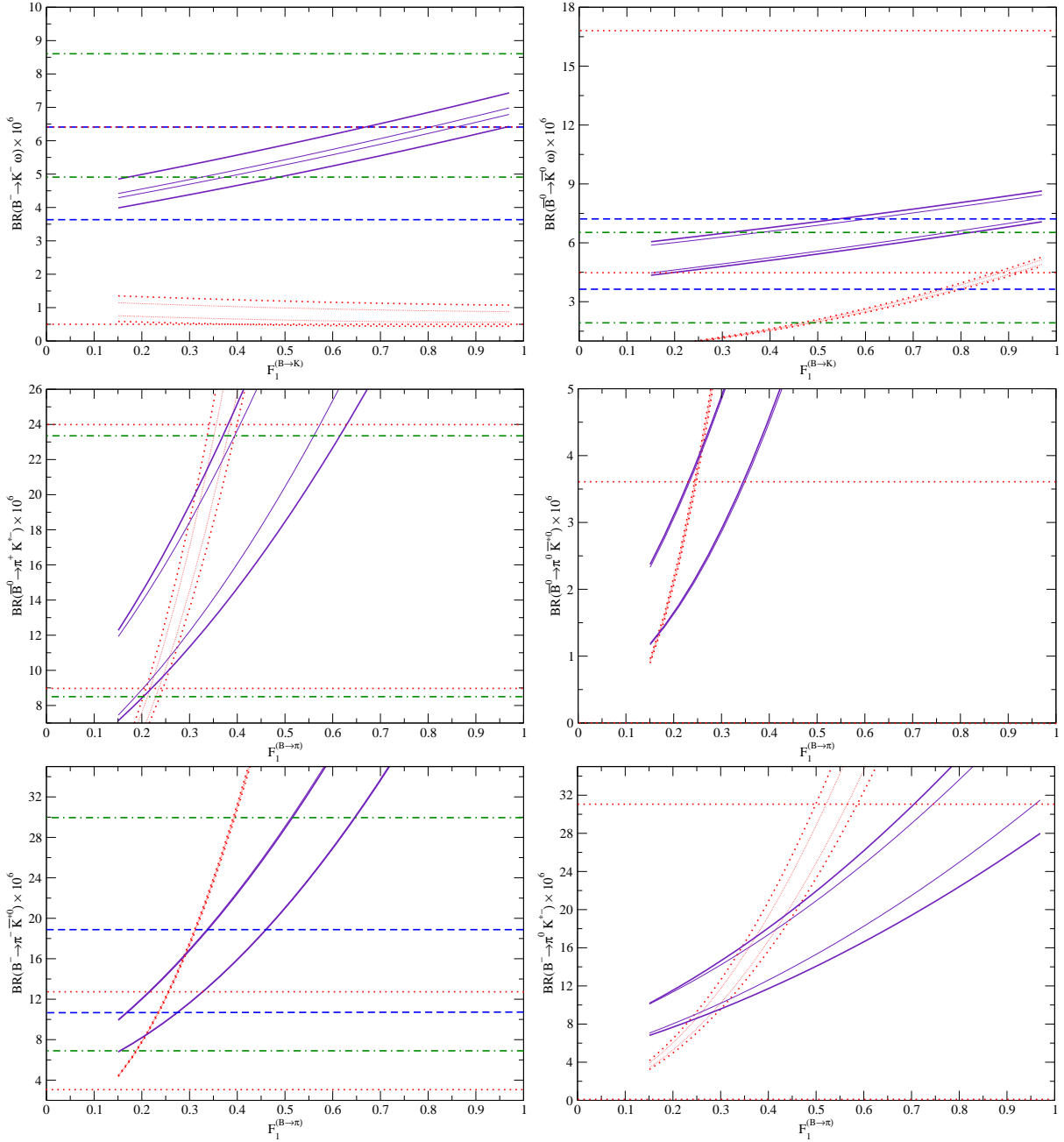

\includegraphics*[width=0.495\columnwidth]{QCDF_frame_BR_kmom.eps}
\includegraphics*[width=0.495\columnwidth]{QCDF_frame_BR_kzbom.eps}
\includegraphics*[width=0.495\columnwidth]{QCDF_frame_BR_ppKsm.eps}
\includegraphics*[width=0.489\columnwidth]{QCDF_frame_BR_pzKszb.eps}
\includegraphics*[width=0.495\columnwidth]{QCDF_frame_BR_pmKszb.eps}
\includegraphics*[width=0.495\columnwidth]{QCDF_frame_BR_pzKsm.eps}
\caption{From top left-handed to bottom right-handed, 
branching ratios  $\mathcal{BR}(B^- \to K^- \omega), \mathcal{BR}(\bar{B}^0 \to \bar{K}^0 \omega), 
\mathcal{BR}(\bar{B}^0 \to \pi^+ K^{\ast -}), \mathcal{BR}(\bar{B}^0 \to \pi^0 \bar{K}^{\ast 0}),
 \mathcal{BR}(B^- \to \pi^- \bar{K}^{\ast 0}), $ $ \mathcal{BR}(B^- \to \pi^0  K^{\ast -})$, as a function of 
CKM matrix elements and form factors $F^{B \to \pi}$ or  $F^{B \to K}$. Solid line(dotted line) for 
QCDF(NF) factorization and  for limiting values of the  CKM matrix element parameters
shown  respectively at 68\%  (thin line)  and 95\%  (thick line) of confidence level. 
Notation: horizontal dotted lines: CLEO data;  horizontal dashed lines:
BABAR data;  horizontal dot-dashed lines: BELLE data. }
\label{fig17}
\end{figure}
\begin{figure}
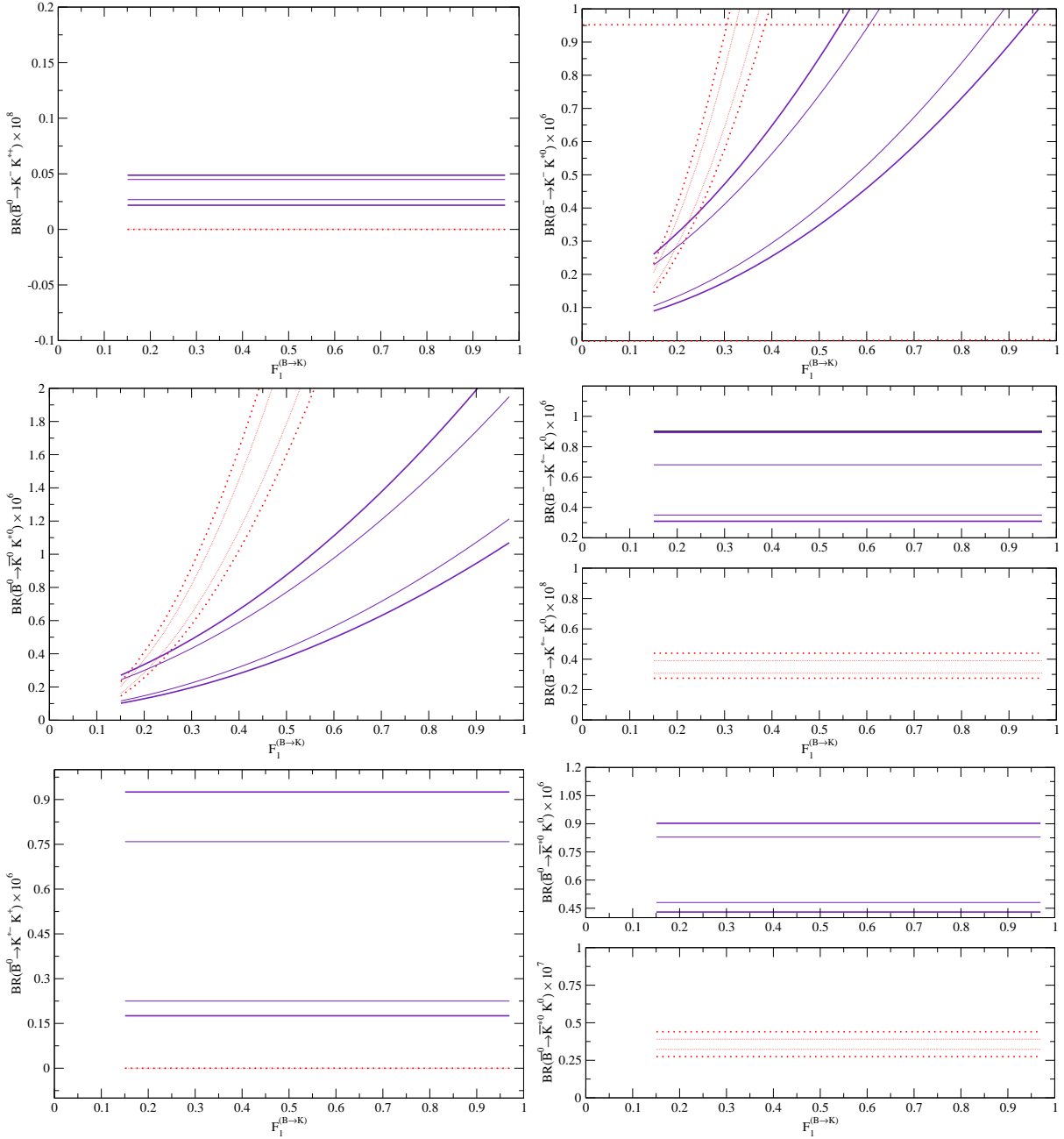

\includegraphics*[height=5.75cm,width=7.85cm,clip=true]{QCDF_frame_BR_kmksp.eps}
\includegraphics*[width=0.495\columnwidth]{QCDF_frame_BR_kmksz.eps}
\includegraphics*[width=0.495\columnwidth]{QCDF_frame_BR_kzbksz.eps}
\includegraphics*[width=0.495\columnwidth]{QCDF_frame_BR_kmskz.eps}
\includegraphics*[width=0.495\columnwidth]{QCDF_frame_BR_kmskp.eps}
\includegraphics*[width=0.495\columnwidth]{QCDF_frame_BR_kszbkz.eps}
\caption{From top left-handed to bottom right-handed, 
branching ratios  $\mathcal{BR}(\bar{B}^0 \to K^{-} K^{\ast +}), \mathcal{BR}(B^- \to K^-  K^{\ast 0}),
\mathcal{BR}(\bar{B}^0 \to \bar{K}^0 K^{\ast 0}), \mathcal{BR}(B^- \to K^{\ast -} K^{0}), 
\mathcal{BR}(\bar{B}^{0} \to K^{\ast -} K^+),  \mathcal{BR}(\bar{B}^{0} \to \bar{K}^{\ast 0} K^0)$,  
as a function of CKM matrix elements and form factors $F^{B \to \pi}$  or $F^{B \to K}$. Solid line(dotted line) for 
QCDF(NF) factorization and  for limiting values of the  CKM matrix element parameters
shown  respectively at 68\%  (thin line) and 95\%  (thick line) of confidence level. 
Notation: horizontal dotted lines: CLEO data;  horizontal dashed lines:
BABAR data;  horizontal dot-dashed lines: BELLE data.
}
\label{fig18}
\end{figure}
\begin{figure}
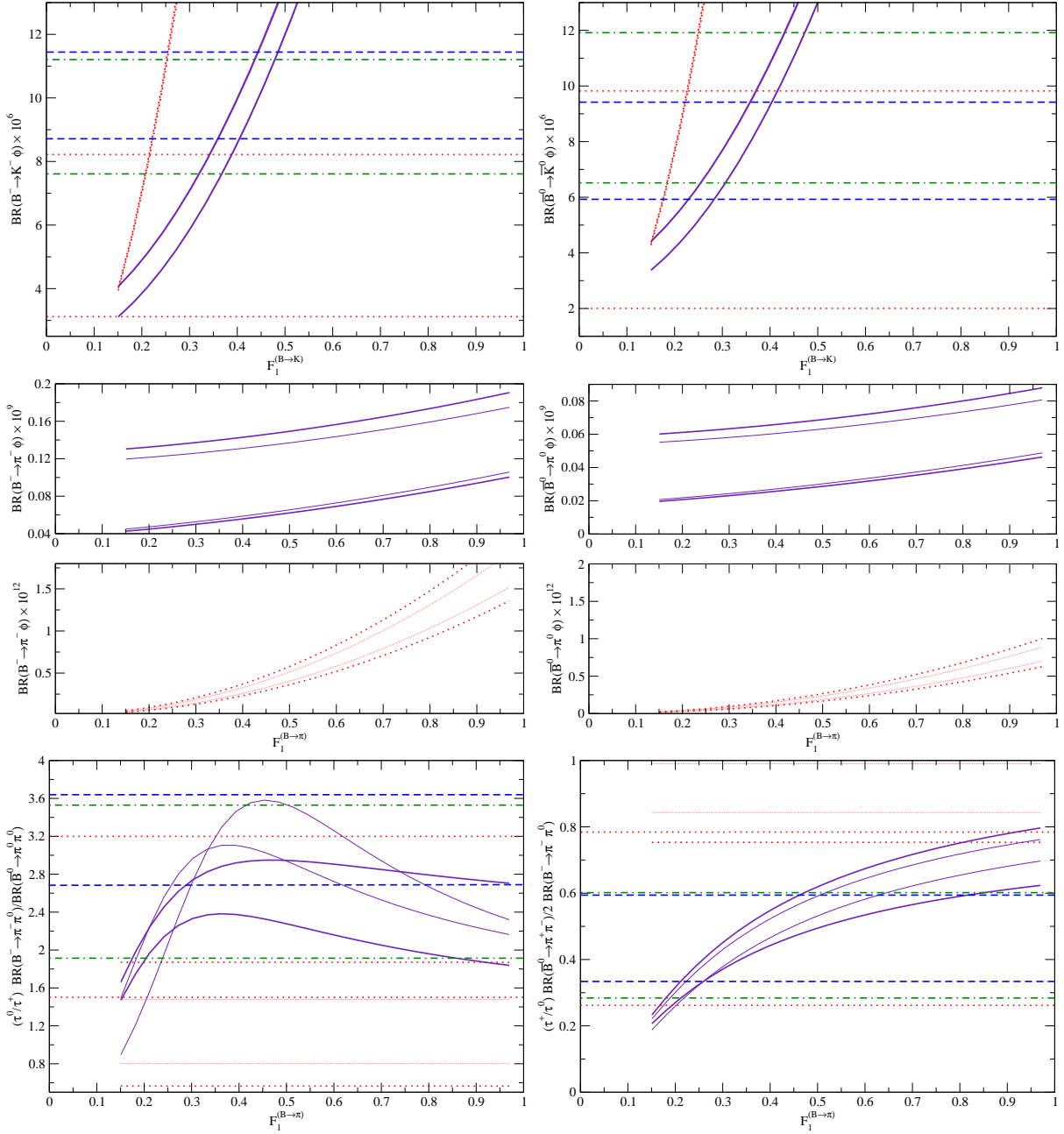

\includegraphics*[width=0.495\columnwidth]{QCDF_frame_BR_kmph.eps}
\includegraphics*[width=0.495\columnwidth]{QCDF_frame_BR_kzbph.eps}
\includegraphics*[width=0.495\columnwidth]{QCDF_frame_BR_pmph.eps}
\includegraphics*[width=0.495\columnwidth]{QCDF_frame_BR_pzph.eps}
\includegraphics*[width=0.495\columnwidth]{QCDF_frame_BR_ratio_pmpz_pzpz.eps}
\includegraphics*[width=0.495\columnwidth]{QCDF_frame_BR_ratio_pppm_pppz.eps}
\caption{From top left-handed to bottom right-handed, 
branching ratios  $\mathcal{BR}(B^- \to K^- \phi), \mathcal{BR}(\bar{B}^0 \to \bar{K}^0 \phi), 
\mathcal{BR}(B^- \to \pi^- \phi), \mathcal{BR}(\bar{B}^0 \to \pi^0 \phi), 
\mathcal{BR}(B^- \to \pi^- \pi^0)/\mathcal{BR}(\bar{B}^0 \to \pi^0 \pi^0)  , 
\mathcal{BR}(\bar{B}^0 \to \pi^+ \pi^-)/ 2\mathcal{BR}(B^- \to \pi^- \pi^0)$,  as a function of 
CKM matrix elements and form factors $F^{B \to \pi}$ or  $F^{B \to K}$. Solid line(dotted line) for 
QCDF(NF) factorization and  for limiting values of the  CKM matrix element parameters
shown  respectively at 68\% (thin line) and 95\%  (thick line) of confidence level. 
Notation: horizontal dotted lines: CLEO data;  horizontal dashed lines:
BABAR data;  horizontal dot-dashed lines: BELLE data.
}
\label{fig19}
\end{figure}

\begin{figure}
\includegraphics*[width=0.495\columnwidth]{QCDF_frame_BR_ratio_pmkp_pzkz.eps}
\includegraphics*[width=0.495\columnwidth]{QCDF_frame_BR_ratio_ppkm_pmkz.eps}
\includegraphics*[width=0.495\columnwidth]{QCDF_frame_BR_ratio_pzkm_pmkz.eps}
\includegraphics*[width=0.495\columnwidth]{QCDF_frame_BR_ratio_pmKszb_ppKsm.eps}
\includegraphics*[width=0.495\columnwidth]{QCDF_frame_BR_ratio_pprm_pmrz.eps}
\includegraphics*[width=0.500\columnwidth]{QCDF_frame_BR_ratio_kmetap_kzetap.eps}
\caption{From top left-handed to bottom right-handed, branching ratios for 
$\mathcal{BR}(\bar{B}^0 \to \pi^+ K^- )/\mathcal{BR}(\bar{B}^0 \to \pi^0 \bar{K}^0), 
\mathcal{BR}(\bar{B}^0 \to \pi^+ K^- )/\mathcal{BR}(B^- \to \pi^- \bar{K}^0), 
2 \mathcal{BR}(B^- \to \pi^0 K^-)/\mathcal{BR}(B^- \to \pi^- \bar{K}^0), 
\mathcal{BR}(B^- \to \pi^- \bar{K}^{\ast 0})/\mathcal{BR}(\bar{B}^0 \to \pi^+ K^{\ast -}), 
\mathcal{BR}(\bar{B}^0 \to \rho^{\pm} \pi^{\mp})/ $ $ \mathcal{BR}(B^- \to \rho^0 \pi^-), 
\mathcal{BR}(B^- \to K^- \eta^{\prime})/\mathcal{BR}(\bar{B}^0 \to \bar{K}^0 \eta^{\prime}),$
as a function of CKM matrix elements and form factors  $F^{B \to \pi}$ or  $F^{B \to K}$. 
Solid line(dotted line) for QCDF(NF) factorization and  for limiting values of the  
CKM matrix element parameters shown  respectively at 68\% (thin line) and 95\% (thick line) 
of confidence level. Notation: horizontal dotted lines: CLEO data; horizontal dashed lines:
BABAR data; horizontal dot-dashed lines: BELLE data.
}
\label{fig20}
\end{figure}
\begin{figure}
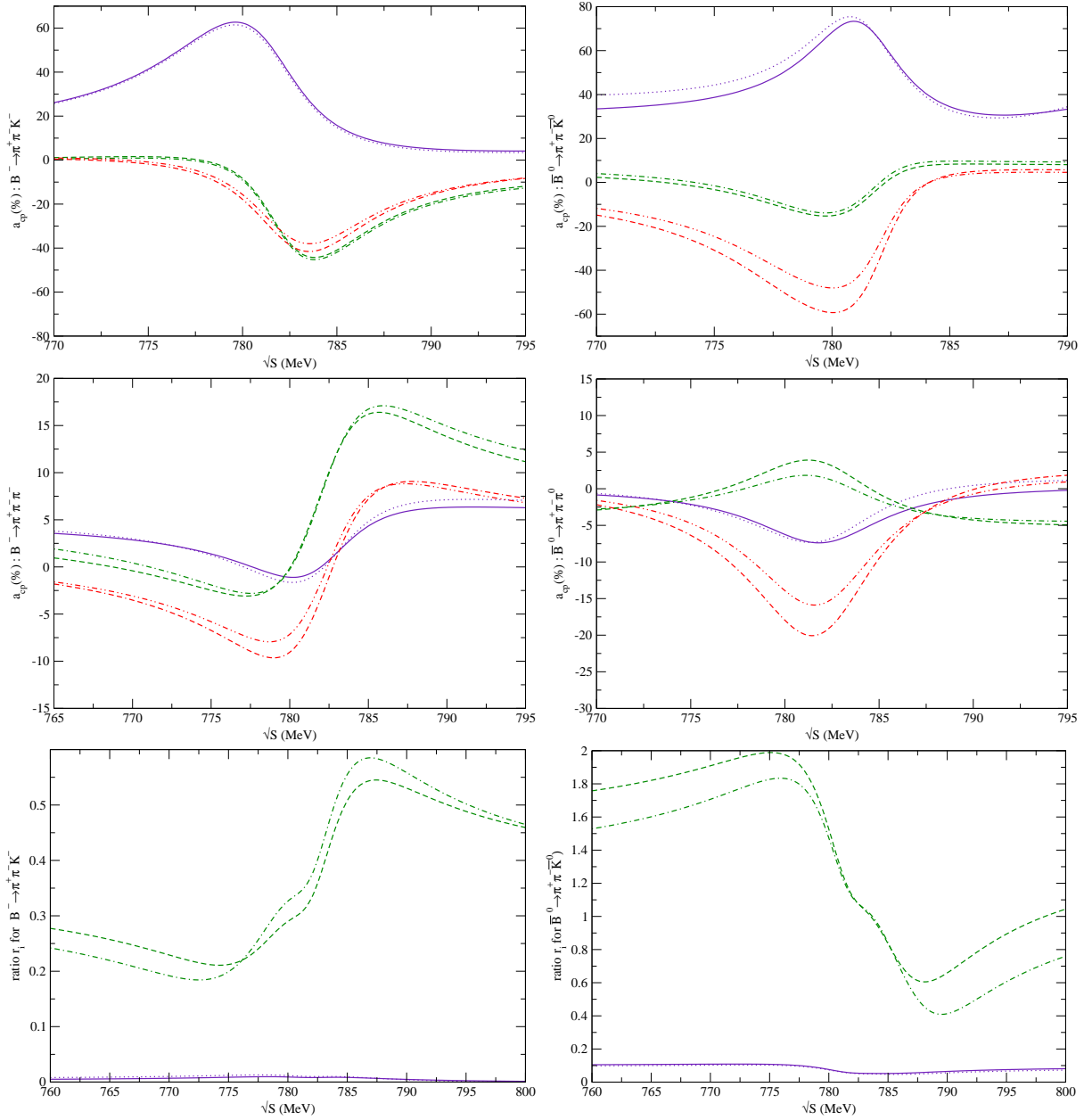

\includegraphics*[width=0.495\columnwidth]{ACP_QCDF_PIPIKM.eps}
\includegraphics*[width=0.495\columnwidth]{ACP_QCDF_PIPIKZ.eps}
\includegraphics*[width=0.495\columnwidth]{ACP_QCDF_PIPIPIM.eps}
\includegraphics*[width=0.495\columnwidth]{ACP_QCDF_PIPIPIZ.eps}
\includegraphics*[width=0.495\columnwidth]{RATIO_QCDF_PIPIKM.eps}
\includegraphics*[width=0.495\columnwidth]{RATIO_QCDF_PIPIKZ.eps}
\caption{First row, $CP$ violating asymmetry, $a_{CP}$, for $B^- \to \pi^+ \pi^- K^-,
\bar{B}^0 \to \pi^+ \pi^- \bar{K}^0$  for max CKM matrix elements. Solid line (dotted line)
for QCDF, dot-dot-dashed line (dot-dash-dashed line) for NF, dot-dashed line (dashed line)
for QCDF with default values and for $F^{B \to K}=0.35(0.42)$. Second row, $CP$ violating
asymmetry, $a_{CP}$, for $B^- \to \pi^+ \pi^- \pi^-, \bar{B}^0 \to \pi^+ \pi^- \pi^0$, for 
max CKM matrix elements. Same notation for lines as in first row with $F^{B \to \pi}=0.27(0.35)$.
Last row, the ratio of penguin to tree amplitudes, $r_i$, for $B \to \pi \pi K$, for max
CKM matrix elements. Solid line (dotted line) for $r_1=p^u/t^u$, dashed line (dot-dashed line)
 for $r_2=p^c/t^u$ and for $F^{B \to K}=0.35(0.42)$. All the figures are given as a function of
$\sqrt{S}$.}
\label{fig14}
\end{figure}
\newpage
\begin{figure}
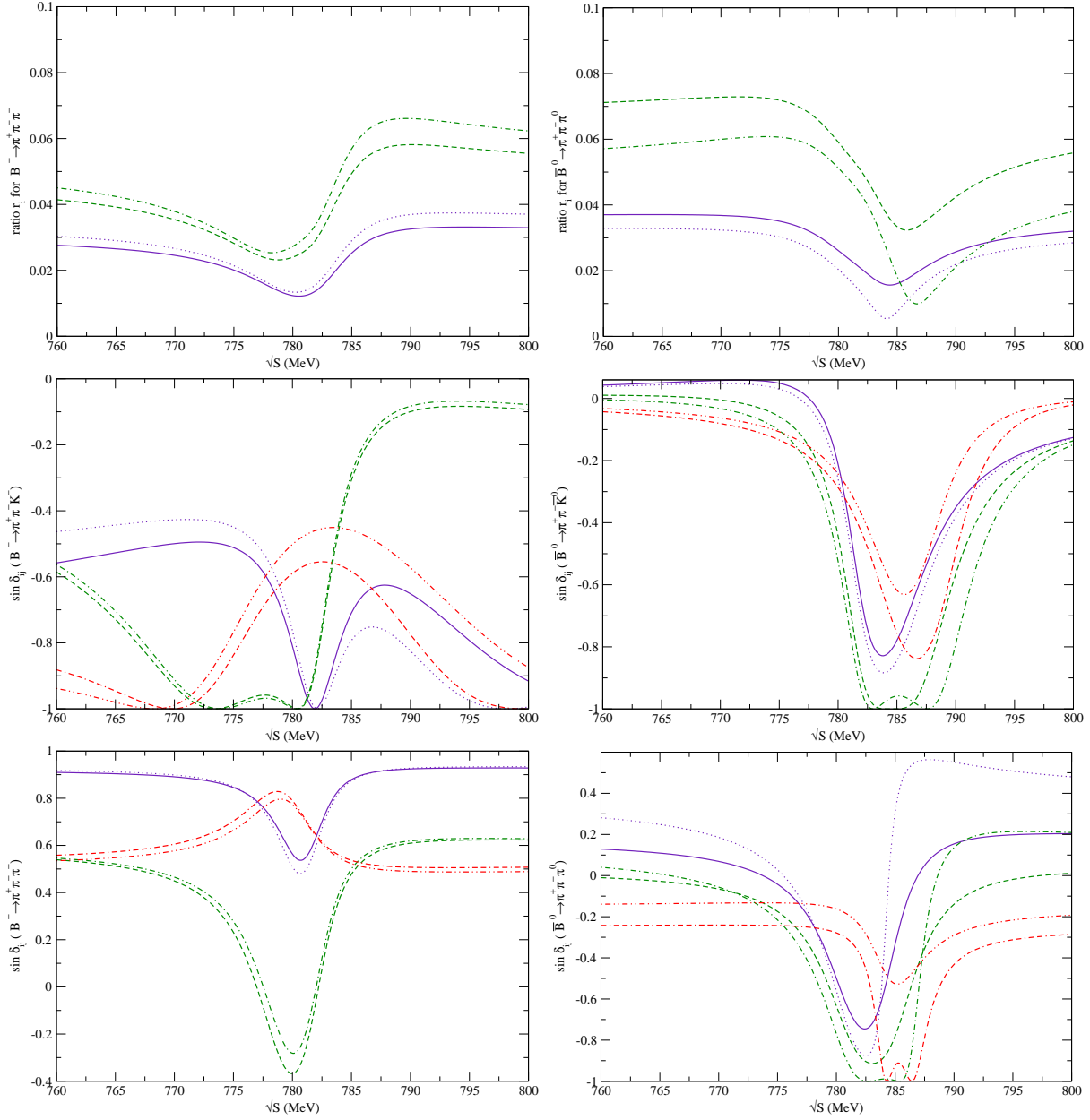

\includegraphics*[width=0.495\columnwidth]{RATIO_QCDF_PIPIPIM.eps}
\includegraphics*[width=0.495\columnwidth]{RATIO_QCDF_PIPIPIZ.eps}
\includegraphics*[width=0.495\columnwidth]{DELTA_QCDF_PIPIKM.eps}
\includegraphics*[width=0.495\columnwidth]{DELTA_QCDF_PIPIKZ.eps}
\includegraphics*[width=0.495\columnwidth]{DELTA_QCDF_PIPIPIM.eps}
\includegraphics*[width=0.495\columnwidth]{DELTA_QCDF_PIPIPIZ.eps}
\caption{First row, the ratio of penguin to tree amplitudes, $r_i$, for $B \to \pi \pi \pi$, for max
CKM matrix elements. Solid line (dotted line) for $r_1=p^u/t^u$, dashed line (dot-dashed line)
 for $r_2=p^c/t^u$ and for $F^{B \to \pi}=0.27(0.35)$.  Second and last row,
solid line (dotted line) for $\sin \delta_{12}$, dot-dot-dashed line (dot-dash-dashed line)
for  $\sin \delta_{23}$, and dashed line (dot-dashed line) for  $\sin \delta_{13}$ for 
$F^{B \to K}=0.35(0.42)$ or  $F^{B \to \pi}=0.27(0.35)$ according to the given decay. All 
the figures are given as a function of $\sqrt{S}$.}
\label{fig15}
\end{figure}
\clearpage
%
%
{\renewcommand\baselinestretch{0.97}
\begin{table}[hpt]
\begin{center}
\begin{tabular}{|ccccccc|} \hline \hline
     NLO      &  $C_{1}$   &  $C_{2}$  & $C_{3}$  &  $C_{4}$ & $C_{5}$ & $C_{6}$   \\
\hline
\hline
\hline
$\mu= m_{b}$ &  1.081      &  -0.190    & 0.014    &  -0.036  &  0.009  & -0.0042   \\
\hline
   NLO       &  $C_{7}/\alpha$   &  $C_{8}/\alpha$  & $C_{9}/\alpha$  &  $C_{10}/\alpha$  &         &  \\
\hline
$\mu= m_{b}$ & -0.011            & 0.060            &  -1.254        &   0.223           &         &   \\
\hline
LO           &                   &                  &  &   & $C_{7\gamma}^{eff}$ & $C_{8g}^{eff}$    \\
\hline
             &                   &                  &  &   & -0.318              & -0.151     \\
\hline
\hline
\hline
     NLO      &  $C_{1}$   &  $C_{2}$  & $C_{3}$  &  $C_{4}$ & $C_{5}$ & $C_{6}$   \\
\hline
\hline
\hline
$\mu= m_{b}$ &  1.150     &  -0.312    & 0.017    &  0.037  &  0.0104  & -0.0045   \\
\hline
   NLO       &  $C_{7}/\alpha$   &  $C_{8}/\alpha$  & $C_{9}/\alpha$  &  $C_{10}/\alpha$  &         &  \\
\hline
$\mu= m_{b}$ & -0.00135            & 0.049            &  -1.302        &   0.252           &         &   \\
\hline
\hline
\end{tabular}
\end{center}
\caption{Upper table: Wilson coefficients $C_i$ in the NDR scheme. 
Input parameters are $\Lambda^{(5)}_{\overline{\rm
MS}}=0.225$\,$\mathrm{GeV}$,  
$m_t(m_t)=167$\,$\mathrm{GeV}$, $m_b(m_b)=4.2$\,$\mathrm{GeV}$, $M_W=80.4$\,$\mathrm{GeV}$, $\alpha=1/129$, 
and $\sin^2\!\theta_W=0.23$. Lower table: Wilson coefficients $C_i$ in naive factorization.}
\label{tab1}
\end{table}}
{\renewcommand\baselinestretch{0.95}
\begin{table}[hpb]
\begin{center}
\begin{tabular}{|l|ccc|}
\hline\hline
\multicolumn{1}{|c|}{Mode} & BABAR & BELLE & CLEO \\
\hline\hline
\hline
$B^- \to \eta K^-$
 & $2.8_{\,-0.7}^{\,+0.8}\pm 0.2$
 & $5.3_{\,-1.5}^{\,+1.8}\pm 0.6$ 
 & $2.2_{\,-2.2}^{\,+ 2.8}$ \\
$\bar{B}^0 \to \eta\bar K^0$
 & $2.6_{\,-0.8}^{\,+0.9}\pm 0.2$ 
 & $<12$ 
 & $<9.3$ \\
$B^- \to \eta' K^-$
 & $76.9\pm 3.5\pm 4.4$ 
 & $78\pm 6\pm 9$ 
 & $80_{\,-\phantom{1}9}^{\,+10}\pm 7$  \\
$\bar{B}^0 \to  \eta'\bar K^0$
 & $55.4\pm 5.2\pm 4.0$ 
 & $68\pm 10_{\,-8}^{\,+ 9}$ 
 & $89_{\,-16}^{\,+18}\pm 9$\\
\hline
$B^- \to K^-\phi$
 & $10.0_{\,-0.8}^{\,+0.9}\pm 0.5$ 
 & $9.4\pm 1.1\pm 0.7$ 
 & $5.5_{\,-1.8}^{\,+2.1}\pm 0.6$ \\
$\bar{B}^0 \to \bar K^0\phi$
 & $7.6_{\,-1.2}^{\,+1.3}\pm 0.5$ 
 & $9.0_{\,-1.8}^{\,+2.2}\pm 0.7$ 
 & $5.4_{\,-2.7}^{\,+3.7}\pm 0.7$  \\
$B^- \to K^- K^{*0}$
 & ---
 & ---
 & $<5.3$ \\
\hline
$B^- \to K^- K^0$
 & $<2.2$ 
 & $<3.4$ 
 & $<3.3$ \\
$\bar{B}^0 \to \bar K^0 K^0$
 & $<1.6$ 
 & $<3.2$ 
 & $<3.3$ \\
$\bar{B}^0 \to K^- K^+$
 & $<0.6$ 
 & $<0.7$ 
 & $<0.8$ \\
\hline
$B^- \to  \bar K^0\rho^-$
 & ---
 & ---
 & $<48$  \\
$B^- \to  K^-\rho^0$
 & $<6.2$ 
 & $<12$ 
 & $<17$  \\
$\bar{B}^0 \to K^-\rho^+$
 & $7.3_{\,-1.2}^{\,+1.3}\pm 1.3$ 
 & $15.1_{\,-3.3\,-1.5}^{\,+3.4\,+1.4}$ 
 & $16.0_{\,-6.4}^{\,+7.6}\pm 2.8$  \\
$\bar{B}^0 \to \bar K^0\rho^0$
 & ---
 & $<12$ 
 & $<39$ \\
\hline
$B^- \to  K^-\omega$
 & $5.0\pm 1.0\pm 0.4$ 
 & $6.7_{\,-1.2}^{\,+1.3}\pm 0.6$ 
 & $3.2_{\,-1.9}^{\,+2.4}\pm 0.8$   \\
$\bar{B}^0 \to \bar K^0\omega$
 & $5.3_{\,-1.2}^{\,+1.4}\pm 0.5$
 & $4.0_{\,-1.6}^{\,+1.9}\pm0.5$  
 & $10.0_{\,-4.2}^{\,+5.4}\pm 1.4$  \\
\hline\hline
\end{tabular}
\end{center}
\caption{Experimental branching ratio data (in units of $10^{-6}$) from the BELLE, CLEO and BABAR $B$-factories for the 
$B \to K  X$ channel where $X$ stands for $\eta^{(\prime)}, \omega, \phi, \rho$ and $K^{(*)}$. }
\label{tab2}
\end{table}}
\newpage
{\renewcommand\baselinestretch{1.1}
\begin{table}
\begin{center}
\begin{tabular}{|l|ccc|}
\hline\hline
\multicolumn{1}{|c|}{Mode} & BABAR & BELLE & CLEO  \\
\hline\hline
\hline
$B^- \to \pi^-\rho^0$
 & $9.3_{\, -1.0}^{\, +1.0}\pm 0.8$ 
 & $8.0_{\,-2.0}^{\,+2.3}\pm 0.7$ 
 & $10.4_{\,-3.4}^{\,+3.3}\pm 2.1$ \\
$B^- \to  \pi^0\rho^-$
 & $11.0_{\, -1.9}^{\, +1.9}\pm 1.9$
 & ---
 & $<43$  \\
$\bar{B}^0 \to \pi^+\rho^-$
 & $13.9 \pm 2.7$
 & ---
 & --- \\
$\bar{B}^0 \to  \pi^-\rho^+$
 & $8.9 \pm 2.5$ 
 & ---
 & --- \\
$\bar{B}^0 \to  \pi^\pm\rho^\mp$
 & $22.6_{\, -1.8}^{\, +1.8}\pm 2.2$ 
 & $29.1_{\,-4.9}^{\,+5.0}\pm 4.0$ 
 & $27.6_{\,-7.4}^{\,+8.4}\pm 4.2$ \\
$\bar{B}^0 \to  \pi^0\rho^0$
 & $<2.5$ 
 & $6.0_{\,-2.3}^{\,+2.9}\pm 1.2$ 
 & $<5.5$\\
\hline
$B^- \to  \pi^-\omega$
 & $5.4_{\, -1.0}^{\,+1.0}\pm 0.5$ 
 & $5.7_{\,-1.3}^{\,+1.4}\pm 0.6$ 
 & $11.3_{\,-2.9}^{\,+3.3}\pm 1.4$\\
$\bar{B}^0 \to  \pi^0\omega$
 & $<3$ 
 & $<1.9$ 
 & $<5.5$  \\
\hline
$B^- \to  \pi^-\phi$
 & $<0.41$ 
 & ---
 & $<5$ \\
$\bar{B}^0 \to  \pi^0\phi$
 & ---
 & ---
 & $<5$\\
\hline
$B^- \to  \pi^-\bar K^{*0}$
 & $15.5\pm 1.8_{\,-3.2}^{\,+1.5}$ 
 & $19.4_{\,-3.9\,-2.1}^{\,+4.2\,+2.1}$ 
 & $7.6_{\,-3.0}^{\,+3.5}\pm 1.6$ \\
$B^- \to  \pi^0 K^{*-}$
 & ---
 & ---
 & $<31$ \\
$\bar{B}^0 \to \pi^+ K^{*-}$
 & ---
 & $14.8_{\,-4.4\,-1.0}^{\,+4.6\,+1.5}$ 
 & $16_{\,-5}^{\,+6}\pm 2$\\
$\bar{B}^0 \to \pi^0\bar K^{*0}$
 & ---
 & ---
 & $<3.6$ \\
\hline
$B^- \to  \pi^-\pi^0$
 & $5.5_{\,-0.9}^{\,+1.0}\pm 0.6$ 
 & $5.3_{\, -1.3}^{\, +1.3}\pm 0.5$ 
 & $4.6_{\,-1.6\,-0.7}^{\,+1.8\,+0.6}$  \\
$\bar{B}^0 \to \pi^+\pi^-$
 & $4.7_{\, -0.6}^{\, +0.6}\pm 0.2$ 
 & $4.4_{\, -0.6}^{\, +0.6}\pm 0.3$ 
 & $4.5_{\,-1.2\,-0.4}^{\,+1.4\,+0.5}$  \\
 $\bar{B}^0 \to \pi^0\pi^0$
 & $1.6_{\,-0.6\,-0.3}^{\,+0.7\,+0.6}$ 
 & $1.8_{\,-1.3\,-0.7}^{\,+1.4\,+0.5}$ 
 & --- \\
\hline
$B^- \to   \pi^-\eta$
 & $4.2_{\,-0.9}^{\,+1.0}\pm 0.3$ 
 & $5.2_{\,-1.7}^{\,+ 2.0}\pm 0.6$ 
 & $1.2_{\,-1.2}^{\,+ 2.8}$ ($<5.7$)  \\
$\bar{B}^0 \to \pi^0\eta$
 & ---
 & ---
 & $<2.9$  \\
$B^- \to   \pi^-\eta'$
 & $<12$ 
 & $<7$ 
 & $<12$ \\
$\bar{B}^0 \to \pi^0\eta'$
 & ---
 & ---
 & $<5.7$ \\
\hline\hline
\end{tabular}
\end{center}
\caption{Experimental branching ratio data (in units of $10^{-6}$) from the BELLE, CLEO and BABAR $B$-factories for 
the $B \to \pi  X$ channel where $X$ stands for $\eta^{(\prime)}, \omega, \phi, \rho, \pi$ and $K^{*}$.}
\label{tab3}
\end{table}}
\begin{table}
\begin{center}
\begin{tabular}{|l|ccc|}
\hline\hline
\multicolumn{1}{|c|}{Mode} & BABAR & BELLE & CLEO \\
\hline\hline
\hline
$B^- \to  \pi^-\bar K^0$
 & $20.0\pm 1.6\pm 1.0$ 
 & $22.0\pm 1.9\pm 1.1$ 
 & $18.8_{\,-3.3\,-1.8}^{\,+3.7\,+2.1}$ \\
$B^- \to  \pi^0 K^-$
 & $12.8_{\,-1.1}^{\,+ 1.2}\pm 1.0$ 
 & $12.8\pm 1.4_{\,-1.0}^{\,+1.4}$ 
 & $12.9_{\,-2.2\,-1.1}^{\,+2.4\,+1.2}$ \\
$\bar{B}^0 \to  \pi^+ K^-$
 & $17.9\pm 0.9\pm 0.7$ 
 & $18.5\pm 1.0\pm 0.7$ 
 & $18.0_{\,-2.1\,-0.9}^{\,+2.3\,+1.2}$ \\
$\bar{B}^0 \to  \pi^0\bar K^0$
 & $10.4\pm 1.5\pm 0.8$ 
 & $12.6\pm 2.4\pm 1.4$ 
 & $12.8_{\,-3.3\,-1.4}^{\,+4.0\,+1.7}$  \\
\hline
\hline
\end{tabular}
\end{center}
\caption{Experimental branching ratio data (in units of $10^{-6}$) from the BELLE, CLEO and BABAR $B$-factories 
for the $B \to \pi K$ channel.}
\label{tab4}
\end{table}
{\renewcommand\baselinestretch{1.70}
\begin{table}
\begin{center}
\begin{tabular}{|l|ccc|}
\hline\hline
\multicolumn{1}{|c|}{Mode} & BABAR & BELLE & CLEO \\
\hline\hline
\hline
$\frac{\tau^{B^+}}{2 \tau^{B^0}} \Bigl[ \frac{B^0 \to \pi^+ \pi^-}{B^+ \to \pi^+ \pi^0}\Bigr]$
 & $0.46 \pm 0.11$ 
 & $0.44 \pm 0.14$ 
 & $0.52 \pm 0.26$ \\
$\frac{\tau^{B^0}}{\tau^{B^+}} \Bigl[ \frac{B^- \to \pi^- \pi^0}{B^0 \to\pi^0 \pi^0}\Bigr]$
 & $3.16\pm 1.68$
 & $2.71 \pm 2.38$
 & $---$  \\
\hline
$\Bigl[ \frac{2 B^{\pm} \to \pi^0 K^\pm}{  B^{\pm} \to \pi^\pm K^0}\Bigr]$
 & $1.28 \pm 0.18$ 
 & $1.16 \pm 0.20$ 
 & $1.36 \pm 0.40$  \\
$\frac{\tau^{B^+}}{\tau^{B^0}} \Bigl[ \frac{B^0 \to \pi^\pm K^\mp}{ B^{\pm} \to \pi^\pm K^0}\Bigr]$
 & $0.94 \pm 0.20$ 
 & $ 0.91 \pm 0.10$ 
 & $ 1.02 \pm 0.25$ \\
$\Bigl[ \frac{B^0 \to \pi^\mp K^\pm}{ B^0 \to \pi^0 K^0}\Bigr]$
 & $1.72 \pm 0.44$ 
 & $1.46 \pm 0.34$ 
 & $1.40 \pm 0.44$  \\
$\frac{\tau^{B^0}}{\tau^{B^+}} \Bigl[ \frac{B^- \to \pi^- \bar{K}^{*0}}{ B^0 \to \pi^+ K^{*-}}\Bigr]$
 & $ --- $ 
 & $ 1.21 \pm 0.58$ 
 & $ 0.44 \pm 0.25$  \\
\hline
$\frac{\tau^{B^0}}{\tau^{B^+}} \Bigl[ \frac{B^- \to K^- \phi}{B^0 \to \bar{K}^0 \phi}\Bigr]$
 &$1.21 \pm 0.24$ 
 &$0.95 \pm 0.26$
 & $0.93 \pm 0.66$\\
$\frac{\tau^{B^0}}{\tau^{B^+}} \Bigl[ \frac{ B^- \to K^- \eta^{\prime} }{B^0 \to \bar{K}^0 \eta^{\prime} }\Bigr]$
 &$1.38 \pm 0.15$ 
 &$1.14 \pm 0.24$
 &$0.89 \pm 0.20$ \\
\hline
$\frac{\tau^{B^0}}{\tau^{B^+}} \Bigl[ \frac{B^- \to K^- \omega}{B^0 \to \bar{K}^0 \omega}\Bigr]$
 & $0.86 \pm 0.29$
 & $1.53 \pm 0.75$
 & $0.29 \pm 0.26$  \\
$\frac{\tau^{B^+}}{\tau^{B^0}} \Bigl[ \frac{B^0 \to \pi^\pm \rho^\mp}{ B^- \to \pi^- \rho^0}\Bigr]$
 & $2.63 \pm 0.47$ 
 & $3.93 \pm 1.37$ 
 & $2.87 \pm 1.52$ \\
\hline\hline
\end{tabular}
\end{center}
{\renewcommand\baselinestretch{1.0}
\caption{Experimental data from the BELLE, CLEO and BABAR $B$-factories for the ratios between branching 
ratios involving pions (first case), pion and kaon (second case), kaon and $\phi$ or $\eta$ (third case) and 
kaon $\omega$ or  pion $\rho$ (fourth case).} }
\end{table}}

\begin{table}
\begin{center}
\begin{tabular}{|l|ccccc||ccccc|}
\hline\hline
\multicolumn{1}{|c|}{Meson} & & & set1 & & & & & set2 &  &   \\
\hline
\multicolumn{1}{|c|}{} & $\varrho_H$ & $\varphi_H$ &$\hspace{-5.0cm}$& $\varrho_A$ & $\varphi_A$  & $\varrho_H$   
&  $\varphi_H$   &&  $\varrho_A$  &  $\varphi_A$ \\
\hline\hline
\hline
$ K \equiv \pi$
 & 1.14
 & -0.75
 & 
 & 2.94
 & 2.38 
 & 2.58
 & 0.12
 & 
 & -1.74
 & 1.62\\
$ K^{*}$
 & 2.82 
 & -1.76
 &
 & -0.66
 & 1.12
 & 2.94
 & -2.01
 &
 & 0.90
 & 1.62\\
\hline
$\rho$
 & 2.94 
 & 2.13
 &
 & 0.18
 & 2.63
 & 2.94
 & -2.64
 &
 & 0.42
 & -2.89\\
$\omega$
 & 1.26 
 & 0.62 
 &
 & -0.78
 & -2.14
 & 2.10
 & 0.37 
 &
 & 1.50 
 & -2.26 \\
\hline
$\eta \equiv \eta^{\prime}$
 & 2.94 
 & 2.46
 &
 & 2.94
 & -2.01 
 & 2.58
 & -0.88 
 &
 & -2.82
 &  -1.26\\
$\phi$
 & 1.00  
 & 0.707
 &
 & -2.82
 &  -2.26
 & 1.00
 & 0.707 
 &
 & +0.66
 & 0.49\\
\hline\hline
\end{tabular}
\end{center}
\caption{Phases (given in radian) $\varphi_{A,H}$ and parameters $\varrho_{A,H}$ for the annihilation and hard-spectator 
scattering contributions, respectively, for $K, K^*, \pi, \rho, \omega, \eta^{(\prime)}, \phi$ and 
 determined for the  $B \to K X$ and $B \to \pi X$ channels. Set1 and set2 
correspond to the maximal and minimal values taken by the CKM parameters $\rho$ and $\eta$.}
\label{tab6}
\end{table}

{\renewcommand\baselinestretch{0.93}
\begin{table}
\begin{center}
\begin{tabular}{|l|ccccc|}
\hline\hline
\multicolumn{1}{|c|}{Mode} & $\langle {\rm Exp.\;  BR.}\rangle$ &  ${\rm BR.}^{NF}$ &  ${\rm BR.}^{QCDF}$ &  ${\rm BR.}^{QCDF,w.a.}$ 
&  ${\rm BR.}^{QCDF,w.h.}$ \\
\hline\hline
\hline
$B^- \to \pi^-\rho^0$
 & $9.2$ 
 & $12.3$ 
 & $19.2$ 
 & $20.9$
 & $10.8$ \\
$B^- \to \pi^0\rho^-$
 & $11.0$
 & $21.2$
 & $12.1$  
 & $12.7$
 & $11.2$\\
$\bar{B}^0 \to \pi^+\rho^-$
 & $13.9$
 & $36.6$
 & $15.8$ 
 & $17.2$
 & $16.1$\\
$\bar{B}^0 \to \pi^-\rho^+$
 & $8.9$ 
 & $19.9$
 & $18.0$
 & $19.8$
 & $19.6$\\
$\bar{B}^0 \to \pi^\pm\rho^\mp$
 & $26.4$ 
 & $28.3$ 
 & $23.9$ 
 & $26.2$
 & $25.6$\\
$\bar{B}^0 \to \pi^0\rho^0$
 & $4.6$ 
 & $0.15$ 
 & $1.8$  
 & $1.0$
 & $0.4$\\
\hline
$B^- \to \pi^-\omega$
 & $7.4$ 
 & $9.4$ 
 & $14.6$ 
 & $14.5$
 & $8.4$\\
$\bar{B}^0 \to \pi^0\omega$
 & $< 3.4$ 
 & $1.3\; 10^{-2}$ 
 & $0.5$  
 & $0.31$
 & $5.4\; 10^{-2}$\\
\hline
$B^- \to \pi^-\phi$
 & $<2.7$ 
 & $1.6\; 10^{-7}$
 & $5.3\; 10^{-4}$ 
 & $5.3\; 10^{-4}$
 & $3.2\; 10^{-5}$\\
$\bar{B}^0 \to \pi^0\phi$
 & $<5.0$
 & $7.5\; 10^{-7}$
 & $2.5\; 10^{-4}$
 & $2.45\; 10^{-4}$
 & $1.4\; 10^{-5}$\\
\hline
$B^- \to \pi^-\bar K^{*0}$
 & $14.1$ 
 & $18.7$  
 & $13.9$
 & $2.9$
 & $15.8$\\
$B^- \to \pi^0 K^{*-}$
 & $<31$
 & $11.0$
 & $11.5$
 & $3.5$
 & $10.2$\\
$\bar{B}^0 \to \pi^+ K^{*-}$
 & $15.4$
 & $15.9$ 
 & $14.4$
 & $2.7$
 & $14.9$\\
$\bar{B}^0 \to \pi^0\bar K^{*0}$
 & $<3.6$
 & $6.5$
 & $3.5$
 & $0.25$
 & $5.1$\\
\hline
$B^- \to \pi^-\pi^0$
 & $5.1$ 
 & $3.9$ 
 & $5.5$
 & $5.5$
 & $3.6$\\
$\bar{B}^0 \to \pi^+\pi^-$ 
 & $4.6$ 
 & $7.8$  
 & $5.1$
 & $6.6$
 & $5.6$\\
$\bar{B}^0 \to \pi^0\pi^0$
 & $1.7$ 
 & $2.9$ 
 & $1.8$
 & $5.2$
 & $0.7$\\
\hline
$B^- \to \pi^-\eta$
 & $3.5$ 
 & $7.4$ 
 & $7.1$
 & $7.4$
 & $5.4$\\
$\bar{B}^0 \to \pi^0\eta$
 & $<2.9$
 & $0.6$
 & $0.43$  
 & $0.2$
 & $0.6$\\
$B^- \to \pi^-\eta'$ 
 & $<10.3$ 
 & $9.9$ 
 & $9.5$
 & $10.1$
 & $6.6$\\
$\bar{B}^0 \to \pi^0\eta'$
 & $<5.7$
 & $0.7$  
 & $0.5$
 & $0.25$
 & $0.7$\\
\hline
\hline
\hline
$B^- \to \eta K^-$
 & $3.4$ 
 & $2.0$ 
 & $2.3$ 
 & $0.7$
 & $3.45$\\
$\bar{B}^0 \to\eta\bar K^0$ 
 & $<7.9$ 
 & $1.1$ 
 & $1.55$
 & $0.3$
 & $2.4$\\
$B^- \to  \eta' K^-$ 
 & $78.3$ 
 & $61.3$  
 & $91.2$
 & $37.8$
 & $71.2$\\
$\bar{B}^0 \to \eta'\bar K^0$ 
 & $70.8$ 
 & $44.0$
 & $75.1$
 & $32.5$
 & $52.0$\\
\hline
$B^- \to K^-\phi$ 
 & $8.3$ 
 & $24.2$ 
 & $8.9$
 & $3.7$
 & $7.4$\\
$\bar{B}^0 \to \bar K^0\phi$
 & $7.3$ 
 & $26.2$  
 & $7.2$
 & $3.2$
 & $6.1$\\
\hline
$B^- \to K^- K^{\ast 0}$
 & $<5.3$
 & $1.0$ 
 & $0.35$
 & $9.6\; 10^{-2}$
 & $0.39$\\
$\bar{B}^0 \to K^- K^{\ast +}$
 & $---$
 & $0.0$ 
 & $2.3\; 10^{-2}$
 & $0.0$
 & $2.1\; 10^{-2}$\\
$\bar{B}^0 \to \bar{K}^0 K^{\ast 0}$
 & $---$
 & $1.0$ 
 & $0.31$
 & $9.6\; 10^{-2}$
 & $0.36$\\
$B^- \to K^{\ast -} K^0$
 & $---$
 & $3.2\; 10^{-2}$ 
 & $0.9$
 & $4.2\; 10^{-2}$
 & $0.9$\\
$\bar{B}^0 \to K^{\ast -} K^+$
 & $---$
 & $0.0$ 
 & $0.75$
 & $0.0$
 & $0.7$\\
$\bar{B}^0 \to \bar{K}^{\ast 0} K^0$
 & $---$
 & $3.2\; 10^{-2}$ 
 & $0.60$
 & $4.2\; 10^{-2}$
 & $0.66$\\
\hline
\hline
\end{tabular}
\end{center}
{\renewcommand\baselinestretch{1.0}
\caption{Theoretical branching ratios (in units of $10^{-6}$) for the $B \to \pi  X$ (top) and $B \to K  X$ (bottom) channels
where $X$ stands for $\eta^{(\prime)}, \omega, \phi, \rho, \pi$ and $K^{*}$. Values are given for naive
factorization, ${\rm BR}^{NF}$, QCD factorization,  ${\rm BR}^{QCDF}$, without annihilation contribution, ${\rm BR}^{QCDF,w.a.}$, 
  without hard scattering spectator contribution, ${\rm BR}^{QCDF,w.h.}$,  and for an average value of the form 
factors and CKM matrix elements.}}
\label{tab7}
\end{table}}
{\renewcommand\baselinestretch{1.3}
\begin{table}
\begin{center}
\begin{tabular}{|l|ccccc|}
\hline \hline
\multicolumn{1}{|c|}{Mode} & $\langle {\rm Exp. \; BR.}\rangle$ &  ${\rm BR.}^{NF}$ &  ${\rm BR.}^{QCDF}$ &  ${\rm BR.}^{QCDF,w.a.}$ 
&  ${\rm BR.}^{QCDF,w.h.}$ \\
\hline\hline
\hline
$B^- \to K^- K^0$
 & $<2.9$  
 & $0.7$ 
 & $0.41$
 & $0.25$
 & $0.45$\\
$\bar{B}^0 \to \bar K^0 K^0$ 
 & $<2.7$ 
 & $0.0$ 
 & $0.27$
 & $0.25$
 & $0.28$\\
$\bar{B}^0 \to K^- K^+$
 & $<0.7$ 
 & $0.7$ 
 & $0.21$
 & $0.0$
 & $0.2$\\
\hline
$\bar{B}^0 \to \bar K^0\rho^-$
 & $<48$
 & $0.6$  
 & $0.47$
 & $0.6$
 & $0.4$\\
$B^- \to K^-\rho^0$
 & $<11.7$ 
 & $0.8$  
 & $1.54$
 & $1.1$
 & $1.0$\\
$\bar{B}^0 \to K^-\rho^+$ 
 & $12.8$ 
 & $3.1$  
 & $9.2$
 & $3.0$
 & $8.2$\\
$\bar{B}^0 \to \bar K^0\rho^0$
 & $<25.5$ 
 & $1.5$ 
 & $5.1$
 & $1.3$
 & $3.1$\\
\hline
$B^- \to K^-\omega$ 
 & $4.9$ 
 & $1.0$   
 & $4.9$
 & $1.6$
 & $5.8$\\
$\bar{B}^0 \to \bar K^0\omega$
 & $6.4$  
 & $9.7\; 10^{-2}$  
 & $6.34$
 & $0.3$
 & $7.8$\\
$B^- \to  \pi^-\bar K^0$
 & $20.2$ 
 & $24.2$ 
 & $20.3$
 & $10.1$
 & $20.8$ \\
$B^- \to  \pi^0 K^-$
 & $12.8$ 
 & $13.1$ 
 & $10.0$  
 & $6.7$ 
 & $11.3$\\
$\bar{B}^0 \to  \pi^+ K^-$
 & $18.1$ 
 & $22.6$  
 & $20.1$  
 & $9.2$
 & $21.7$ \\
$\bar{B}^0 \to  \pi^0\bar K^0$
 & $11.9$ 
 & $9.4$  
 & $12.0$ 
 & $3.1$ 
 & $11.6$ \\
\hline\hline\hline
$\frac{\tau^{B^+}}{2 \tau^{B^0}} \Bigl[ \frac{B^0 \to \pi^+ \pi^-}{B^+ \to \pi^+ \pi^0}\Bigr]$
 & $0.47$ 
 & $0.9$ 
 & $0.4$ 
 & $0.55$ 
 & $0.75$\\
$\frac{\tau^{B^0}}{\tau^{B^+}} \Bigl[ \frac{B^- \to \pi^- \pi^0}{B^0 \to\pi^0 \pi^0}\Bigr]$
 & $2.91$
 & $1.5$ 
 & $3.1$
 & $0.95$
 & $5.2$\\
\hline
$\Bigl[ \frac{2 B^{\pm} \to \pi^0 K^\pm}{  B^{\pm} \to \pi^\pm K^0}\Bigr]$
 & $1.26$ 
 & $1.1$ 
 & $0.65$  
 & $1.2$
 & $0.71$\\
$\frac{\tau^{B^+}}{\tau^{B^0}} \Bigl[ \frac{B^0 \to \pi^\pm K^\mp}{ B^{\pm} \to \pi^\pm K^0}\Bigr]$
 & $0.95$ 
 & $0.8$  
 & $0.65$
 & $0.8$
 & $0.6$\\
$\Bigl[ \frac{B^0 \to \pi^\mp K^\pm}{ B^0 \to \pi^0 K^0}\Bigr]$
 & $1.52$ 
 & $2.2$ 
 & $1.32$  
 & $2.3$
 & $1.4$\\
$\frac{\tau^{B^0}}{\tau^{B^+}} \Bigl[ \frac{B^- \to \pi^- \bar{K}^{*0}}{ B^0 \to \pi^+ K^{*-}}\Bigr]$ 
 & $0.82$ 
 & $1.3$  
 & $1.04$
 & $1.15$
 & $1.2$\\
\hline
$\frac{\tau^{B^0}}{\tau^{B^+}} \Bigl[ \frac{B^- \to K^- \phi}{B^0 \to \bar{K}^0 \phi}\Bigr]$
 & $1.03$ 
 & $1.0$
 & $1.0$
 & $1.0$
 & $1.0$\\
$\frac{\tau^{B^0}}{\tau^{B^+}} \Bigl[ \frac{ B^- \to K^- \eta^{\prime} }{B^0 \to \bar{K}^0 \eta^{\prime} }\Bigr]$
 & $1.13$ 
 & $1.5$
 & $1.34$ 
 & $1.3$
 & $1.5$\\
\hline
$\frac{\tau^{B^0}}{\tau^{B^+}} \Bigl[ \frac{B^- \to K^- \omega}{B^0 \to \bar{K}^0 \omega}\Bigr]$
 & $0.89$
 & $11.7$  
 & $0.83$
 & $5.3$
 & $0.8$\\
$\frac{\tau^{B^+}}{\tau^{B^0}} \Bigl[ \frac{B^0 \to \pi^\pm \rho^\mp}{ B^- \to \pi^- \rho^0}\Bigr]$ 
 & $3.14$ 
 & $2.1$ 
 & $1.1$
 & $1.16$
 & $2.1$\\
\hline
\hline
\end{tabular}
\end{center}
{\renewcommand\baselinestretch{1.0}
\caption{Theoretical branching ratios for the $B \to K  X$ (top, in units of $10^{-6}$) channel and  
ratios (bottom) between
$B \to K  X$ and $B \to \pi  X$ channels.  Values are given for naive factorization, ${\rm BR}^{NF}$, QCD 
factorization,  ${\rm BR}^{QCDF}$, without annihilation contribution, ${\rm BR}^{QCDF,w.a.}$, 
  without hard scattering spectator contribution, ${\rm BR}^{QCDF,w.h.}$, and for an average value of the form 
factors and CKM matrix elements.}}
\label{tab8}
\end{table}}
\end{document}